\pdfoutput=1
\documentclass[11pt,twoside,a4paper,cmspaper,final,collab]{cms-tdr}
\def\svnVersion{0de5b5f}\def\svnDate{2023/10/02}\def\cmsCernNoTag{CERN-EP-2022-028}\def\cmsCernDate{\today}\def\cmsMessage{Published in Physical Review D as \href{http://dx.doi.org/10.1103/PhysRevD.108.052002}{\doi{10.1103/PhysRevD.108.052002}.}}
\begin{document}\cmsNoteHeader{EGM-20-001}

\renewcommand{\Pa}{\ensuremath{\mathcal{A}}\xspace}
\newcommand{\Pepem}{\Pep\Pem}
\newcommand{\ptagen}{\ensuremath{p_{\mathrm{T,\,\Pa}}}\xspace}
\newcommand{\ptG}{\ensuremath{p_{\mathrm{T},\,\Gamma}}\xspace}
\newcommand{\etaG}{\eta_{\Gamma}}
\newcommand{\etaagen}{\ensuremath{\eta_{\Pa}}\xspace}
\newcommand{\agg}{\Pa\to\gamma\gamma}
\newcommand{\hgg}{\PH\to\gamma\gamma}
\newcommand{\haa}{\PH\to \Pa\Pa}
\newcommand{\hag}{\PH\to \Pa\Pa \to 4 \gamma}
\newcommand{\pgg}{\Pgpz\to\gamma\gamma}
\newcommand{\egg}{\Pgh\to\gamma\gamma}
\newcommand{\boost}{\gamma_{\mathrm{L}}}
\newcommand{\boostav}{\langle \boost \rangle}
\newcommand{\dphi}{\Delta\varphi}
\newcommand{\deta}{\Delta\eta}
\newcommand{\Ea}{E_{\Pa}}
\newcommand{\ma}{m_{\Pa}}
\newcommand{\mG}{m_{\Gamma}}
\newcommand{\tbtit}{\ensuremath{3{\times}3}\xspace}
\newcommand{\tbt}{\ensuremath{3{\times}3}\xspace}
\newcommand{\fbf}{\ensuremath{5{\times}5}\xspace }
\newcommand{\imgwindow}{\ensuremath{32{\times}32}\xspace}
\newcommand{\magen}{m_{\Pa}}
\newcommand{\mapred}{m_{\Gamma}}
\newcommand{\maroi}{\ensuremath{m_{\Pa}\text{-ROI}}\xspace}
\newcommand{\zee}{\PZ \to \Pepem}
\newcommand{\gjet}{\ensuremath{\gamma+\text{jet}}\xspace}
\newcommand{\scale}{\ensuremath{s_{\text{scale}}}\xspace}
\newcommand{\smear}{\ensuremath{s_{\text{smear}}}}

\hyphenation{had-ron-i-za-tion}
\hyphenation{cal-or-i-me-ter}
\hyphenation{de-vices}
\hyphenation{de-tec-tor}
\hyphenation{extra-po-la-tion}

\cmsNoteHeader{EGM-20-001}
\title{Reconstruction of decays to merged photons 
using {end-to-end} deep learning with domain continuation 
in the CMS detector
}

\date{\today}

\abstract{
A novel technique based on machine learning is introduced to reconstruct the decays of highly Lorentz-boosted particles. Using an \textit{end-to-end} deep learning strategy, the technique bypasses existing rule-based particle reconstruction methods typically used in high energy physics analyses. It uses minimally processed detector data as input and directly outputs particle properties of interest. The new technique is demonstrated for the reconstruction of the invariant mass of particles decaying in the CMS detector. The decay of a hypothetical scalar particle $\Pa$ into two photons, $\agg$, is chosen as a benchmark decay. Lorentz boosts $\boost = 60$--600 are considered, ranging from regimes where both photons are resolved to those where the photons are closely merged as one object. A training method using domain continuation is introduced, enabling the invariant mass reconstruction of unresolved photon pairs in a novel way. The new technique is validated using $\pgg$ decays in LHC collision data.
}

\hypersetup{
pdfauthor={CMS Collaboration},
pdftitle={Reconstruction of decays to merged photons using end-to-end deep learning with domain continuation in the CMS detector},
pdfsubject={CMS},
pdfkeywords={CMS,  machine learning, merged photons, boosted decays, mass regression}}

\maketitle 

\section{Introduction}
\label{sec:Intro}

Since the standard model (SM) of particle physics continues
to show excellent agreement with measurements performed at particle colliders, such as the CERN LHC, the search for physics beyond the standard model (BSM) has led experiments, such as CMS, to look beyond simple decay topologies toward more exotic, potentially overlooked ones. An important class of such exotic signatures is composed of Lorentz-boosted particles that are sufficiently energetic to induce collimation of their decay products~\cite{boostedrev}. An important consideration in developing a robust search program for such phenomena is understanding whether existing particle reconstruction algorithms are sensitive to the decays of boosted particles.

Particle-flow (PF) algorithms~\cite{pf,atlaspf} are widely used
frameworks for the reconstruction of particle decays in the LHC
detectors. These algorithms reconstruct and identify each individual
particle in an event using an optimized combination of information from
the various detector elements. Although PF algorithms have been
successfully used in previous CMS analyses, they lean heavily on the
assumption that particle decays are well resolved and well isolated, a feature not generally true for boosted particle decays that often exhibit some degree of merging.

In its loosest sense, merging may refer to the collimation of particles that are otherwise individually resolved but are in overlapping ensembles. For instance, in the hadronic decay of a boosted high-mass resonance to jets, the jets overlap to form a single, large-radius jet. In such cases, there is an ambiguity in how to group the decays so as to consistently reconstruct the properties of the parent particle. Although such \textit{ensemble merging} may be mitigated, the corresponding analyses require a fine-tuned strategy for clustering, often at the cost of losing reconstruction efficiency in some regions of phase space. Therefore, attempts to directly probe the mass of exotic resonances are rarely pursued in the boosted regime, unlike for the reconstruction of known SM resonances~\cite{cmsboostedtop,boostedhcms,atlasboostedtop,boostedhatlas}.

At higher boosts, a more experimentally challenging form of merging arises when the separation between the decay products approaches the detector resolution. Consider, for instance, decay products interacting with the CMS electromagnetic calorimeter (ECAL). At separations approaching the Moli\`ere radius of the calorimeter material, the particle showers of the decay products begin to overlap. This effect can cause the merged decay products to be misreconstructed as a single-particle candidate. Such \textit{shower merging}, if within a few Moli\`ere radii, might still be discernible by current dedicated shower clustering tools, even in regimes inaccessible to PF. The \tbt clustering algorithm, used for reconstructing low-energy $\pgg$ decays~\cite{CMS:EGM-14-001}, is a notable example. However, at separations approaching the Moli\`ere radius, such clustering tools are unable to discern distinct clusters and begin to lose their sensitivity. Thus, they are difficult to adapt to exotic searches that typically probe a wide range of particle masses and Lorentz boosts.

A special case of shower merging occurs when the decay products are collimated to the point of depositing their energy within the same calorimeter cell, or less than the Moli\`ere radius. In this limit of fully overlapping particle showers, the resulting shower pattern is nearly indistinguishable from that of a true, single-particle shower. The only distinction in such cases is the slightly greater spreading in the particle shower along the principal axis connecting the points where the decay products enter the calorimeter. Reconstructing such an \textit{instrumentally merged} decay remains a challenge for existing experimental tools. Such a tool must discern subtle differences in the energy distribution of the fully merged particle shower.

As a concrete example of such a signature, we consider the exotic decay
$\PH/\PX\to \Pa\Pa$ of the Higgs boson $\PH$ with a mass of about
125\GeV, or some new resonance $\PX$, to a pair of hypothetical scalar
particles $\Pa$. Such decays are well motivated in extended Higgs
sectors or BSM models involving two-Higgs doublets, an additional
singlet, or axionlike particle production~\cite{hexo,exo,axioncollider}. Depending on the mass and decay mode of the $\Pa$, one or more forms of shower or instrumental merging, as described above, may occur in its decay. In particular, for decays to diphotons $\agg$~\cite{bogdanhaa}, since photons are massless, the particle $\Pa$ can be arbitrarily light and the resulting diphoton system correspondingly merged. For the mass of the $\Pa$ satisfying $\ma \lesssim 0.4\GeV$, the $\hag$ signal is increasingly dominated by events where both $\agg$ decays experience shower or instrumental merging. When this occurs, each $\agg$ decay is misreconstructed as a single photonlike cluster, burying the $\hag$ signal in existing SM $\hgg$ events~\cite{hgg,hcoupling}. Moreover, in this $\ma$ regime, the decay modes of the $\Pa$ into more massive particles become inaccessible. This feature further emphasizes the importance of the diphoton decay mode at low $\ma$.

In this paper, we introduce a novel particle reconstruction strategy based on a modern machine learning (ML) approach also known as deep learning. We show that such a strategy addresses the above merging challenges associated with reconstructing boosted particle decays over a wide range of boosts. We describe the concept of \textit{end-to-end} particle reconstruction using ML algorithms that bypass PF~\cite{hgg_e2e} and train directly on minimally processed detector data to reconstruct a particle property of interest. To benchmark our technique, we reconstruct the invariant mass of simulated $\agg$ decays for masses in the range $\ma = 0.1$--1.0\GeV and for boost regimes where the diphotons are barely resolved to instrumentally merged. For particle $\Pa$ boosts $\boost = \Ea/\ma$, where $\Ea$ is the energy of the $\Pa$, these correspond to boosts of $\boost=60$--600. A single end-to-end ML mass regressor is used, where regressor refers to the output mass of the ML algorithm. For instrumentally merged diphotons, we further develop a novel ML technique called domain continuation. This technique involves extending the training domain to nonphysical values, to include the learning of masses below the detector resolution. Using both end-to-end particle reconstruction and domain continuation, we are able to reconstruct the invariant mass of instrumentally merged diphotons in the CMS detector, a first for a particle reconstruction method. The technique is validated using $\pgg$ decays in LHC collision data collected by the CMS experiment in 2017. We achieve a substantial sensitivity gain over existing benchmark algorithms, allowing previously inaccessible boost regimes to be probed.

\section{The CMS detector}
\label{sec:Detector}

The central feature of the CMS apparatus is a superconducting solenoid
of 6\unit{m} internal diameter, providing a magnetic field of
3.8\unit{T}. Within the solenoid volume are a silicon pixel and strip
tracker, a lead tungstate crystal ECAL, and a brass and scintillator
hadron calorimeter, each composed of a barrel and two endcap
sections. Forward calorimeters extend the pseudorapidity ($\eta$)
coverage provided by the barrel and endcap detectors. Muons are detected in gas-ionization chambers embedded in the steel flux-return yoke outside the solenoid.

The ECAL consists of 75\,848 lead tungstate crystals, which provide a
coverage of $\abs{\eta} < 1.48 $ in the barrel region (EB) and $1.48 <
\abs{\eta} < 3.00$ in the two endcap regions. The full azimuthal range $\abs{\phi} <\pi$ is covered for all regions.

In the barrel section of the ECAL, which is the focus of this paper, an energy resolution of about 1\% is achieved for
photons in the tens of \GeV energy range, for those that do not convert to an $\Pepem$ pair before reaching the EB. The energy resolution for photons converting in the EB is about 1.3\% up to $\abs{\eta} = 1$, rising to about 2.5\% at $\abs{\eta} = 1.48$~\cite{CMS:EGM-14-001}. Each EB crystal has dimensions of $2.18\times2.18\times23\,\cm^3$. This corresponds to an angular resolution of $\deta \times \dphi = 0.0174 \times 0.0174$ in the direction of the collision. The crystal Moli\`ere radius is comparable to the crystal width,
leading to more than 90\% of the energy of photons converting only in the ECAL to be laterally contained within a \tbt crystal matrix~\cite{CMS:EGM-14-001}.

The material structure of the inner tracking detectors changes with pseudorapidity. As a result, the number of radiation lengths ($X_0$) that a photon traverses before reaching the EB surface varies from $0.4\,X_0$ along the central plane ($\eta \approx 0$) up to $1.8\,X_0$ on the forward edge ($\eta \approx 1.4$). Correspondingly, the amount of energy that a photon loses, and its probability to convert into an $\Pepem$ pair before reaching the ECAL crystals, varies with $\eta$.

Finally, an important feature of the proton collisions at the LHC is the presence of additional, soft collisions concurrent with the primary hard-scattering event. Such additional proton-proton interactions within the same or nearby bunch crossings, known as pileup (PU), can contribute additional energy depositions in the ECAL. Since these energy deposits may potentially be misidentified as photons, appropriate pileup mitigation techniques are applied~\cite{puppi}.

A more detailed description of the CMS detector, together with a definition of the coordinate system used and the relevant kinematic variables, is reported in Ref.~\cite{cms}.

\section{End-to-end ML particle reconstruction}
\label{sec:E2E}

A number of LHC-related studies have attempted to use ML techniques to reconstruct boosted particle decays, particularly for heavy-resonance decays exhibiting ensemble merging. These included the use of jet images~\cite{toptag,toptagatlas}, graphs~\cite{gnn-Nachman,gnn-DeepSet,gnn-ParticleNet}, and other physics-inspired structures~\cite{deepcsv,benergy,cmsllp,lola,kyle}. The large majority of these, however, rely on PF candidates as inputs. Although such attempts show promise over traditional rule-based selection algorithms, they are ultimately constrained by the PF information itself. In their current form, such PF-based techniques would not be sensitive to shower-merged decays since information about the individual decay products is not present.

The powerful feature-learning capabilities of modern ML algorithms provide an alternative. Using inputs that are as unfiltered and informationally rich as possible, such algorithms have been shown to outperform cut-based or even ML-based methods utilizing input variables modeled by hand~\cite{alexnet,skinclass,alphazero,alphafold,neutrino,microboone,lartpc}. This is expected if the input variables are difficult to model by hand or if their correlations are difficult to extract. Moreover, by encompassing the majority of the reconstruction workflow into the functionality of the ML algorithm, potential information loss due to stepwise optimizations is avoided. Such end-to-end ML algorithms are thus trained directly on minimally processed detector data with the objective of predicting the desired quantity of interest. This motivates an end-to-end ML-based particle reconstruction strategy for the challenges of shower merging. We accomplish this by transforming the detector data into high-fidelity images~\cite{qvg_e2e,hgg_e2e,atlasEM}, and using these to train a convolutional neural network (CNN) that outputs the final parent particle property of interest. Similar applications in the literature have employed order-invariant graph-based ML models as well~\cite{gnn-CMS,gravnet}.

The end-to-end ML approach has specific advantages for boosted particle decays. The first is the gain in information granularity offered by detector data for features that cannot easily be reduced to a particle-level, or shower-shape-type representation. Second, the choice of a CNN, or any similar hierarchical ML architecture like a graph-based network, allows detector features to be learned across several length scales. Features from the crystal level to the cluster level and beyond are learned in a complementary way. This is particularly advantageous for boosted particle decays that may exhibit merging at multiple scales. Third, by training on minimally processed data rather than heavily filtered or clustered data, the ML algorithm may learn to adapt to more varied, higher-dimensional changes in the data, potentially developing a greater robustness to evolving data-taking conditions. Finally, there is the computational simplicity of using a single ML algorithm versus a hierarchy of algorithms each with its own intermediate optimizations. Admittedly, transitioning to a fully end-to-end ML-based particle reconstruction framework would involve an extensive overhaul of the workflows employed by the LHC experiments such as CMS. Moreover, to what extent the entire particle reconstruction workflow can or should be reduced to a single ML algorithm remains an open question.

Although a comprehensive end-to-end particle reconstruction framework is a long-term undertaking, its benefits can already be demonstrated through targeted applications. In this paper, we test the potential of this technique by investigating a previously inaccessible boost regime, namely the merged $\agg$ decay. We train an ML regression algorithm to reconstruct the generated mass $\ma$ using only the electromagnetic shower pattern of the merged diphoton decay in the CMS ECAL. This enables us to exploit subtle differences in the energy distribution of the ECAL shower pattern~\cite{acat} to analyze even extremely merged decays.

\begin{figure*}[!htbp]
\centering
\includegraphics[height=.38\linewidth]{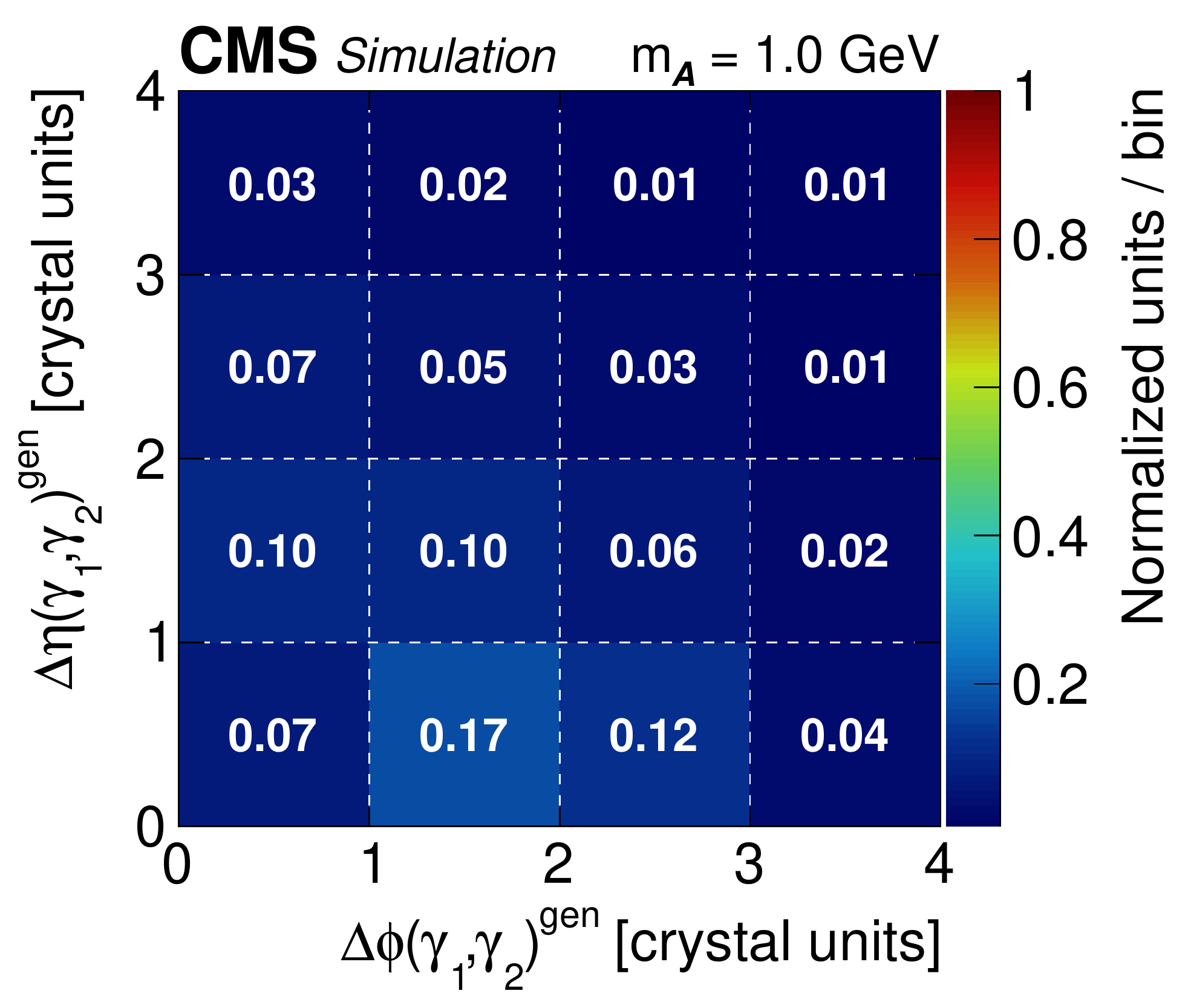}
\quad
\includegraphics[height=.38\linewidth]{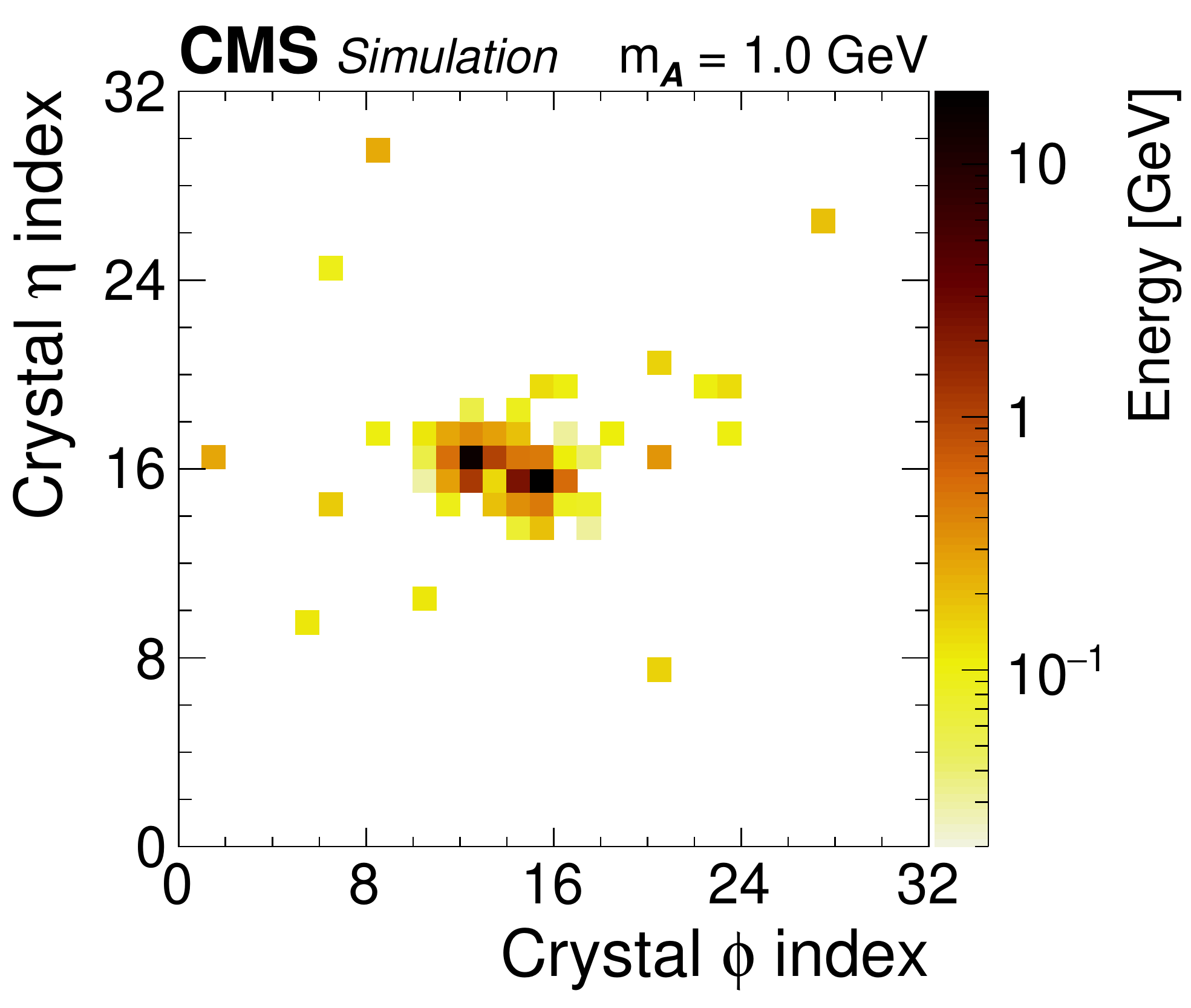}
\\
\includegraphics[height=.38\linewidth]{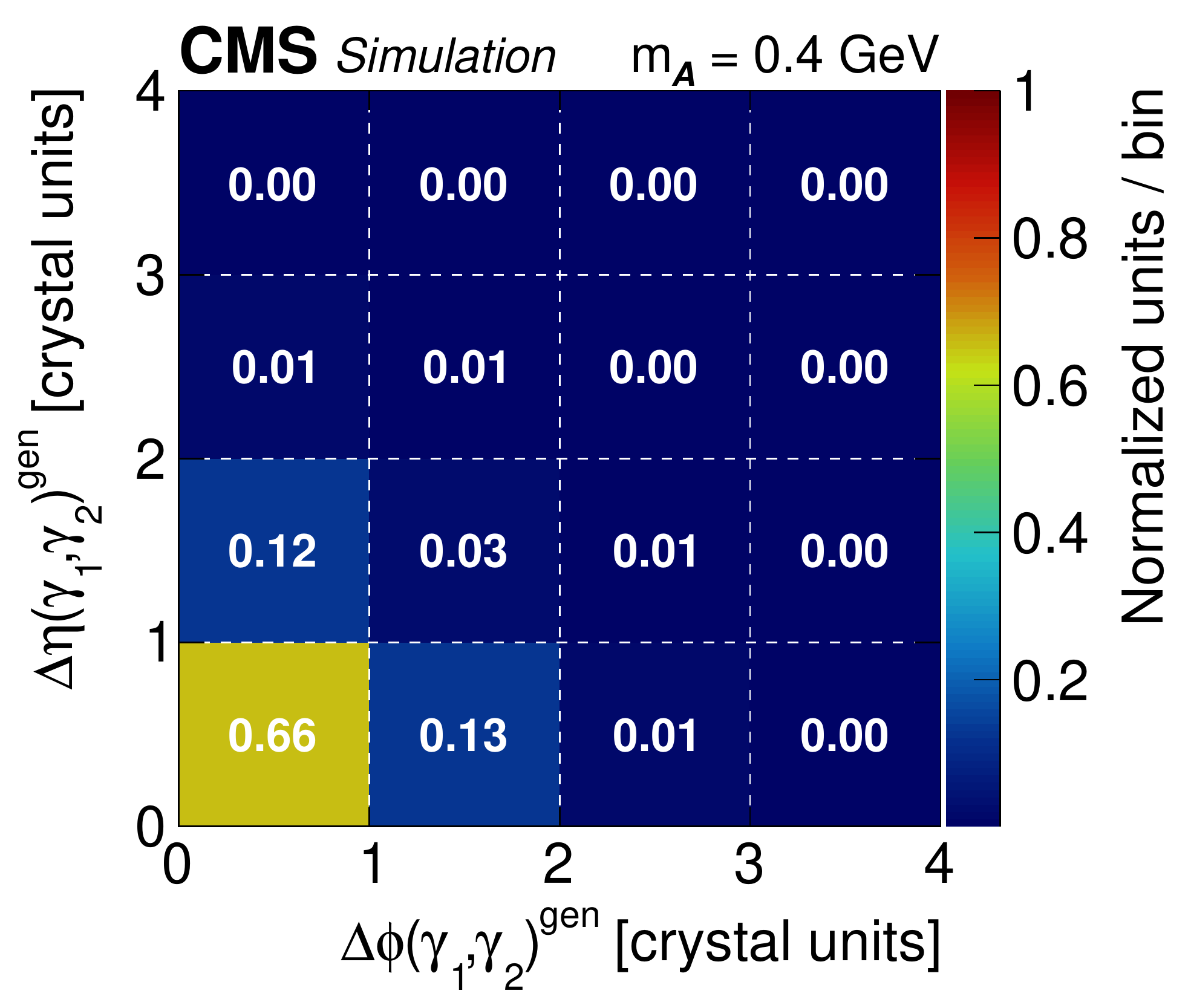}
\quad
\includegraphics[height=.38\linewidth]{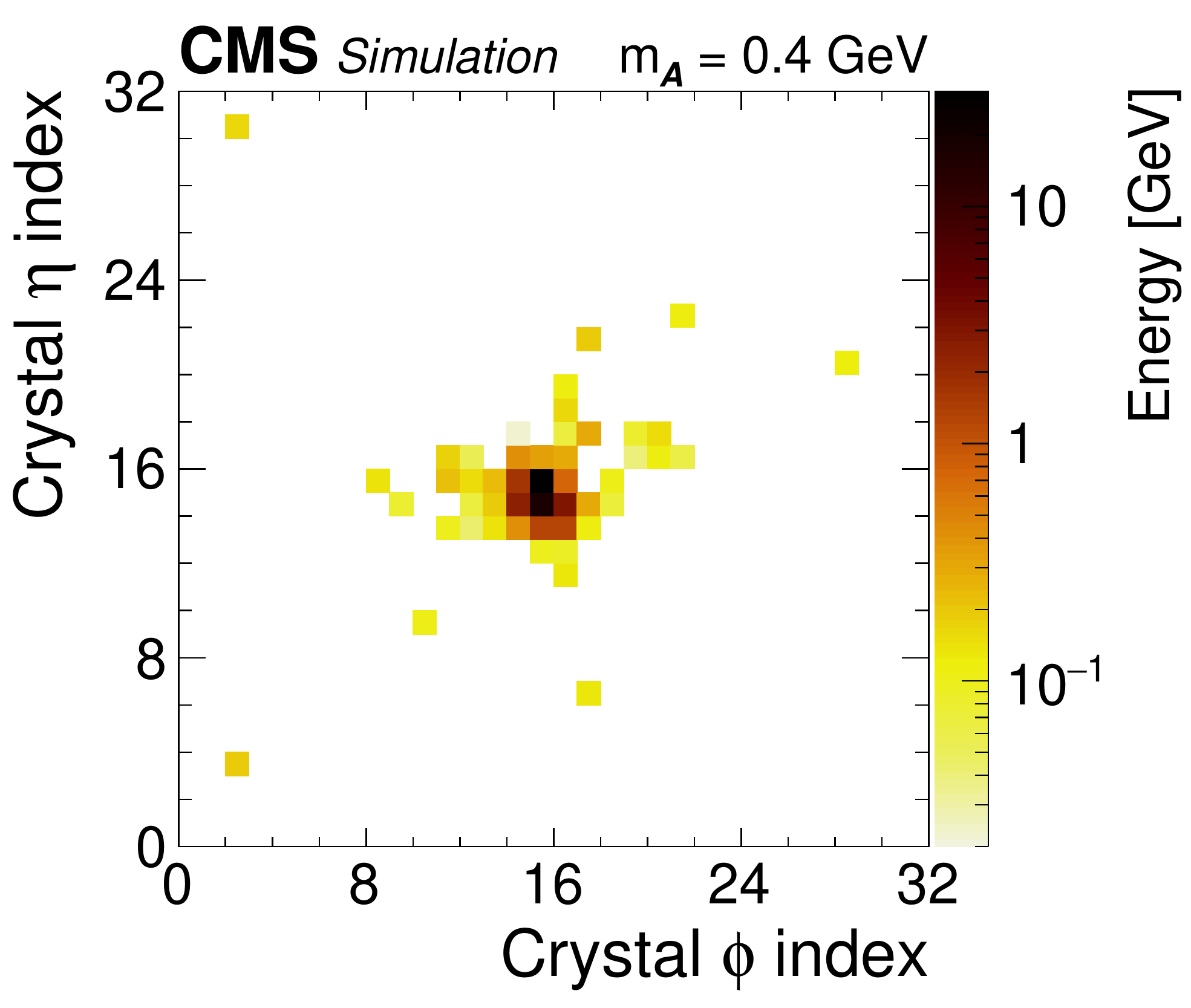}
\\
\includegraphics[height=.38\linewidth]{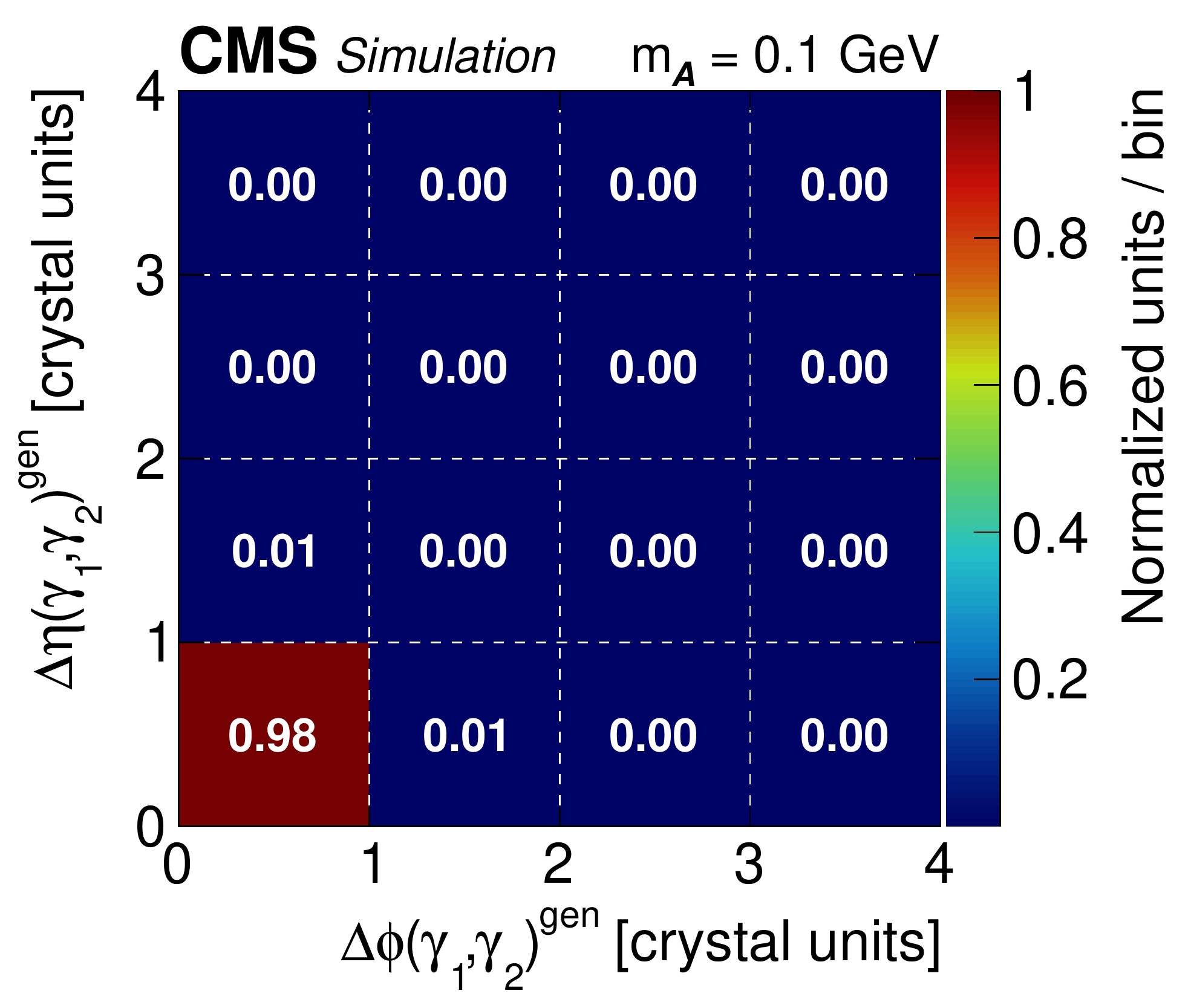}
\quad
\includegraphics[height=.38\linewidth]{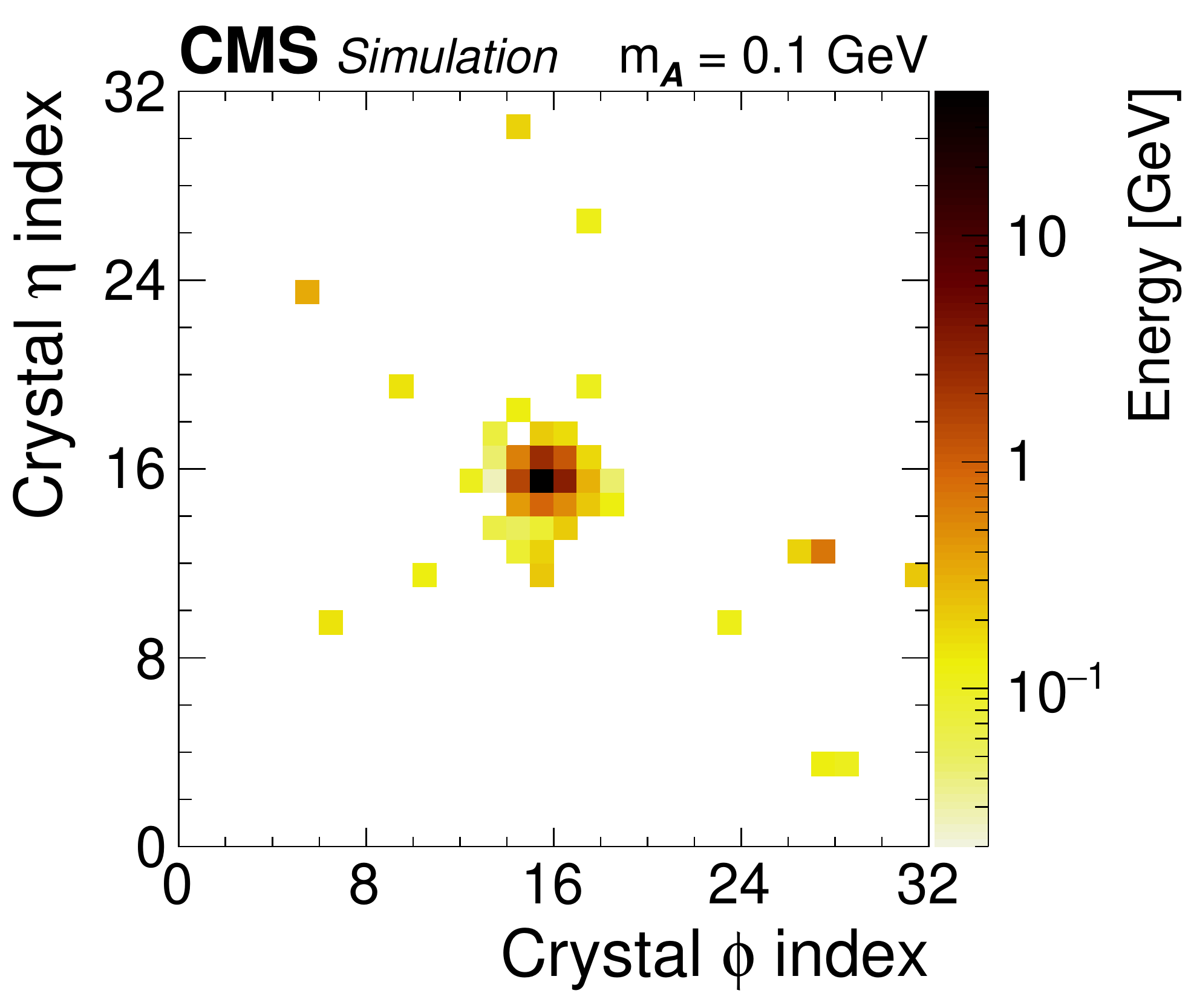}
\\
\caption{Simulation results for the decay chain $\haa$, $\agg$
at various boosts: (upper plots) barely resolved, $\ma = 1.0\GeV$, $\boost = 50$; (middle plots) shower merged, $\ma = 0.4\GeV$, $\boost = 150$; and (lower plots) instrumentally merged, $\ma = 0.1\GeV$, $\boost = 625$. The left column shows the normalized distribution of opening angles between the leading ($\gamma_1$) and subleading ($\gamma_2$) photons from the particle $\Pa$ decay, expressed by the number of crystals in the $\eta$ direction, $\Delta\eta(\gamma_1, \gamma_2)^{\mathrm{gen}}$, versus the $\phi$ direction, $\Delta\phi(\gamma_1, \gamma_2)^{\mathrm{gen}}$. Note that the distributions include contributions outside of the plotted ranges and thus may not sum to unity within the displayed ranges. The right column displays the ECAL energy shower pattern for a single $\agg$ decay, plotted in relative ECAL crystal index coordinates and color-coded by energy. In all cases, only decays reconstructed as a single PF photon candidate passing selection criteria are used.
}
\label{fig:genDR}
\end{figure*}

We parametrize the diphoton merging in terms of the Lorentz boost of the particle $\Pa$. Figure~\ref{fig:genDR} illustrates the typical merging regimes at three different boosts. Samples of $\agg$ decays misreconstructed by PF as single photons passing candidate selection criteria (discussed in Sec.~\ref{sec:Dataset}) are used. The $\agg$ samples are derived from simulated $\haa$ events in which the true positions of the diphoton decays are known. The distribution of the opening angles between the higher-$\pt$ (leading, $\gamma_1$) and lower-$\pt$ (subleading, $\gamma_2$) photons from the simulated particle $\Pa$ decay is shown in Fig.~\ref{fig:genDR} (left column). The angles are expressed in number of ECAL crystals in the $\eta$ direction, $\Delta\eta(\gamma_1, \gamma_2)^{\mathrm{gen}}$, versus the $\phi$ direction, $\Delta\varphi(\gamma_1, \gamma_2)^{\mathrm{gen}}$. Typical ECAL energy deposition patterns from a single $\agg$ decay are also displayed in Fig.~\ref{fig:genDR} (right column). The diphoton particle showers are resolved in the ECAL for boosts of $\boost \lesssim 50$ (Fig.~\ref{fig:genDR}, upper row), they are shower merged for $50 \lesssim \boost \lesssim 250$ (Fig.~\ref{fig:genDR}, middle row), and instrumentally merged for $\boost \gtrsim 250$ (Fig.~\ref{fig:genDR}, lower row). Note that the same $\boost$ value can lead to considerably different merging, depending on the opening angle between the diphotons.

The barely resolved decays ($\boost \approx 50$) represent the challenge of ensemble merging. First, rule-based methods may fail to categorize the $\agg$ shower pattern as either one or two photon objects, potentially causing the event to be discarded. Second, resolved $\agg$ decays can resemble both a photon conversion to an $\Pepem$ pair and low-energy deposits from pileup. For these latter two possibilities, track information, if present, can potentially be exploited to distinguish between energy deposits from $\agg$ decays where the photons do not convert before reaching the ECAL and those arising from single photons converting before reaching the ECAL and from pileup, as is done in PF. For the shower-merged regime, not even dedicated shower clustering tools have the sensitivity to reconstruct the invariant mass of the $\agg$ system. As we will show in this paper, only the end-to-end ML approach can effectively operate in all three regimes. The $\agg$ decay thus provides a simple but comprehensive test case for all the merging regimes occurring in the decays of boosted particles. 

\section{Detector images}
\label{sec:Images}

Since the photons from $\agg$ decays deposit energy primarily in the ECAL, for simplicity, we use an image construction strategy that consists of only ECAL information. We take a \imgwindow matrix of ECAL crystals around the most energetic (seed) crystal of the reconstructed photon candidate and create an image array. This corresponds to an angular cone of $\sqrt{\smash[b]{(\Delta\eta)^2+(\Delta\phi)^2}} \approx 0.3$, and ensures the subleading photon in the $\agg$ decay is fully contained in the image. Although generous, our results are not strongly sensitive to the size of the image. Each pixel in the image exactly corresponds to the energy deposited in a single ECAL crystal. These energy depositions represent the actual interaction of the incident photon with the detector material (Fig.~\ref{fig:genDR}, right column). This approach is distinct from those that use PF candidates, in which case the $\agg$ decay, reconstructed as a single particle, would be displayed as a single image pixel, by construction. No rotation is performed on the images since electromagnetic showers are not rotationally symmetric. In addition to the $\eta$-$\phi$ symmetry being broken by the CMS magnetic field, general rotations of square pixels are destructive operations that distort the particle shower pattern.

For simplicity, only photons depositing energy in the barrel section of the ECAL are used in this paper. For general particle decays, the ECAL images can be combined with additional subdetector images~\cite{qvg_e2e}, or parsed into a multi-subdetector graph if one is using such an architecture. Including such tracking images, even as a complement to the $\agg$ ECAL images, could enable a better accounting of contributions from $\Pepem$ conversions and pileup. However, they were not included in this study.

As noted in Sec.~\ref{sec:Detector}, because of the $\eta$-dependent material structure of the inner tracker, electromagnetic shower development varies significantly with $\eta$. Once the \imgwindow crystal matrix of the particle shower is taken, the resulting image no longer contains explicit information about where in the ECAL the shower was located. We thus perform two modifications to recover this information during the training of our ML algorithm. The first is to split the ECAL images described above into a two-layer image that contains the transverse and longitudinal components of the crystal energy, which are defined as $E\sin\theta$ and $E\abs{\cos\theta}$, respectively, where $E$ is the crystal energy and $\theta$ is the polar angle of the crystal energy deposit position. The second is to include the crystal seed coordinates.

The calibrated ECAL detector data exist at various levels of processing: from minimally processed to the filtered and clustered version more easily accessible at the analysis level. Since the clustered detector data are historically optimized for the needs of PF, it is worth revisiting this choice for the end-to-end ML technique. As discussed in Appendix~\ref{app:aodvminiaod}, training on minimally processed instead of clustered data significantly improves our results. We thus emphasize our earlier statement that using minimally processed detector data is critical in realizing the full potential of ML. We make exclusive use of minimally processed detector data in this paper. One caveat is that accessing such data is becoming logistically challenging because of the trend toward more compact CMS file formats, necessitated by the growing volume of LHC data.

\section{Candidate selection and datasets}
\label{sec:Dataset}

To address the most challenging diphoton particle decays, we require that the diphoton system be misreconstructed by the PF algorithm as a single photon candidate, labeled $\Gamma$. As a consequence of the PF ambiguities in categorizing low-boost $\agg$ decays $(\boost \lesssim 50)$ as either one- or two-photon objects, this requirement does not necessarily imply that the diphotons are merged. For diphotons that are barely resolved, if one of the photons is too low in energy ($\pt < 10\GeV$), it is not reconstructed by PF as a separate photon even if it is sufficiently separated from the other photon.
Acceptance of this information results in a range of boost regimes. Reconstructed photons are also required to pass shower shape and isolation criteria that would accept single photons with $>90$\% efficiency. These photon selection criteria are similar to those used by CMS in the SM $\hgg$ analysis~\cite{hgg}, which emphasizes how a merged signal might be buried in such events. For each selected photon candidate $\Gamma$, a single detector image is constructed (described in Sec.~\ref{sec:Images}), which is then passed to the ML algorithm for either training or inference.

For training, we use a sample of simulated $\agg$ decays in which the particle $\Pa$ decays promptly. The sample includes pileup but otherwise no additional particles. An ensemble of continuously distributed masses $\magen$ is used, such that the ML model described in Sec.~\ref{sec:Train} does not learn the discretization of the mass points. These samples are generated with $\ptagen =$ 20--100\GeV, $\magen =$ 0--1.6\GeV, and $\abs{\etaagen} < 1.4$. This corresponds to Lorentz boosts with $\boost$ in the approximate range 10--1000 and adequately covers the regimes of interest (discussed in Sec.~\ref{sec:E2E}). After applying photon selection criteria, samples with closely merged photons ($\boost \gtrsim 150$) are preferentially selected over samples with resolved photons ($\boost \lesssim 50$), since the latter are more likely to fail single-photon shower-shape criteria. This tends to sculpt the phase space in $(\ptagen,\,\magen)$, which can result in the mass regressor preferring to output masses in regions with higher populations. To prevent this from happening, the training samples are generated to ensure that the $(\ptagen,\,\magen)$ phase space is uniformly distributed \textit{after} the photon selection criteria have been applied. After applying these requirements, approximately 780,000 $\agg$~decays are available for training, a sample of about 26,000 decays for validation, and another 26,000 for testing. In addition, as explained in Sec.~\ref{subsec:Train-continue}, the training samples are augmented with true, single-photon samples with similar kinematic properties. After the selection requirements, there are 150,000, 26,000, and 26,000 simulated single photons for training, validation, and testing, respectively. The validation set is used to optimize the ML model hyperparameters, and the test set is used to assess the performance of the chosen model on a statistically independent sample.

To benchmark the ML technique, we use both simulated and actual CMS collision data. The benchmark studies are described further in Sec.~\ref{subsec:Train-eval}. In addition to pileup, underlying events are included in the simulated $\agg$ samples to make them more realistic. The simulated samples are obtained from $\hag$ events generated at fixed masses of $\magen = 0.1$, 0.4, and 1.0\GeV, all with prompt $\Pa$ decays. The decays correspond to median boosts of $\boostav = 600$, 150, and 60, respectively, covering a similar boost range as used for training. The collision data contain events enriched in $\pgg$ decays selected from the 2017 and 2018 CMS data-taking periods~\cite{cmstrigger}, corresponding to an integrated luminosity of 41.5 and 56.9 $\fbinv$, respectively.
To enrich the data sample with $\pgg$ decays from hadronic jets, as explained in Sec.~\ref{subsec:Benchmarks:Data}, only reconstructed photon candidates with $\pt =$ 20--35\GeV are used. Shower-shape and isolation requirements, which are slightly tighter than for the $\agg$ analysis described above, are applied.

To study the robustness of the ML technique as a function of various parameters, as described in Sec.~\ref{sec:Stability},
events enriched with electrons from $\zee$ decays are used. The electrons are obtained from 2017 data and simulation using the ``tag-and-probe'' method~\cite{tnp}. The selected electron is required to coincide with a photon candidate passing the photon selection criteria described earlier.

For all simulated samples, events are generated with corresponding 2017 data-taking pileup conditions and fully simulated CMS geometry and detector response~\cite{cms}.

\section{Mass regression model and training}
\label{sec:Train}

A \textsc{ResNet} CNN architecture~\cite{resnet} forms the basis of our mass regression model. Although other ML architectures exist, including those that use graph-based techniques, the emphasis in this paper is placed on the general reconstruction method rather than the optimization of the ML model.

The energy deposits in the ECAL detector images are first scaled by a sample-wide constant so that the sample-wide energy distribution is approximately within the $[0,1]$ interval. Each image associated with a reconstructed photon candidate $\Gamma$ is then passed to the \textsc{ResNet} model, which outputs to a global maximum pooling layer. A more detailed description of the \textsc{ResNet} model used is presented in Ref.~\cite{hgg_e2e}. The outputs are then concatenated with the crystal seed coordinates of the photon candidate. The seed coordinates are also scaled to lie within the $[0,1]$ interval. The concatenated outputs are then fully connected to a final output node that gives the predicted or \textit{regressed} value of the scalar-particle mass $\ma$ for that photon candidate. The above ML mass regressor contains 89k trainable parameters.

To train the mass regressor, the regressed mass $\mapred$ is compared
with the true (generated) $\magen$ by calculating the absolute error
loss function $\abs{\mapred - \magen}$, averaged over the training
batch. Other loss functions are equally performant. This loss function
is then minimized using the \texttt{ADAM} optimizer~\cite{adam}. To facilitate convergence, the mass values are also scaled to lie within $[0,1]$ before they are passed to the optimizer. This procedure represents our basic training strategy.

\subsection{Performance limitations when \texorpdfstring{$\ma \to 0$}{mA->0}}

\begin{figure*}[tbp]
\centering
  \includegraphics[height=.4\linewidth]{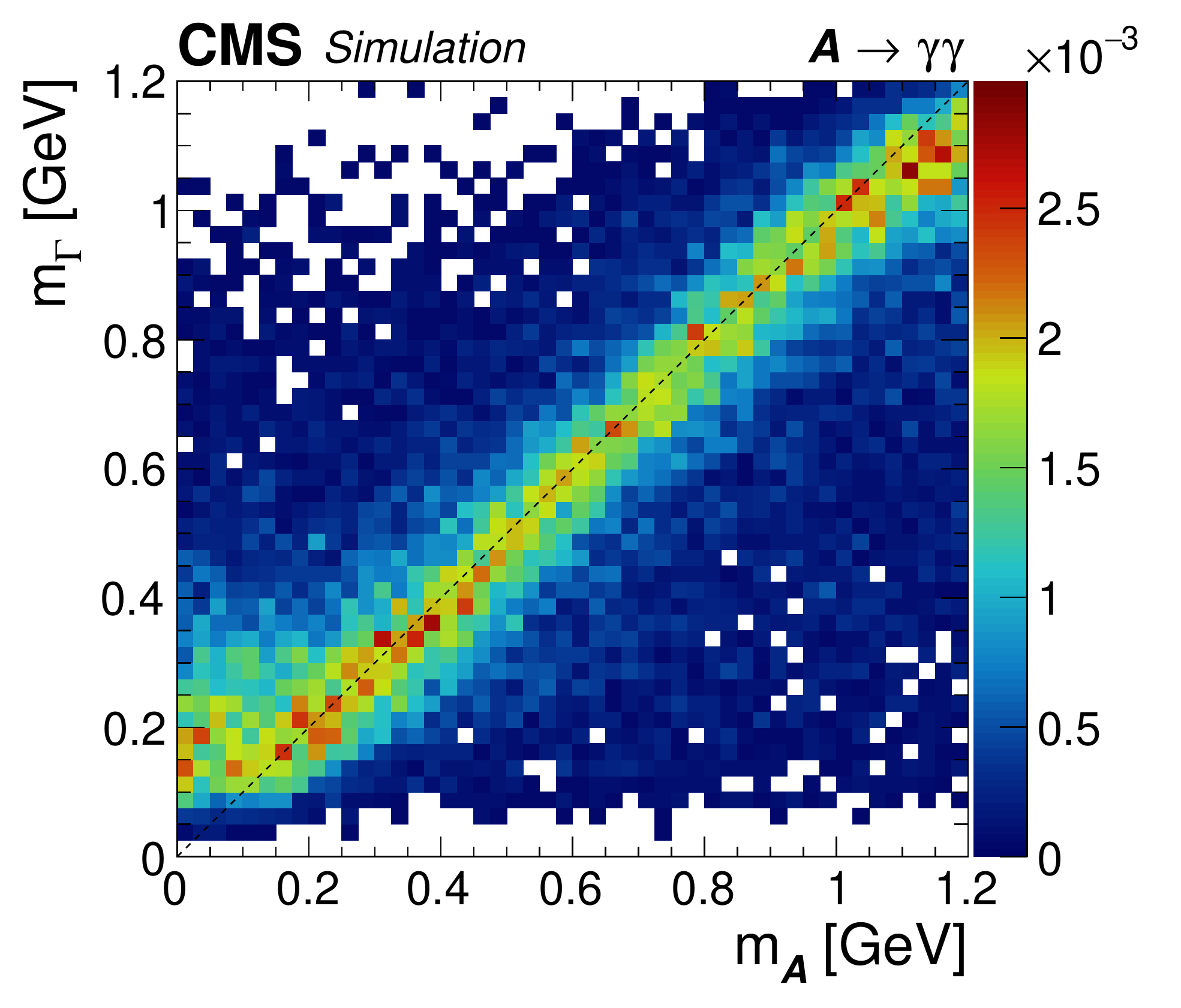}
\qquad
  \includegraphics[height=.4\linewidth]{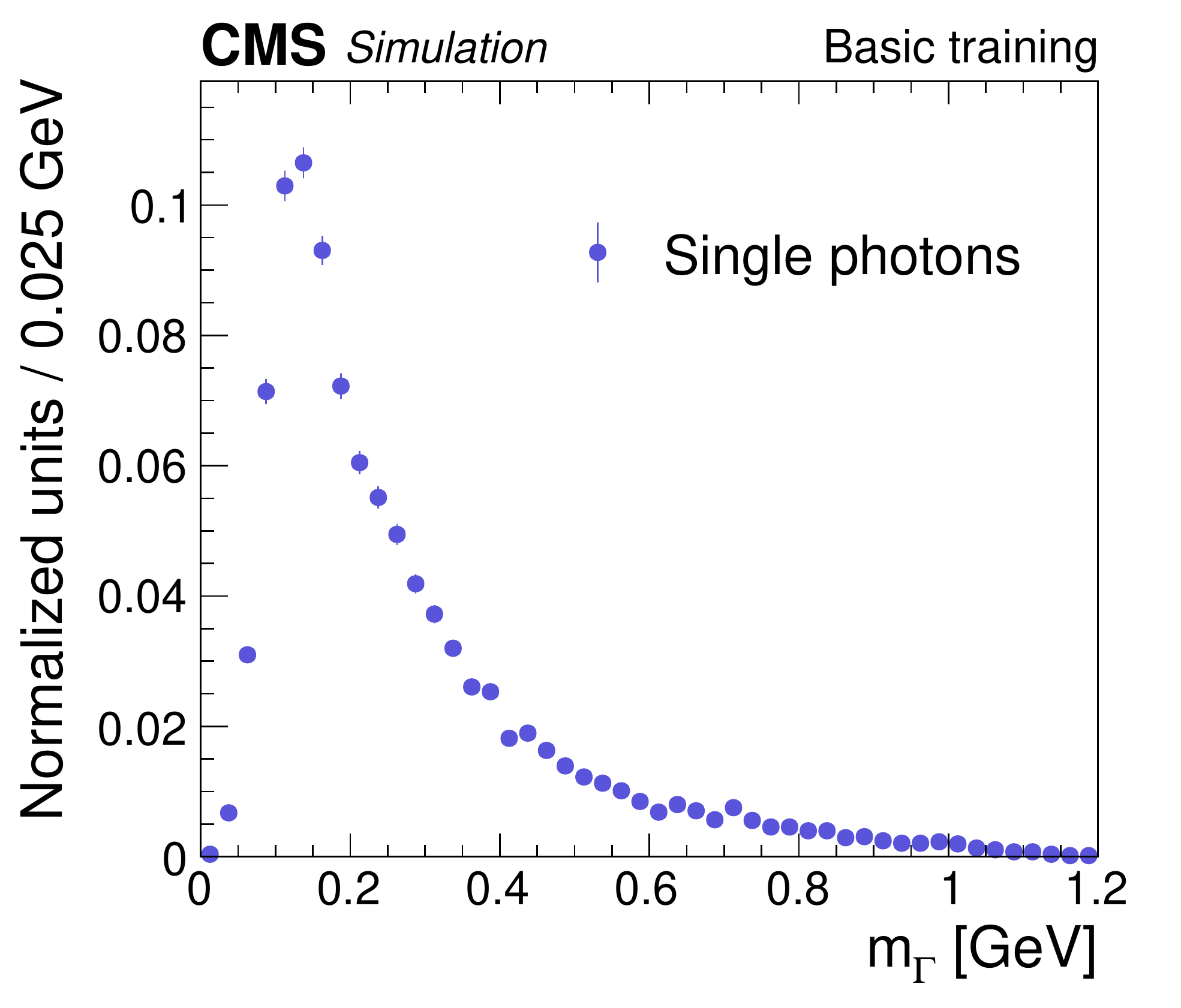}
\caption{Left: the regressed mass $\mapred$ vs. the generated $\ma$ value for simulated $\agg$ decays generated uniformly in $(\pt,\,\ma)$ before domain continuation is implemented. The regressed $\mapred$ is normalized in 0.025\GeV vertical slices of the generated $\ma$. The color scale to the right of the plot gives the normalized number of events per vertical slice in 0.025\GeV bins of $\mG$. Right: the regressed $\mG$ distribution for simulated single-photon samples only, before domain continuation, resulting in a distinct peak in the low-$\mG$ region. The distribution is normalized to unity with the vertical bars on the points indicating the statistical uncertainty.}
\label{fig:naivereg}
\end{figure*}

Relying solely on this basic training strategy has significant performance limitations. As shown in Fig.~\ref{fig:naivereg} (left), naively training the mass regressor as described above results in a nonlinear response near either boundary of the mass regression range. At the high-$\ma$ boundary, this issue can be resolved by trivially extending the training mass range. Thus, if one wanted to obtain a usable regression range up to $\ma \approx 1.2\GeV$, one would train with a sample including higher masses, as we have, up to $\ma \approx 1.6\GeV$, depending on the mass resolution at the upper mass range. Predictions in the extended mass range are then discarded during the physics analysis. This approach cannot, however, be used in any obvious way for the low-$\ma$ boundary, since it is constrained by the physical requirement $\ma > 0$. The mass region $\mapred \lesssim 200\MeV$, of considerable theoretical interest for the diphoton decay mode in BSM models (discussed in Sec.~\ref{sec:Intro}), would therefore be inaccessible. The use of the mass regressor as a tool for reconstructing $\pgg$ decays would also be lost. Moreover, significant biases in the regressed masses of true photons would arise. As illustrated in Fig.~\ref{fig:naivereg} (right), photons would be regressed as a peak around $\mapred \lesssim 200\MeV$, reducing even further the usable range of the mass regressor when single-photon backgrounds are included.

\subsection{Domain continuation to negative masses}
\label{subsec:Train-continue}

\begin{figure*}[tbp]
\centering
\includegraphics[height=.25\linewidth]{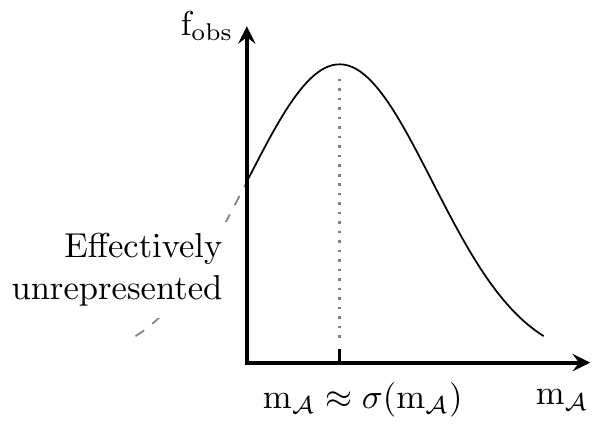}
\includegraphics[height=.25\linewidth]{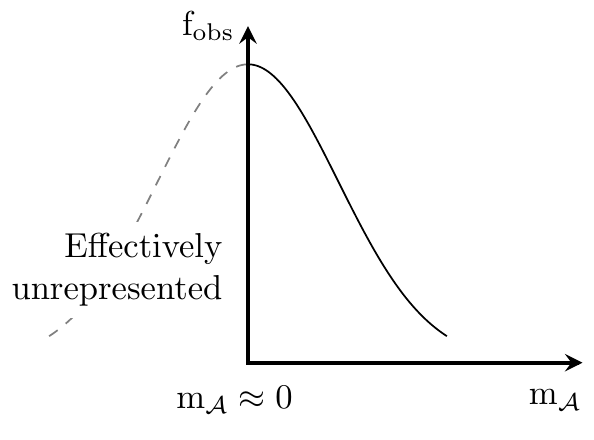}
\includegraphics[height=.25\linewidth]{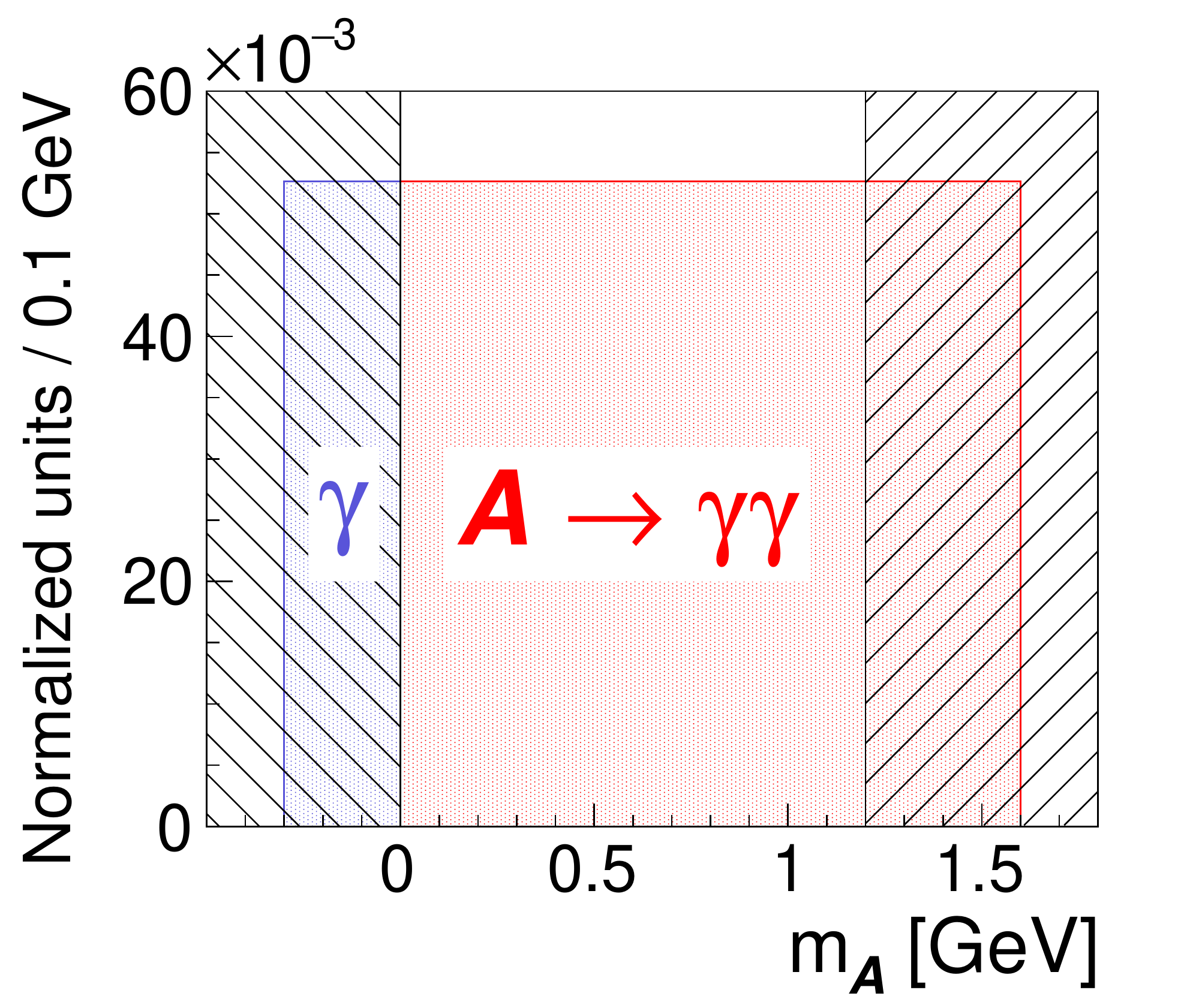}
\caption{Pictorial representation of the $\ma \to 0$ boundary problem occurring when attempting to regress below the mass resolution. Left: the distribution of physically observable $\agg$ invariant masses ($\mathrm{f}_{\mathrm{obs}}$) vs. the generated $\ma$. When $\ma \approx \sigma(\ma)$, the left tail of the mass distribution becomes underrepresented in the training set. Middle: As $\magen \to 0$, only half of the mass distribution is represented. The regressor subsequently defaults to the last full mass distribution at $\magen \approx \sigma(\ma)$. Right: with domain continuation, the generated mass distribution of the original training samples ($\agg$, red region) is augmented with topologically similar samples that are randomly assigned nonphysical masses ($\gamma$, blue region). This allows the regressor to see a full mass distribution over the entire region of interest (unhatched region). Predictions in the black hatched regions are discarded.}
\label{fig:continuation}
\end{figure*}

Fundamentally, the above boundary problem arises because, when training the mass regressor, the physically observable $\agg$ invariant mass distribution becomes underrepresented for samples with $\magen < \sigma(\ma)$, where $\sigma(\ma)$ is the mass resolution. This issue is illustrated in Fig.~\ref{fig:continuation} (left). For samples where $\magen \approx \sigma(\ma)$, the full, physically observable mass distribution ($\mathrm{f}_{\mathrm{obs}}$), visualized as a Gaussian distribution, is barely represented in the training set. As $\magen \to 0$, shown in Fig.~\ref{fig:continuation} (middle), only half of the mass distribution is now observable. For these underrepresented samples, the mass regressor defaults to the closest mass value whose observable distribution is well represented, $\magen \approx \sigma(\ma)$, since this will appear most similar.
This results in a gap above $\mapred \approx 0$ and an accumulation of masses at $\mapred \approx 200\MeV$. More generally, this boundary problem manifests itself when regressing a quantity $q$, with resolution $\sigma(q)$, over the range $(a,b)$, for samples with $q \lesssim a+\sigma(q)$ or $q \gtrsim b-\sigma(q)$. This effect only becomes negligible at either boundary in the limit $\sigma(q) \ll a,b$.

This motivates a solution for the low-$\ma$ boundary problem by extending the regression range below $\magen = 0$, into the nonphysical domain. These are then populated with ``topologically similar'' samples. We thus augment the training set with samples artificially and randomly labeled with negative masses. During inference, we remove the nonphysical predictions $\mapred < 0$. As a topologically similar sample, either a sample of decays with a fixed mass $\magen \approx 0.001\MeV$ or a sample of single photons can be used. In this paper, we use the latter, although we find either method works well. If we require that the ``negative mass'' samples have the same mass density as the ``positive mass'' samples in the training set
(cf. Fig.~\ref{fig:continuation}, right), then only a single hyperparameter is needed, the minimum artificial mass value, $\mathrm{min}(\magen)$. This can be set by choosing the negative value with the smallest magnitude that closes the low-$\ma$ gap (cf. Fig.~\ref{fig:naivereg}, left) and linearizes the mass response in the physical domain, $\mapred > 0$. We find a value of $\mathrm{min}(\magen) = -0.3\GeV$ to be sufficient. Other applications may seek to optimize both the minimum artificial mass value and the number density of the augmented samples. Note that having the augmented samples carry negative mass values is simply a consequence of the low-$\ma$ boundary coinciding with $\mapred=0$. Had the boundary coincided with a positive mass, positive artificial mass values would be involved as well.

The above procedure effectively tricks the mass regressor into seeing a full invariant mass distribution for all physical $\agg$ decays, even when they reside below the detector mass resolution. As a result, the full low-$\ma$ regime becomes accessible. In addition, this procedure provides a simple way for suppressing single-photon backgrounds by requiring $\mG > 0$. Since single photons will tend to be regressed with negative masses, removing samples with $\mapred < 0$ reduces single-photon contributions in a mass decorrelated way. The only trade-off is that low-$\ma$ samples incur a selection efficiency to be regressed within $\mapred>0$. However, this is expected for most $\agg$ merged cases that cannot be distinguished from true photons.

By analogy to complex analysis, we denote the above procedure as \textit{domain continuation}. Similar procedures, however, all occur in the statistical tests used in high energy physics~\cite{lhoodlhc}. Our final training strategy implements domain continuation on top of the basic training strategy described at the beginning of this section.

\subsection{Out-of-sample response}

An important feature of ML-based regression algorithms is that their
predictions are bound by the regression range on which they were
trained. This is true even when out-of-sample decays, or decays not
represented in the training set, are given to the mass regressor. If
this fact is not addressed, unexpected peaks and features in the
regressed mass spectrum can potentially appear. Although hadronic jets
are indeed out of sample, it is desirable not to reject them at the stage of the mass regressor in order to enable the reconstruction of, \eg, embedded $\pgg$ decays. If desired, these can instead be suppressed by modifying the photon selection criteria, for instance, as described in Sec.~\ref{subsec:Benchmarks:Data}.

Furthermore, $\agg$ decays from more massive $\Pa$ particles than those used during training are a potential issue. These will be regressed as a false mass peak near the upper $\ma$ boundary. For this and other reasons stated earlier, when addressing the boundary problem at the upper mass range, we ignore predictions above $\mapred > 1.2\GeV$ (Fig.~\ref{fig:continuation}, right). Lastly, to suppress single photons, as noted above, we ignore predictions with $\mapred < 0$. During inference, the use of the mass regressor is thus limited to samples regressed within the region of interest (ROI): $\maroi$ $\in [0, 1.2]\GeV$. The impact of this method on the sample selection efficiency is estimated in Sec.~\ref{sec:Validation}.

\subsection{Evaluation}
\label{subsec:Train-eval}

In Sec.~\ref{sec:Validation}, we validate the ML training using a number
of figures of merit. We primarily use the mean absolute error, $\mathrm{MAE} = \langle \abs{\magen-\mapred} \rangle$ over the test set, but also consider its normalized counterpart, the mean relative error, $\mathrm{MRE} = \langle \abs{\magen-\mapred}/\magen \rangle$. The MAE (MRE) approximately corresponds to the mean absolute (relative) mass resolution. If treating the mass distribution at a fixed $\ma$ as a signal model, a lower MRE, similar to a lower relative mass resolution, implies a better signal significance assuming some fixed background contribution. The regression efficiency, or 
the number of samples regressed within $\maroi\in [0, 1.2]\GeV$, divided
by the total number of samples considered, is also used as a figure of merit. Similarly, if treating the mass distribution at a fixed $\ma$ as a signal model, a higher regression efficiency implies a higher signal significance for a given background contribution.

In Sec.~\ref{sec:Benchmarks}, we benchmark the physics performance of the end-to-end ML mass regressor (\textit{end-to-end}) by comparing its response with both a traditional neural network-based mass regressor (\textit{photon NN}) and a shower cluster-based algorithm used for mass reconstruction ($\tbtit$ \textit{algorithm}). The photon NN is trained on a mix of 11 shower-shape and isolation variables, identical to those used by CMS for multivariate photon tagging~\cite{hgg}. These variables are passed to a fully connected neural network of 128 nodes and three hidden layers that regresses the scalar-particle mass. For a fair comparison, it is also trained with domain continuation. The resulting performance was insensitive to small increases or decreases in network size. The \tbt algorithm is similar to that used by CMS for the reconstruction of low-energy $\pgg$ decays in the calibration of the ECAL~\cite{CMS:EGM-14-001}. It first identifies a local crystal energy maximum (seed). If the seed energy is above some energy threshold, then a \tbt crystal matrix (cluster) around the seed is formed. If a pair of nearby clusters is found, the reconstructed mass is calculated as the invariant mass of the two clusters. If a pair of clusters cannot be found, a default mass value outside of the $\maroi$ range is passed.

Lastly, in Sec.~\ref{sec:Stability}, to assess the robustness and generalizability of the end-to-end mass regressor, we compare its mass response versus various kinematic and detector quantities of interest. We conclude with a comparison of the regressed mass spectrum in data versus simulation.

\section{Training validation}
\label{sec:Validation}

\begin{figure*}[!htbp]
\centering
\includegraphics[height=.42\linewidth]{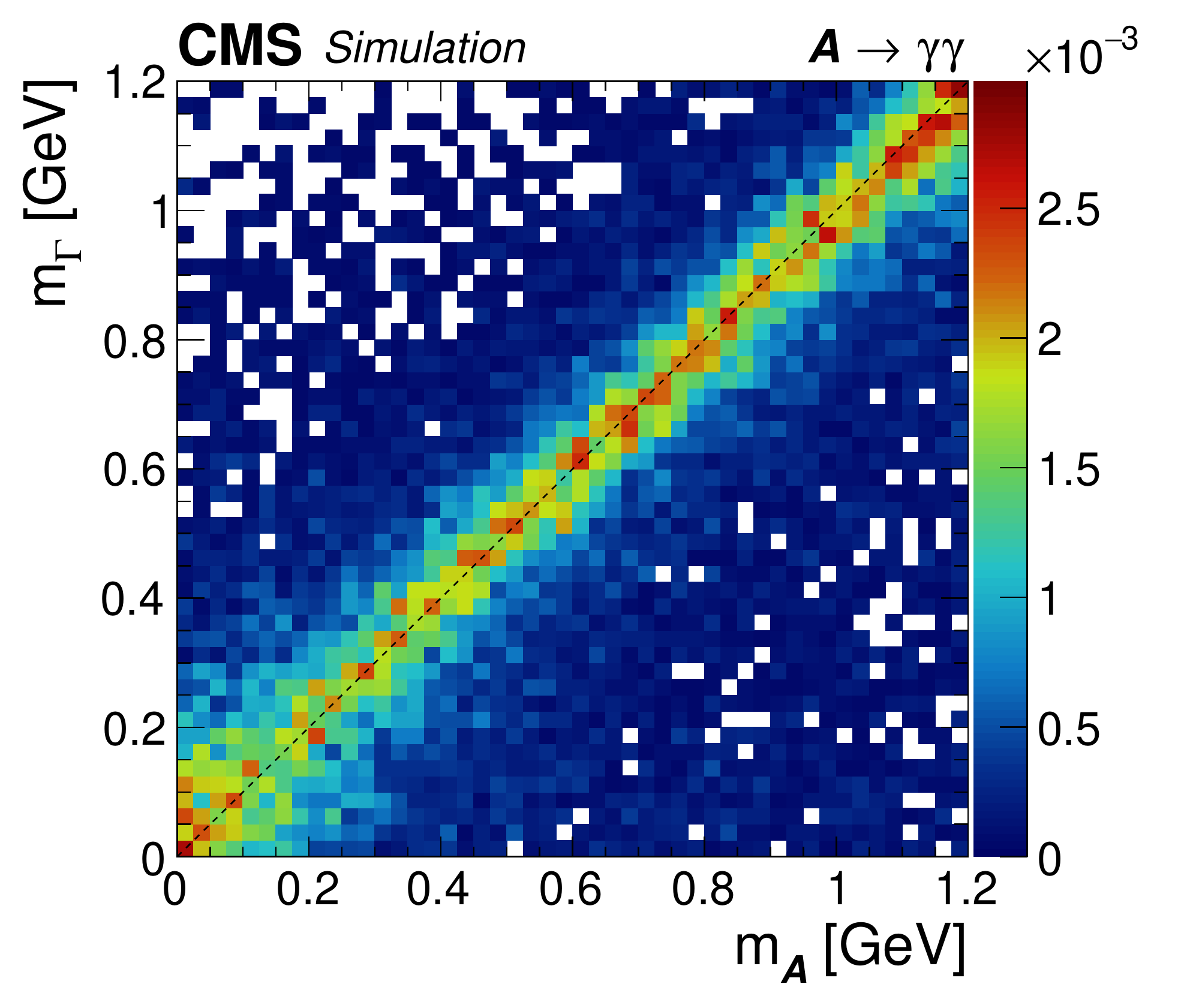}
\\
\includegraphics[height=.4\linewidth]{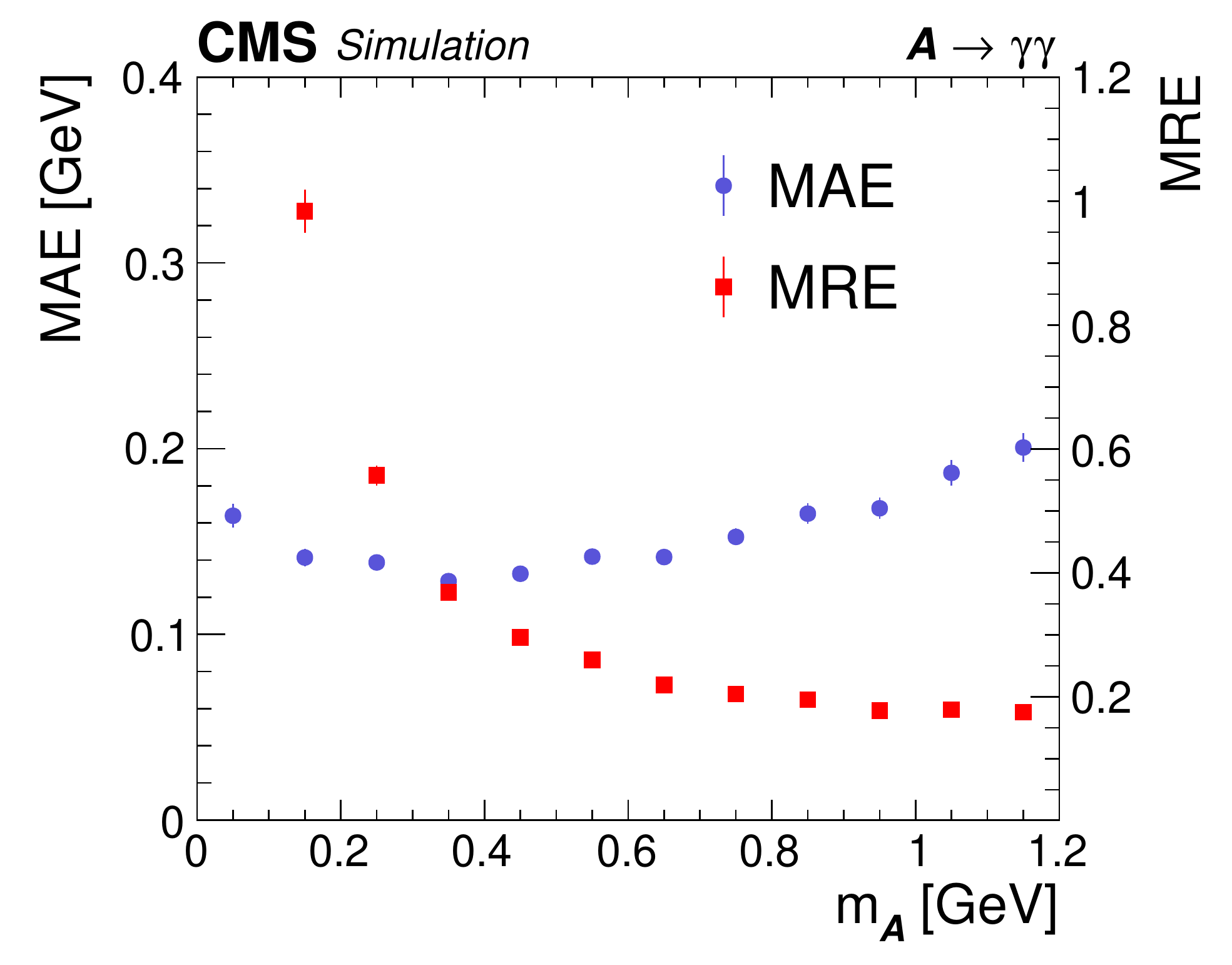}
\;
\includegraphics[height=.4\linewidth]{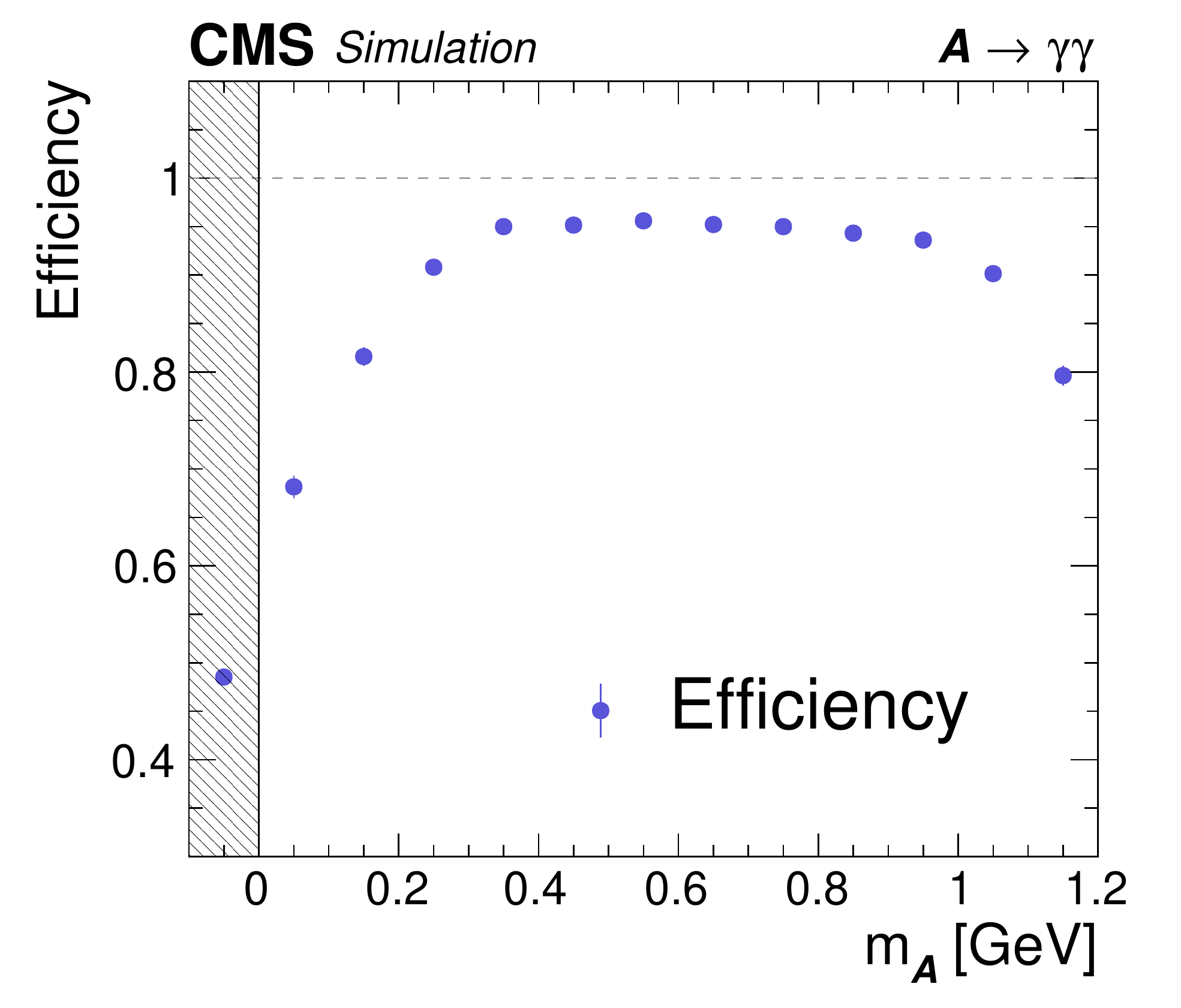}
\caption{Mass regression performance for simulated $\agg$ samples generated uniformly in $(\pt,\ma)$, corresponding to mean boosts in the range $\boostav =$ 600--50 for $\magen =$ 0.1--1.2 GeV. Upper: regressed $\mG$ vs. generated $\ma$. The regressed $\mapred$ is normalized in 0.025\GeV vertical slices of the generated $\ma$. The color scale to the right of the plot gives the normalized number of events per vertical slice in 0.025\GeV bins of $\mG$. Lower left: the MAE (blue circles, use left scale) and MRE (red squares, use right scale) vs. the generated $\ma$. For clarity, the MRE for $\ma < 0.1\GeV$ is not shown since its value diverges as $\magen \to 0$. Lower right: the $\ma$ regression efficiency as a function of the generated $\ma$. The hatched region shows the efficiency for single photons. The vertical bars on the points show the statistical uncertainty in the simulated sample.}
\label{fig:pgun-response}
\end{figure*}

To validate the training of the mass regressor and to characterize its performance, we use a test sample of $\agg$ decays with a continuous uniform $\ma$ distribution. The regressed versus generated mass is shown in Fig.~\ref{fig:pgun-response} (upper). We observe a linear and well-behaved mass response throughout the $\maroi$ range. Since the regressed mass range spans more than an order of magnitude, some variation can be seen in the mass resolution, and thus in the shape of the regressed $\mG$ versus $\ma$ response. Notably, the mass regressor is able to probe the low-$\ma$ regime where it exhibits a gentle and gradual loss in resolution upon approaching the $\mapred = 0$ boundary. As discussed in Sec.~\ref{sec:Train}, this behavior is the result of using domain continuation. This performance confirms the ability of the end-to-end ML technique to access the highest boost regimes, where shower and instrumental merging are present, yet maintain performance into the high-$\ma$ regime, where the particle showers become resolved.

The MAE and MRE as functions of the generated mass are displayed in Fig.~\ref{fig:pgun-response} (lower left). The MAE varies from 0.14 (0.20) \GeV for $\magen$ in the range 0.1 (1.2) \GeV, corresponding to mean boosts of $\boostav =$ 600 (50), respectively. In general, the absolute mass resolution worsens with increasing $\ma$, as reflected in the MAE trend. However, the relative mass resolution tends to improve with mass, as is evident in the MRE values, converging to about $20\%$ for $\ma = 1.2\GeV$. If the $\agg$ mass distribution is used as a signal model, then, for a fixed regression efficiency, a lower relative resolution implies better signal significance, given some fixed background contribution. Notably, as shown in Fig.~\ref{fig:pgun-response} (lower left), for $\magen\lesssim 0.3\GeV$, the MAE starts worsening with decreasing mass. This can be attributed to the gradual deterioration of the mass regressor below the detector's mass resolution.

These figures of merit are achieved with a regression efficiency between 70 and 95\%, as shown in Fig.~\ref{fig:pgun-response} (lower right). The regression efficiency for a given sample as a function of $\magen$ is defined as the number of events in a particular $\magen$ bin that has $\mG$ within $\maroi$, divided by the total number of samples in that bin. If the $\agg$ mass distribution is used as a signal model, then, for a fixed mass resolution, a higher regression efficiency implies better signal significance, for some fixed background contribution. The efficiency is primarily driven by how much of the mass peak can fit within $\maroi$. Thus, it is highest at the midway point of $\maroi$ and falls off to either side. The relatively poorer mass reconstruction at low $\ma$ causes the efficiency to fall off more steeply than it does at high $\ma$. About $50\%$ of single photons are rejected by the $\maroi$ requirement, as seen by the hatched region in Fig.~\ref{fig:pgun-response} (lower right). Photons with $\mapred > 0$ are primarily due to $\Pepem$ conversions.

\section{Physics validation}
\label{sec:Benchmarks}

To validate the physics performance of the end-to-end mass regressor, we
compare it with two traditional reconstruction strategies: a photon
NN-based mass regressor trained on shower-shape and isolation variables,
and a \tbt shower clustering algorithm (described in
Sec.~\ref{subsec:Train-eval}). Although the figures of merit from the previous section provide a high-level characterization of the reconstruction performance, the ultimate physical quantity of interest is the regressed mass spectrum itself, which is the focus of this section.

\subsection{Simulation}
\label{subsec:Benchmarks:MC}

\begin{figure*}[!htbp]
\centering
  \includegraphics[width=.31\linewidth]{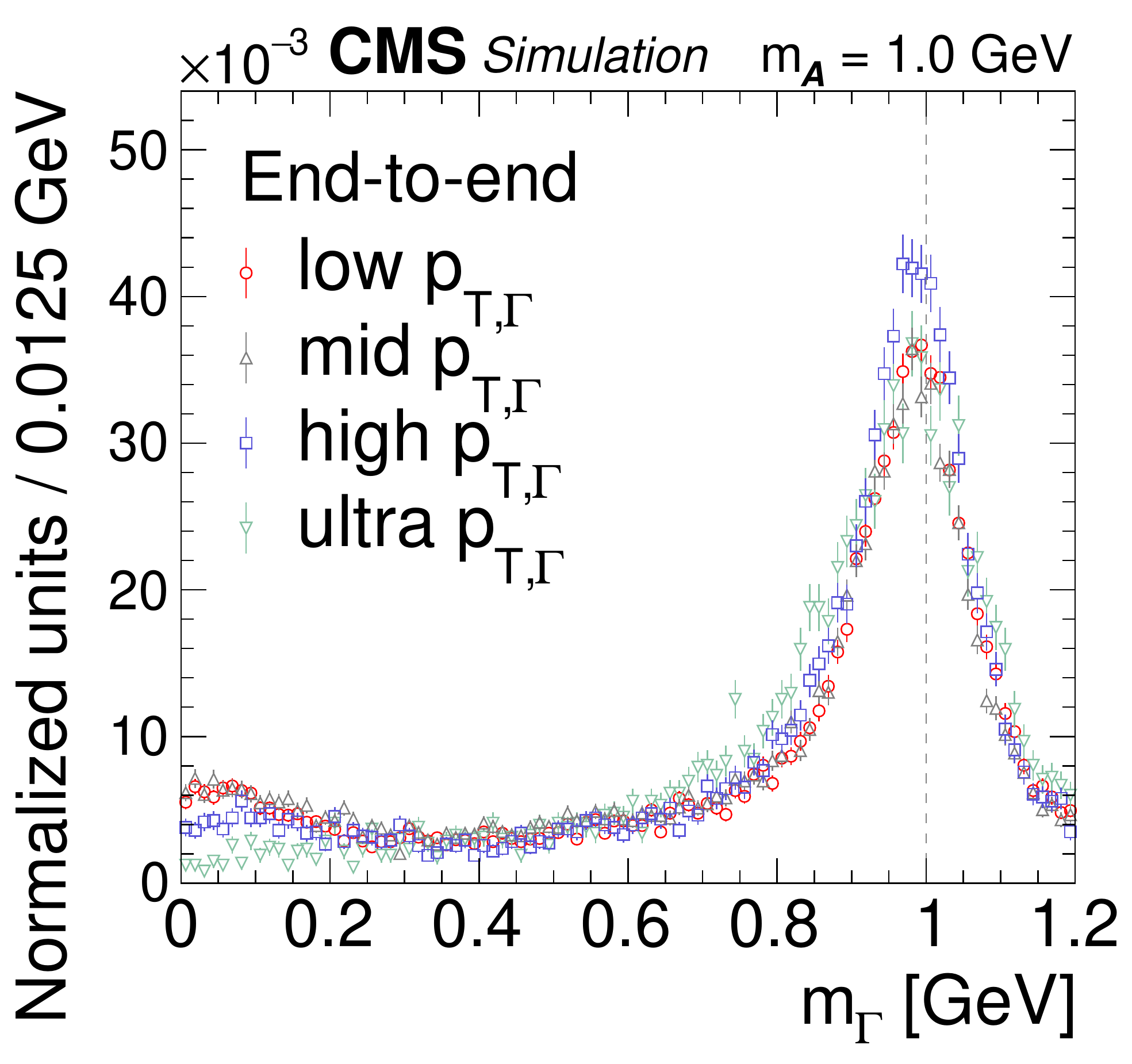}
  \includegraphics[width=.31\linewidth]{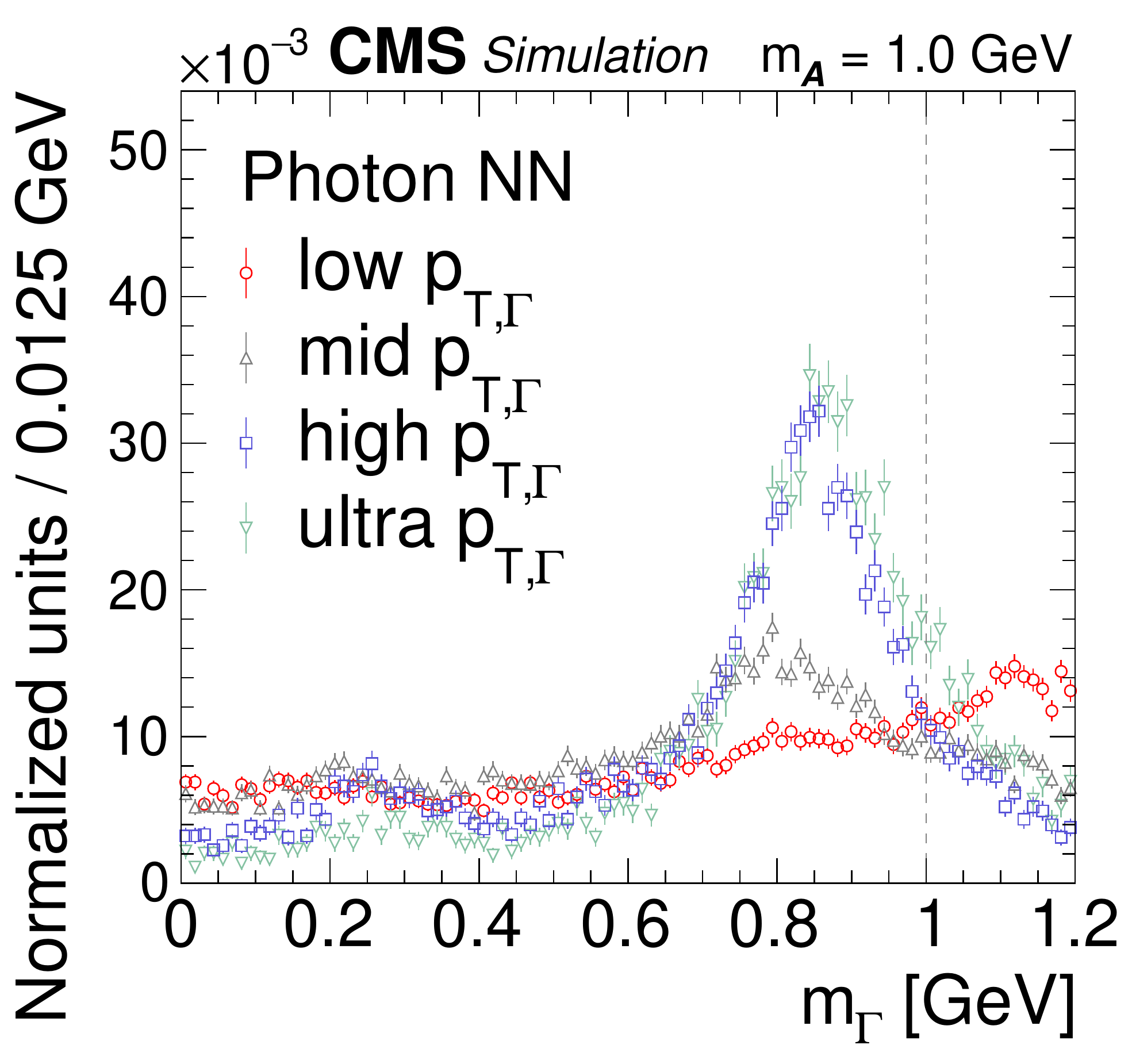}
  \includegraphics[width=.31\linewidth]{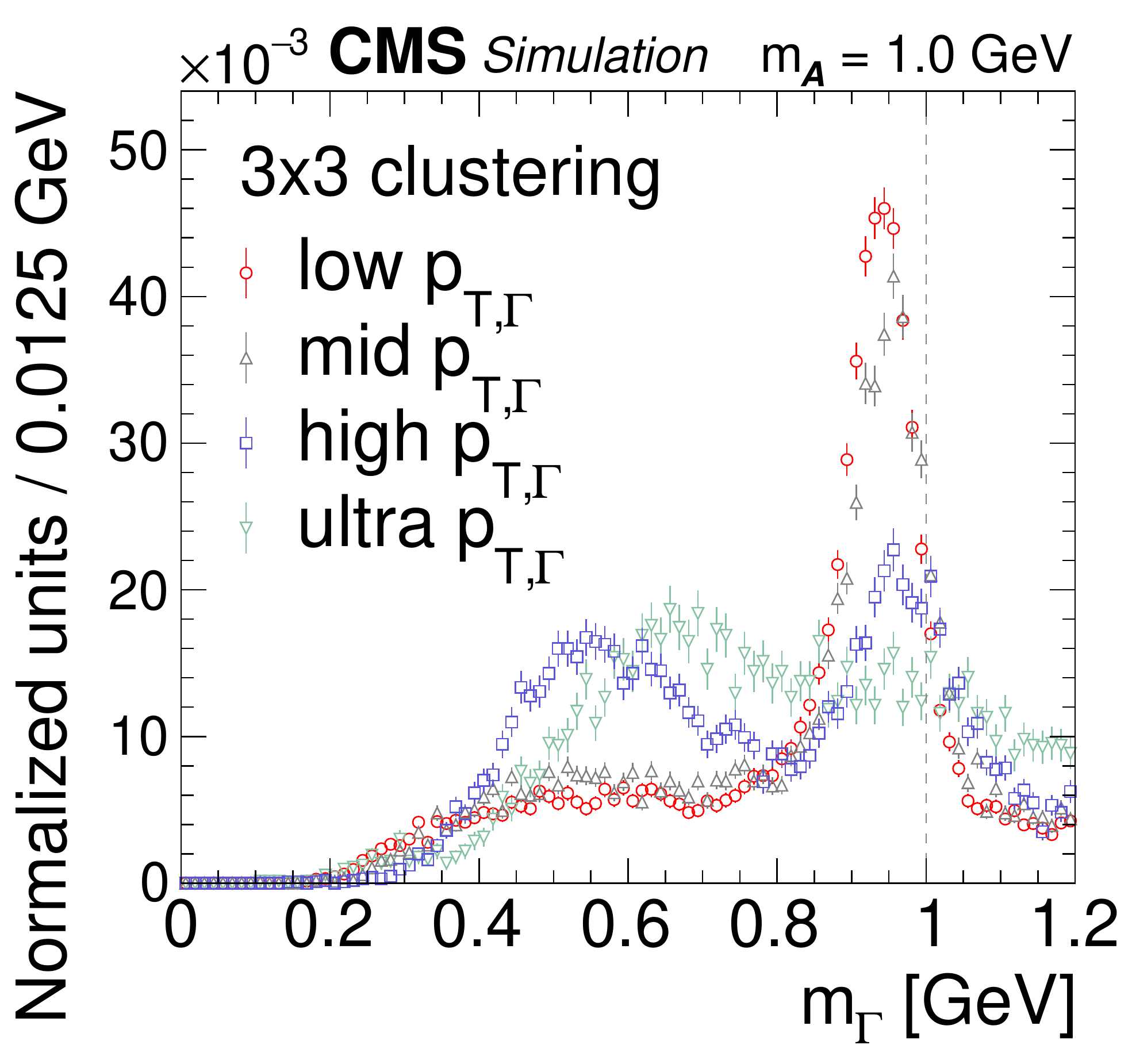}
\\
  \includegraphics[width=.31\linewidth]{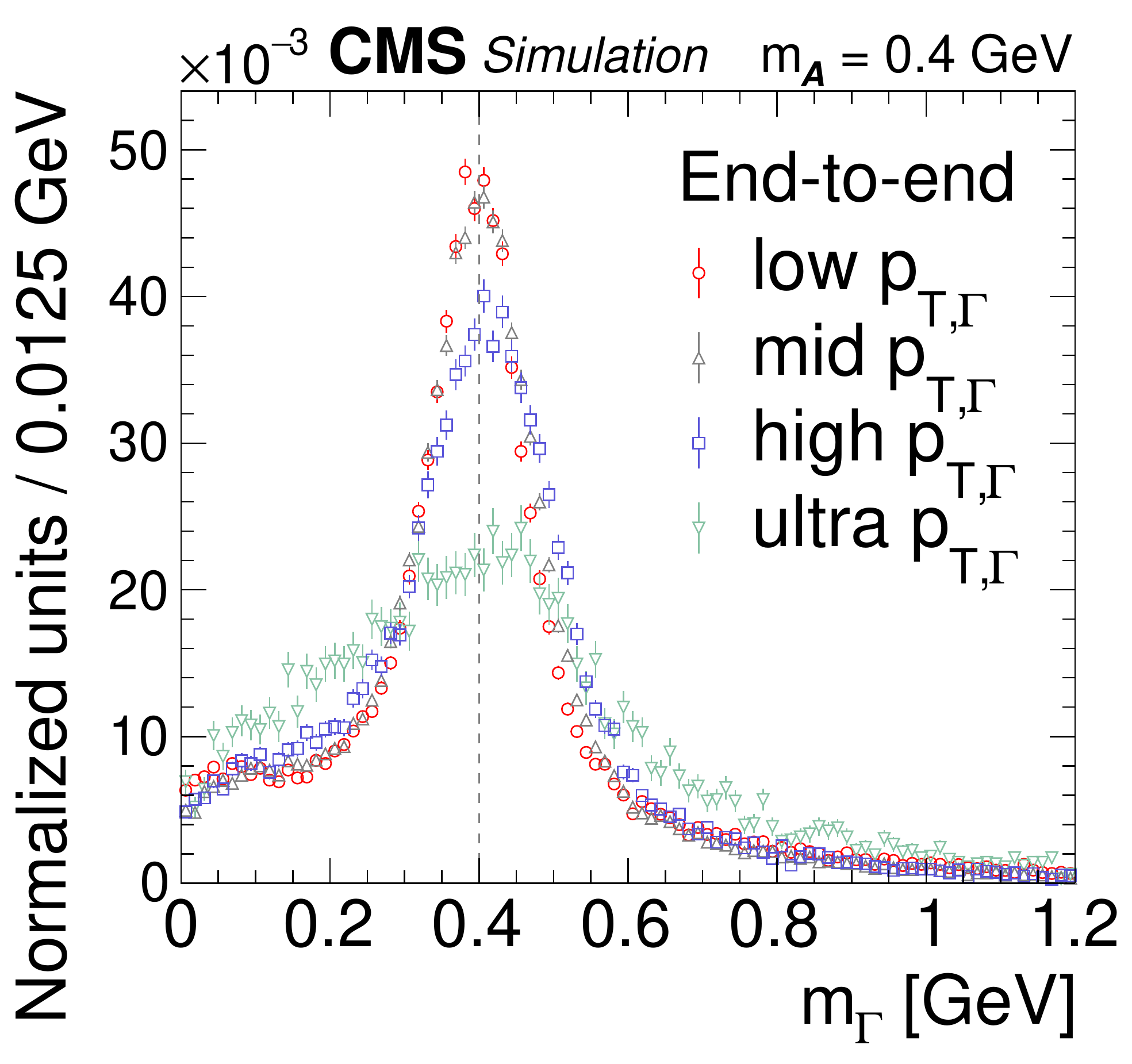}
  \includegraphics[width=.31\linewidth]{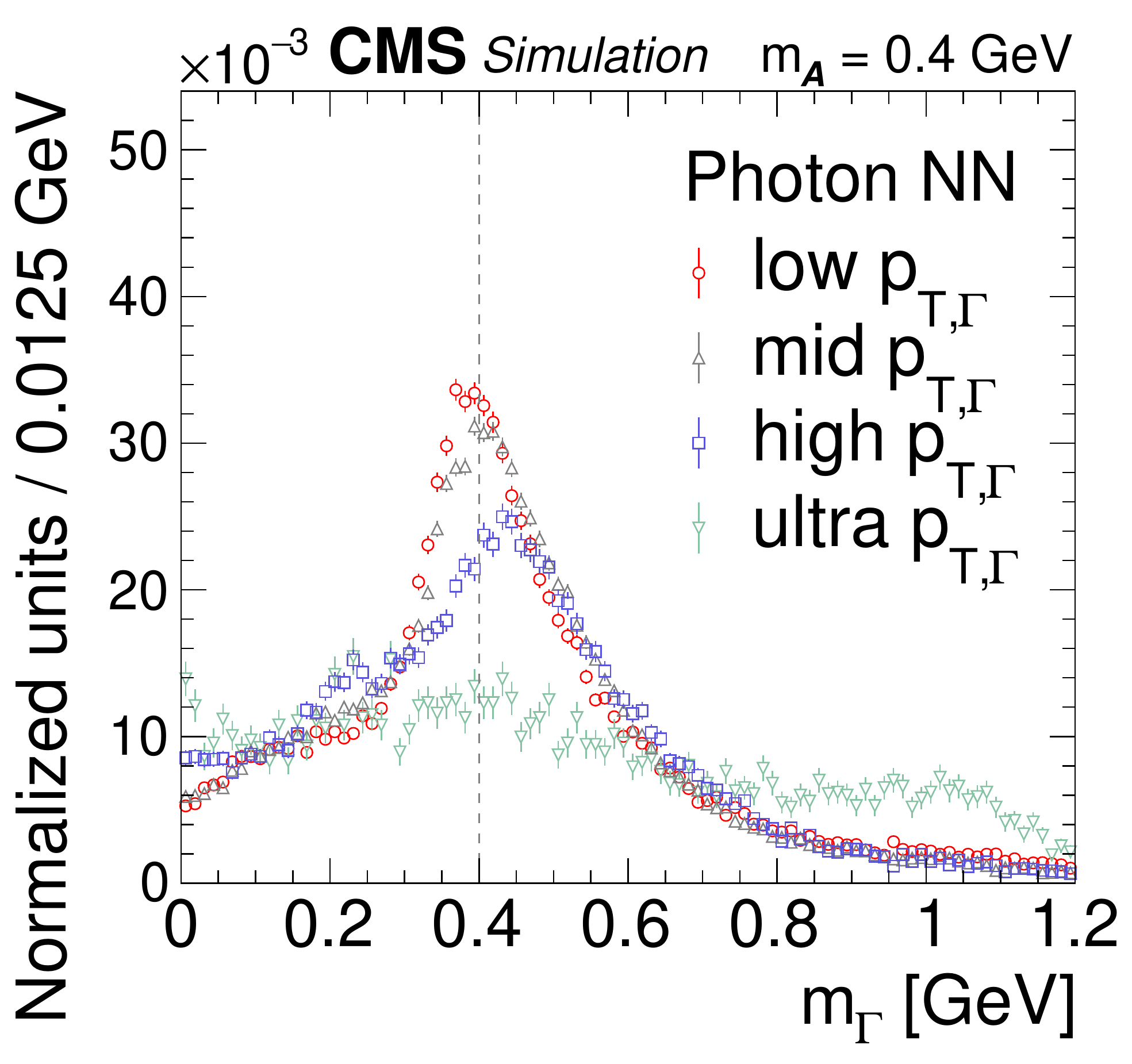}
  \includegraphics[width=.31\linewidth]{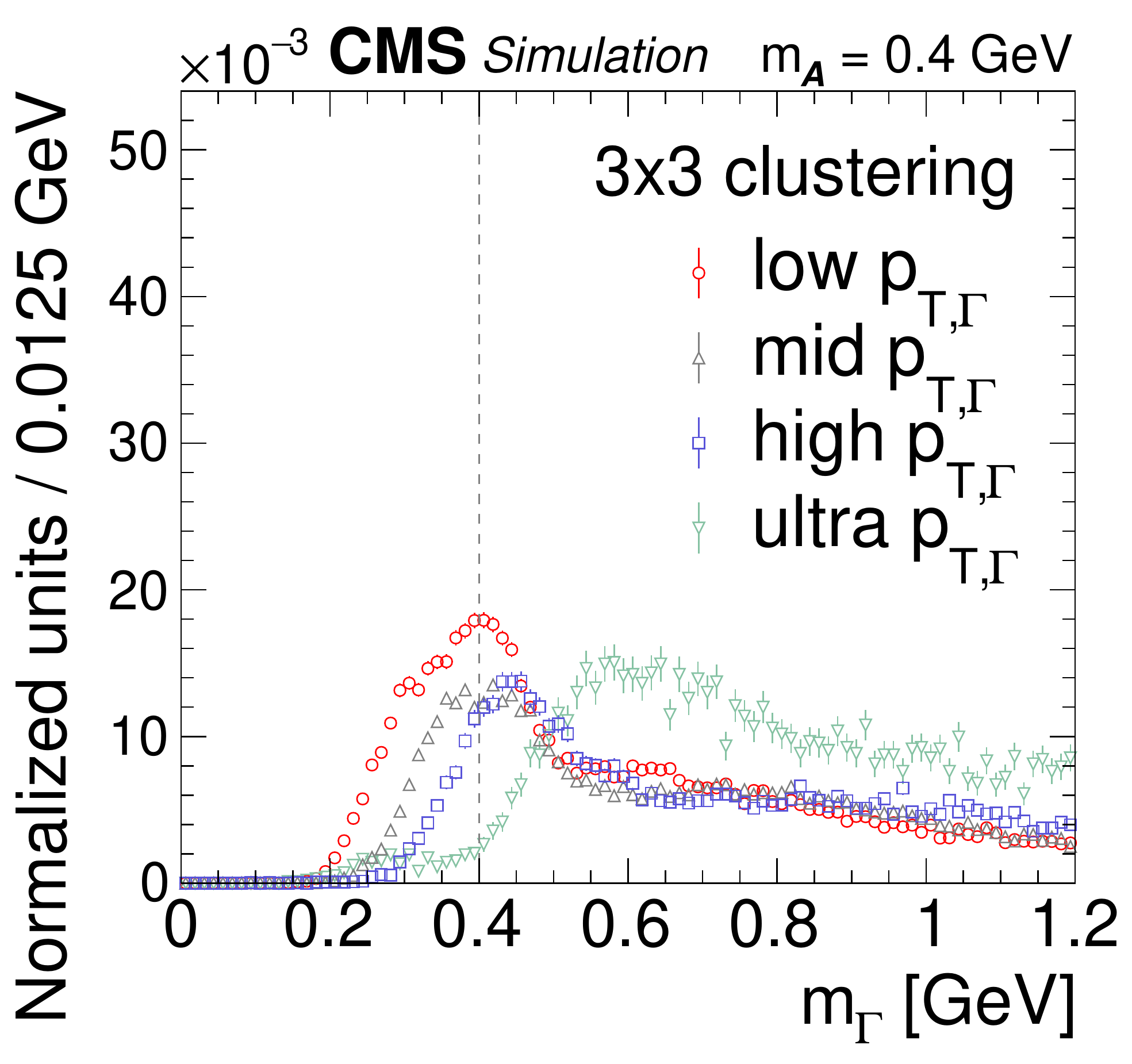}
\\
  \includegraphics[width=.31\linewidth]{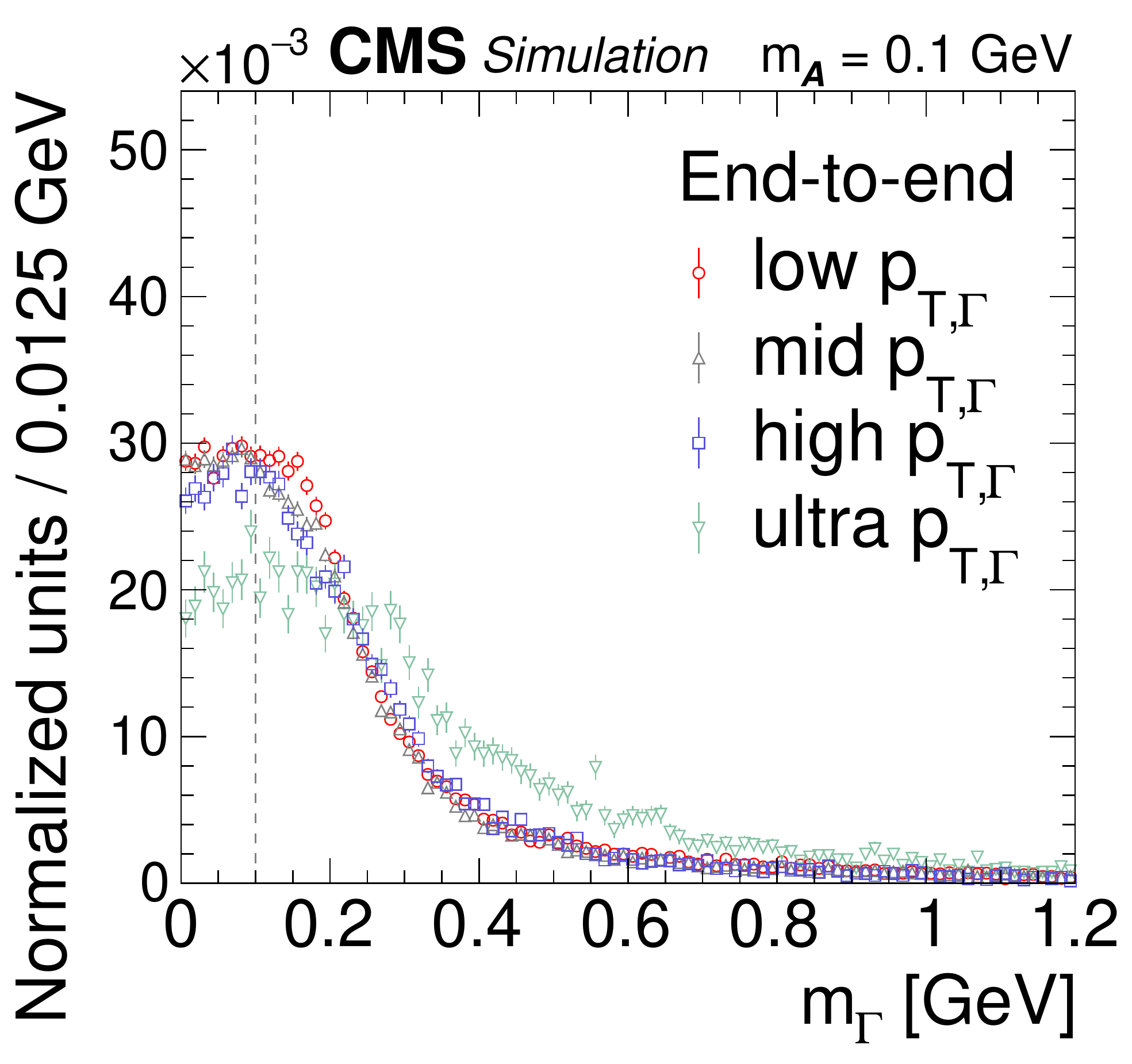}
  \includegraphics[width=.31\linewidth]{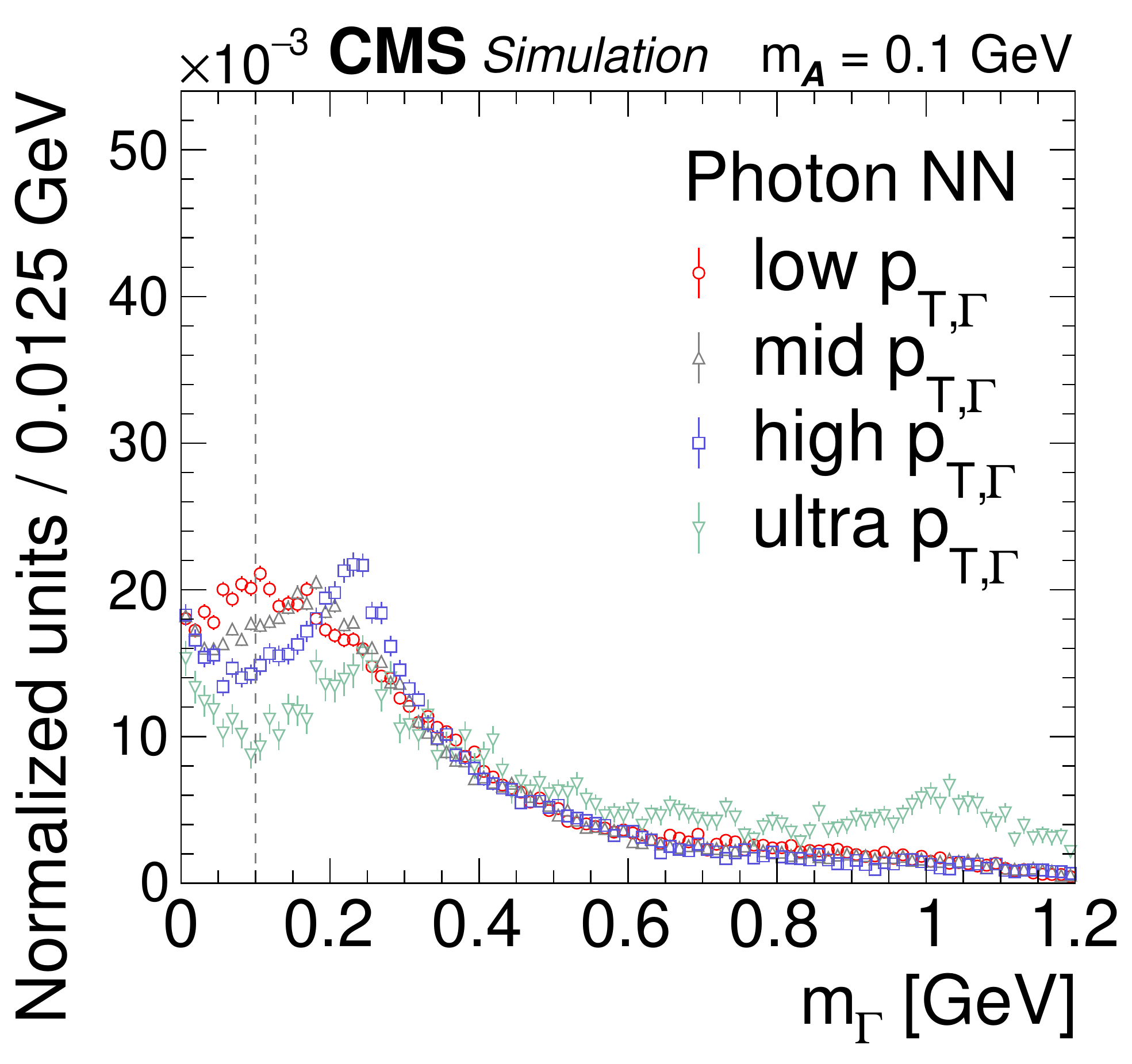}
  \includegraphics[width=.31\linewidth]{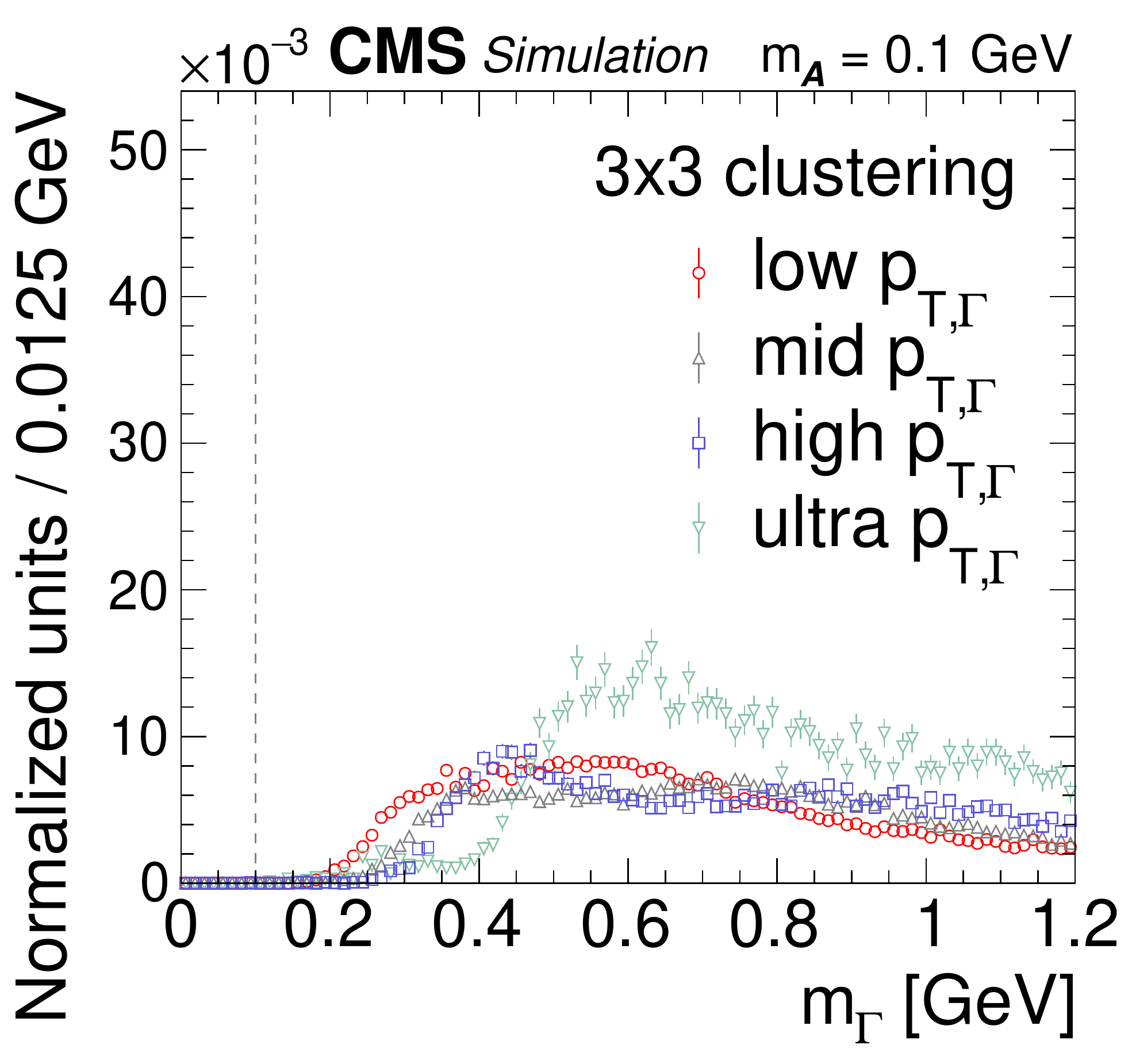}
\\
  \includegraphics[width=.31\linewidth]{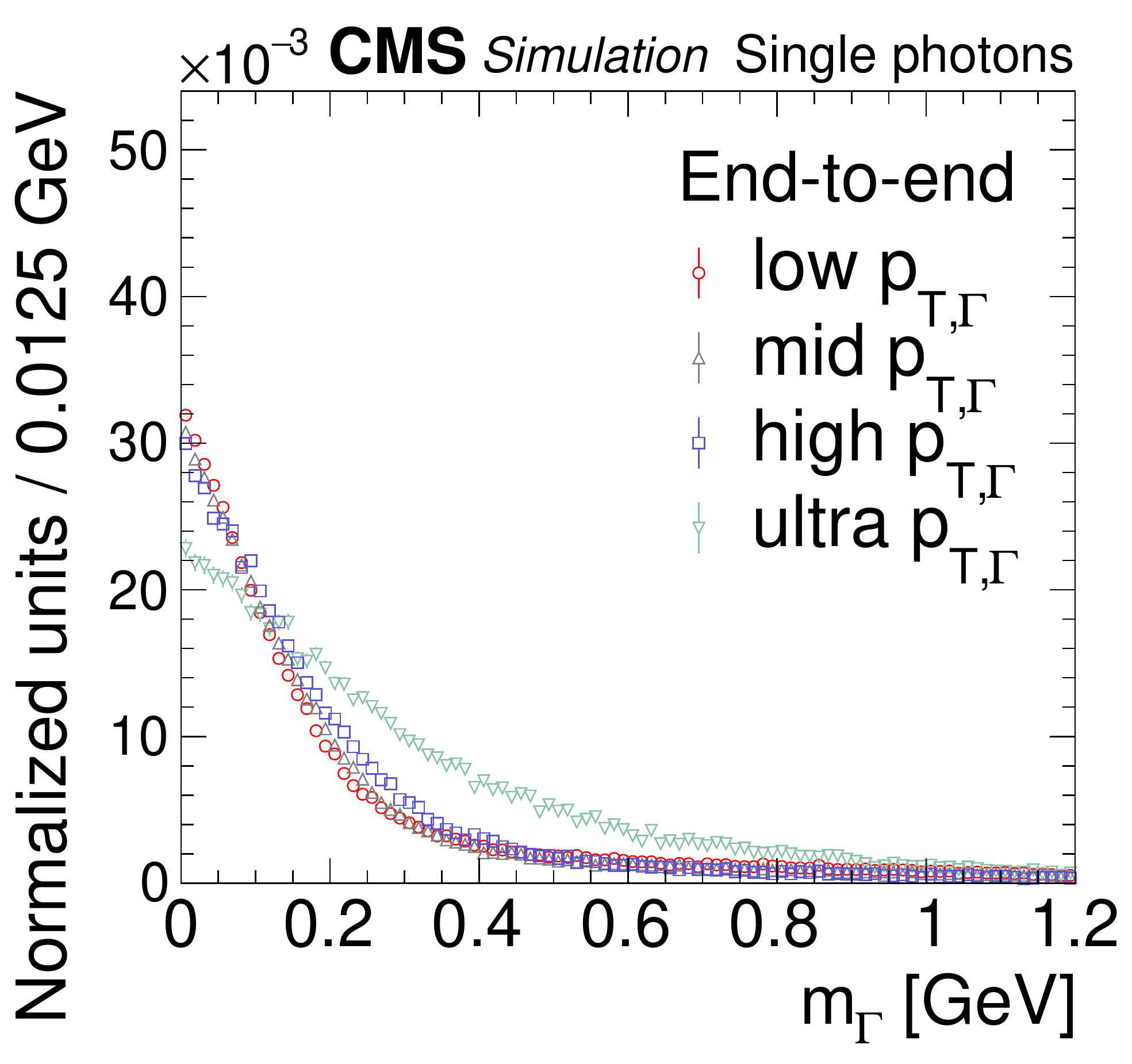}
  \includegraphics[width=.31\linewidth]{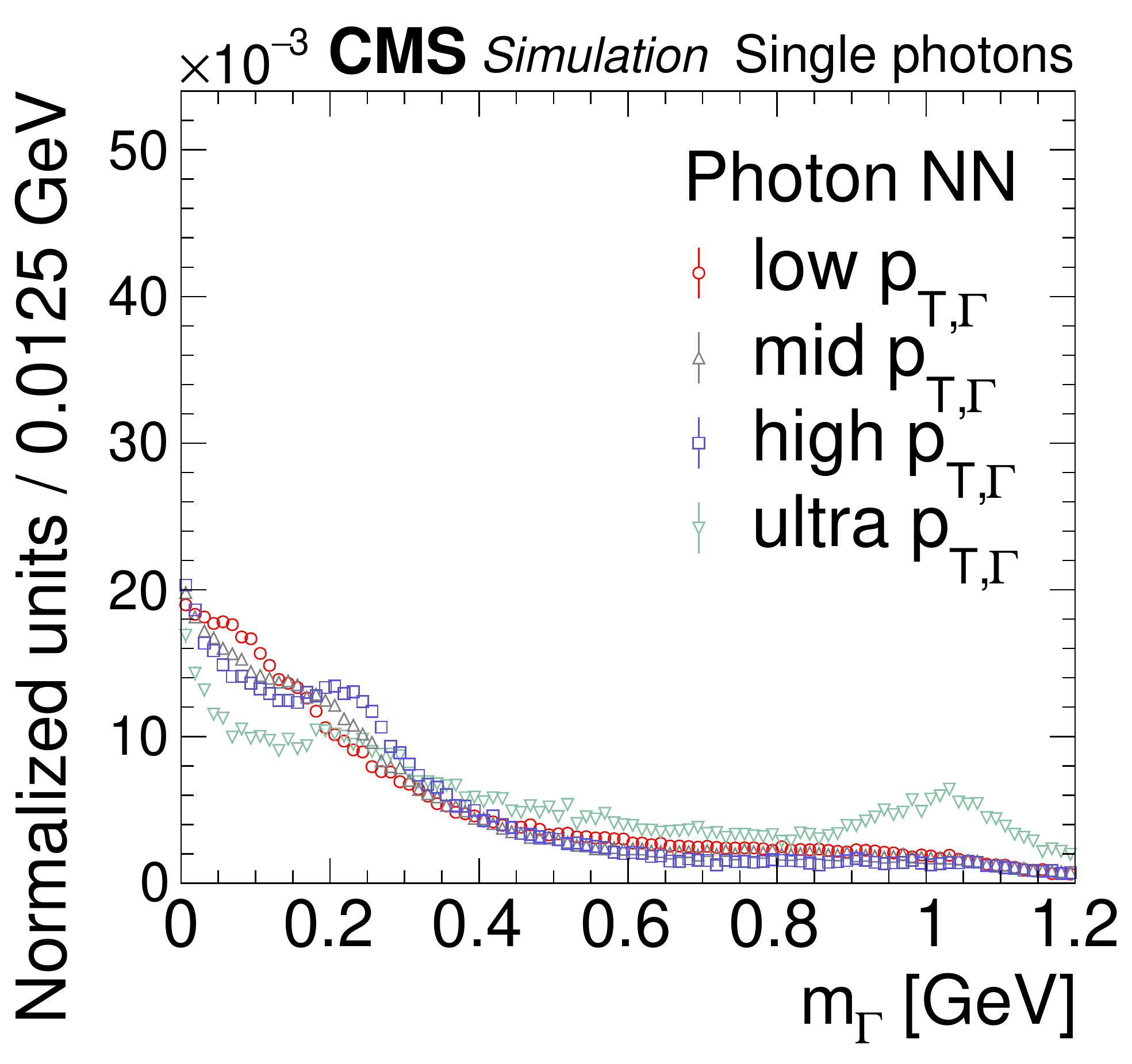}
  \includegraphics[width=.31\linewidth]{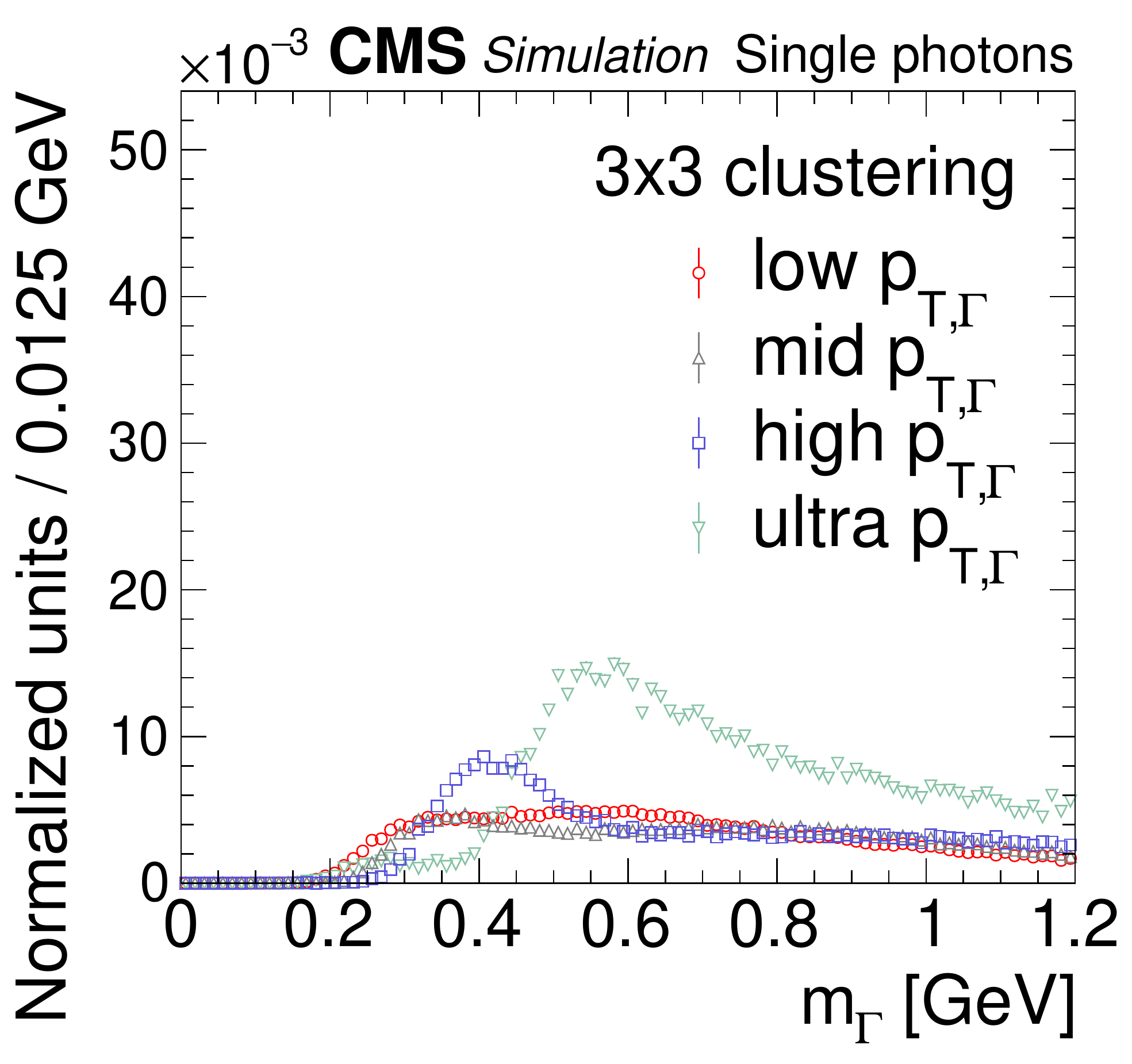}
\\
\caption{Reconstructed mass spectra for end-to-end (left column), photon NN (middle column), and \tbt algorithms (right column) for $\agg$ decays with $\ma = 1.0\GeV$ (upper row), $\ma = 0.4\GeV$ (second row), $\ma = 0.1\GeV$ (third row), and for isolated single photons (lower row). For each panel, the mass spectra are separated by reconstructed $\ptG$ value into ranges of 30--55 GeV (red circles, low-$\ptG$), 55--70 GeV (gray triangles, mid-$\ptG$), 70--100\GeV (blue squares, high-$\ptG$), and $>$100\GeV (green inverted triangles, ultra-$\ptG$). The vertical bars on the points give the statistical uncertainties. All the mass spectra are normalized to unity, including samples outside $\maroi$. The vertical dotted line shows the input $\ma$ value.}
\label{fig:h4g-benchmark}
\end{figure*}

The validation on simulated data compares the mass spectra for $\agg$ decays for various fixed-mass values. As described in Sec.~\ref{sec:Dataset}, we use $\agg$ samples obtained from simulated $\hag$ events with masses $\magen = 0.1$, 0.4, and 1.0\GeV. In these events, the particle $\Pa$ energy is distributed around a median of $\Ea$ $\approx m_{\PH}/2 \approx 60\GeV$, corresponding to median boosts of $\boostav \approx 600$, $150$, and $60$ for the respective $\ma$~masses. In addition to the photon selection criteria, event selection criteria similar to those used in the CMS $\hgg$ analysis are applied. The reconstructed mass spectra are shown in Fig.~\ref{fig:h4g-benchmark} for the different algorithms and $\ma$ mass values. For each mass value, representing a different median boost regime, the samples are further broken down by ranges of reconstructed $\ptG$, to highlight the stability of the mass spectrum with energy. These $\ptG$~ranges are:
\begin{itemize}
    \item[(i)] low-$\ptG$: $30 < \ptG < 55\GeV$,
    \item[(ii)] mid-$\ptG$: $55 < \ptG < 70\GeV$,
    \item[(iii)] high-$\ptG$: $70 < \ptG < 100\GeV$, and
    \item[(iv)] ultra-$\ptG$: $\ptG > 100\GeV$.
\end{itemize}
Some overlap in the boost values across different $\ma$ masses is to be expected. Note that the mass regressor has been trained on samples with $\ptG < 100\GeV$ and that reconstructed candidates only in the range $\abs{\etaG} < 1.44$ are used.

For boosts $\boostav \approx 60$ and $\magen = 1.0\GeV$, only the end-to-end algorithm (Fig.~\ref{fig:h4g-benchmark}, upper left) consistently reconstructs a mass peak for all $\ptG$ ranges. The position of the mass peak remains stable, with the resolution improving in the high-$\ptG$ category. The end-to-end regression performs best when the $\agg$ decay products are moderately merged, and neither fully resolved nor fully merged. The mass peak in the ultra-$\ptG$ category is well reconstructed despite being outside the trained phase space. This demonstrates that the phase space extrapolation is effective for internally learned quantities like $\ptG$.

For $\magen = 1.0\GeV$, the photon NN (Fig.~\ref{fig:h4g-benchmark}, upper middle) has difficulty reconstructing the mass peak, except at the higher-$\ptG$ categories. This can be understood in terms of the information content of the input variables on which the algorithm was trained. At higher $\ptG$, the two photons are more likely to be moderately merged so that their showers are contained within the \fbf crystal block where the shower-shape variables are defined. At lower $\ptG$, the two photons are more often resolved so that the lower-energy photon shower falls outside the \fbf crystal block. The photon NN must then rely on the isolation variables, which are defined much more coarsely over a cone of $\sqrt{\smash[b]{(\Delta\eta)^2+(\Delta\phi)^2}} < 0.3$ about the seed crystal. Since these variables have much less discriminating power, this results in a steep falloff in reconstruction performance. To improve the performance, the photon NN could be augmented with the momentum components of the lower-energy photon, in instances where the PF is able to reconstruct it.

Lastly, for $\magen = 1.0\GeV$, the \tbt algorithm (Fig.~\ref{fig:h4g-benchmark}, upper right) is the only one competitive with the end-to-end method for lower $\ptG$. As the photon clusters become resolved, the \tbt method thus becomes an effective tool for mass reconstruction. However, as soon as the clusters begin to merge at higher $\ptG$, a sudden dropoff in reconstruction efficiency occurs since the \tbt algorithm is unable to compute a mass for a single cluster. A spurious peak develops at $\mapred \approx 0.5\GeV$ for decays with sufficient showering prior to the ECAL. The \tbt method is thus only useful for a limited range of low boosts. The comparatively weaker performance of the end-to-end method to the \tbt one at low-$\ptG$ is attributable to the underrepresentation of relevant boosts in the training sample. The optimization of the end-to-end technique for these lower boosts is beyond the scope of this paper.

For boosts $\boostav \approx 150$ and $\magen = 0.4\GeV$, the end-to-end method (Fig.~\ref{fig:h4g-benchmark}, second row, left) is able to reconstruct the mass peak with full sensitivity across most of the $\ptG$ ranges. Only at the highest $\ptG$ range does the mass peak significantly degrade, although it is still reasonably well behaved. Training with higher $\ptG$ could potentially improve this behavior. The photon NN performs its best in this regime (Fig.~\ref{fig:h4g-benchmark}, second row, middle) because a majority of the photon showers fall within the \fbf crystal block. However, the mass resolution is still significantly worse compared with the end-to-end method. The \tbt algorithm (Fig.~\ref{fig:h4g-benchmark}, second row, right) is barely able to reconstruct a mass peak for these boosts. We recall that, if the \tbt algorithm is unable to find a pair of clusters, an invariant mass cannot be calculated, and a default mass value outside $\maroi$ is instead passed.

For boosts $\boostav \approx 600$ and $\magen = 0.1\GeV$, the end-to-end method (Fig.~\ref{fig:h4g-benchmark}, third row, left) reaches the limits of its sensitivity, although it is still usable. We attribute the performance of the end-to-end method to its ability to discern subtle differences in the energy distribution of the fully merged particle shower, which we expect to have more smearing along the principal axis connecting the $\agg$ diphotons. Notably, even at these boosts, the position of the mass peak measured with the end-to-end method remains stable with $\ptG$. This is not the case for the photon NN (Fig.~\ref{fig:h4g-benchmark}, third row, middle) whose peak becomes erratic and displaced with increasing $\ptG$. The \tbt method is not able to calculate a mass at this level of merging for the same reasons stated earlier.

For reference, the regressed mass spectrum for single, isolated photons from $\hgg$ decays is shown in Fig.~\ref{fig:h4g-benchmark} (lower row). Both the end-to-end (left) and photon NN (middle) methods are able to regress to the $\mapred \approx 0\GeV$ boundary, with a smoothly falling distribution since they were trained with domain continuation (cf. Fig.~\ref{fig:naivereg}, right). The remaining photons within $\maroi$ come from photon conversions that acquire an effective mass because of nuclear interactions.

Summarizing the validation performance on simulated data, the end-to-end ML technique is the only one that is able to robustly and consistently probe boost regimes ranging from resolved to instrumentally merged showers.

\subsection{Data}
\label{subsec:Benchmarks:Data}

To validate the findings described above from simulated events, we perform a cross-check using $\gjet$ events from CMS data, where the jet is reconstructed as a photon candidate. The CMS dataset was acquired in 2017 at the LHC with proton-proton collisions at $\sqrt{s} =$ 13\TeV. In Sec.~\ref{subsec:Stability:conditions}, we also compare the results from this dataset with a similar one acquired in the 2018 LHC running. These two datasets correspond to integrated luminosities of 41.5 and $56.9\fbinv$, respectively.
If the jet contains an energetic, collimated neutral-meson decay ($\ptG \gtrsim 20\GeV$), typically a $\pgg$ or $\egg$, it will be misreconstructed as a single photon $\Gamma$ and the event will pass a diphoton trigger. Since the energy of the jet will, in general, be shared among several constituent particles, the $\Pgpz$ is more likely to be reconstructed as the lower-energy photon in the event. Thus, a data sample enriched in merged photons is obtained by selecting events passing a diphoton trigger and selecting the lower-energy reconstructed photon, which we additionally require to pass our photon selection criteria. The selected sample is then given to the $\mG$ regressor, whose output we study below. Further details about the trigger and photon selection criteria can be found in the $\hgg$ analysis~\cite{hgg}. We emphasize that the mass regressor is being used to regress the $\mapred$ of individual reconstructed photon candidates, which we assume to be merged photons, not the invariant mass of the reconstructed diphoton event itself. As before, only reconstructed candidates in the range $\abs{\etaG} < 1.44$ are used.

An important caveat in regressing the $\mapred$ of energetic photons
($\ptG \gtrsim 20\GeV$) within jets is the presence of other hadrons
within the jet. At these energies, we emphasize that it is no longer the
case that neutral-meson decays are well isolated in jets, a main point of distinction compared with the isolated $\agg$ decays used to train the mass regressor. In general, the neutral-meson decay will be collimated with other hadrons, including, potentially, several merged $\Pgpz$ decays. The effect of these additional hadrons is to smear and distort the resulting $\mapred$ spectrum and introduce an energy dependence in the $\mapred$ value. We therefore restrict our study to $20 < \ptG < 35\GeV$ and require tighter shower-shape criteria, to increase the contribution from well-isolated $\pgg$ decays that more closely resemble the $\agg$ decay. The low-$\pt$ threshold is dictated by the chosen diphoton trigger. Although these tighter criteria mitigate the stated effects, the impact of these effects remains visible.

Within the above restricted $\ptG$ range, the $\Pgpz$ is boosted to the approximate range $\boost =$ 150--250, putting its invariant mass reconstruction out of reach of all but the end-to-end mass regressor. Also present is the higher-mass $\Pgh$, which, though produced with a much lower cross section, is boosted to only about the range $\boost =$ 30--60, just within reach of the \tbt algorithm.

\begin{figure}[!htbp]
\centering
\includegraphics[width=.47\textwidth]{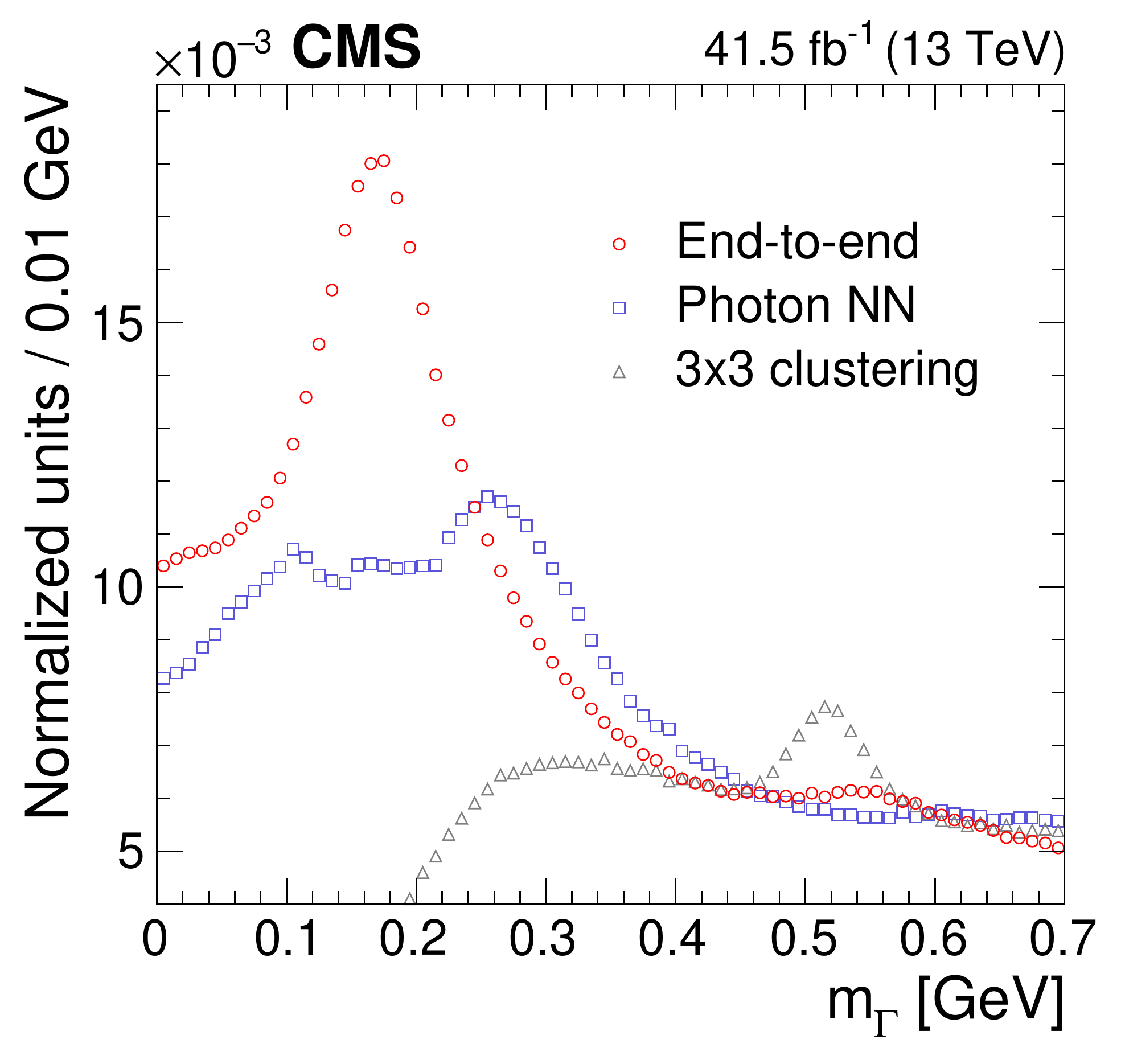}
\hfill
\caption{Reconstructed mass $\mG$ for end-to-end (red circles), photon NN (blue squares), and \tbt (gray triangles) algorithms for hadronic jets from data enriched with $\pgg$ decays. All distributions are normalized to the same number of events, including those outside $\maroi$. The statistical uncertainties in the distributions are negligible.
}
\label{fig:pi0-benchmark}
\end{figure}

As clearly seen in Fig.~\ref{fig:pi0-benchmark}, the end-to-end method is able to reconstruct a prominent $\Pgpz$ peak. Indeed, it is the only algorithm able to do so. The $\Pgpz$ peak appears more prominent than the corresponding $\Pa$ peak at $\ma=0.1\GeV$ (cf. Fig~\ref{fig:h4g-benchmark}, third row, left) because of the higher mass and lower boost of the $\Pgpz$ in this case. The photon NN exhibits an erratic response, suggesting it does not have the data granularity needed to probe this regime. Likewise, the \tbt method is unable to reconstruct the $\Pgpz$ peak at all. It is, however, able to reconstruct the $\Pgh$ peak, as expected. We attribute the weaker $\Pgh$ peak in the end-to-end method to the aforementioned smearing effect of additional hadrons in the jet, as discussed further in Appendix~\ref{app:shower3b3}. The \tbt method is less sensitive to this effect because of the restriction on the ECAL energy clusters being in the smaller \tbt window.

Whether the sensitivity of the end-to-end method to the effects of jet hadronization represents an advantage or disadvantage depends on the application. For an analysis searching for isolated $\agg$ decays~\cite{gluexagg}, background processes from neutral mesons in jets will be smeared in mass, providing a distinct advantage for separating their mass spectra from that of true $\agg$ decays peaking at similar masses. The optimization of the end-to-end technique for the mass regression of neutral mesons in jets is beyond the scope of this paper.

The unique capability of the end-to-end technique to reconstruct highly boosted particle decays thus opens the door to physics searches in boost regimes previously inaccessible to existing reconstruction algorithms. Additionally, because of the difficulty of obtaining low-energy $\Pgpz$ decays ($E\approx1\GeV$) with increasing luminosity, the ability to reconstruct the more abundantly available high-energy ($E\approx10\GeV$) $\Pgpz$ decays instead offers the possibility of improving the reach of existing CMS ECAL intercrystal calibration methods, which rely on such decays~\cite{CMS:EGM-14-001}.

\section{Robustness of the algorithm}
\label{sec:Stability}

To further assess the robustness and generalizability of the end-to-end ML-based mass regressor, we study how the regressed mass varies with respect to a number of key quantities of interest. Such studies are useful in revealing potential biases of the mass regressor technique to kinematic regions and detector conditions for which it was not trained. These mass dependence studies are performed on data using both $\pgg$ events and electrons from events enriched with $\zee$~decays. 

\subsection{Mass dependence on kinematic quantities}
\label{subsec:Stability:kinematics}

\begin{figure*}[!htbp]
\centering
\includegraphics[height=.38\linewidth]{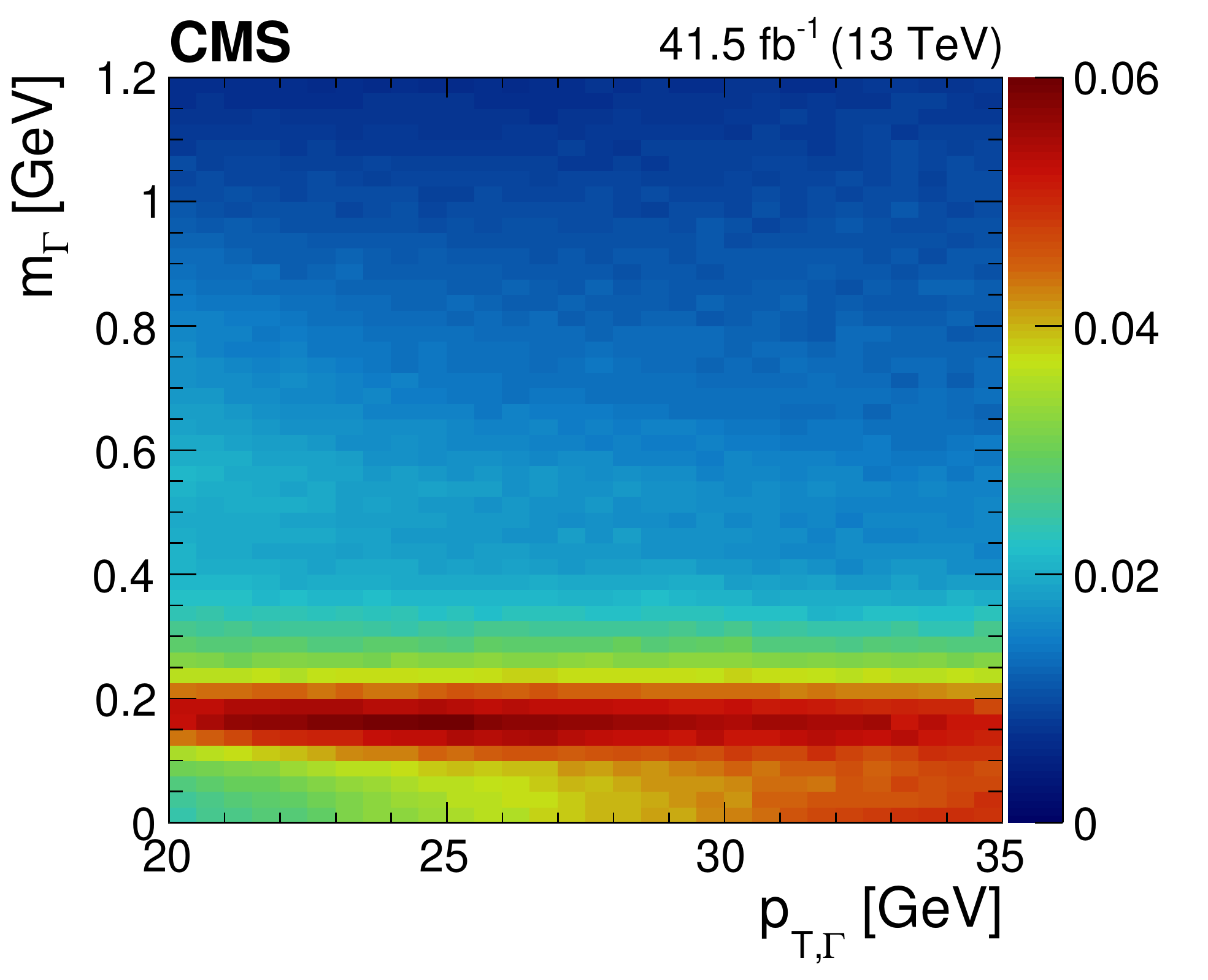}
\quad
\includegraphics[height=.38\linewidth]{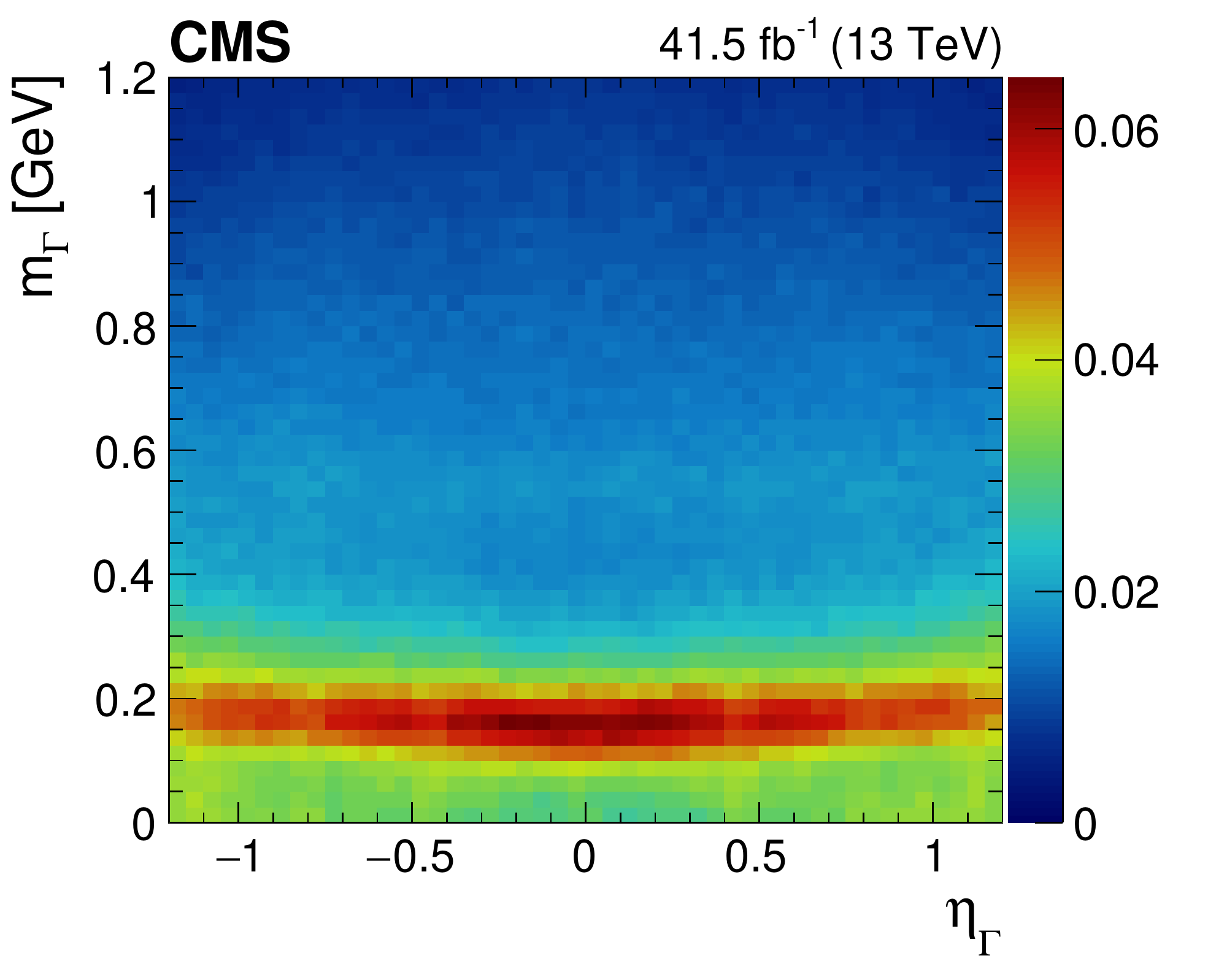}
\caption{Regressed mass from $\pgg$ data events vs. $\ptG$ (left) and $\etaG$ (right). For both plots, the regressed mass distributions are normalized in vertical slices of the accompanying kinematic quantity to highlight the intrinsic dependence on the quantity. The relative contribution over each vertical slice is given by the color scale to the right of each plot.}
\label{fig:pi0-stability-kinematics}
\end{figure*}

We first measure the dependence of the regressed mass on reconstructed kinematic quantities such as $\ptG$ and $\etaG$. These studies have the caveat outlined in Sec.~\ref{subsec:Benchmarks:Data} concerning the distortions in the regressed $\Pgpz$ invariant mass distribution coming from jet hadronization. Figure~\ref{fig:pi0-stability-kinematics} (left) shows a two-dimensional plot of the regressed mass versus $\ptG$ for $20 < \ptG < 35\GeV$. A clear band is observed that is independent of $\ptG$. We attribute this band to well-isolated $\pgg$ decays, which are more prominent in this relatively low-$\ptG$ range. This is consistent with our earlier results from simulated $\agg$ decays, shown in Fig.~\ref{fig:h4g-benchmark}, albeit for a narrower $\ptG$ range. A broadening of the mass distribution for $\ptG \gtrsim 30\GeV$ is also visible in Fig.~\ref{fig:pi0-stability-kinematics} (left). However, this is likely an artifact of the discontinuous increase in the photon population of the selected sample, due to the turn-on of the leading-$\pt$ threshold for the diphoton trigger.

Next, we study the dependence of the regressed mass on the reconstructed $\etaG$. As discussed earlier, the number of radiation lengths traversed by an electromagnetic particle entering the ECAL barrel at $\eta \approx 1.4$ is nearly fivefold that at $\eta \approx 0$ because of differences in the underlying tracker material structure (described in Sec.~\ref{sec:Detector}). Thus, this study is a useful check on the sensitivity of the regressed mass to the level of electromagnetic shower development in the $\agg$ energy deposits. As seen in Fig.~\ref{fig:pi0-stability-kinematics} (right), the regressed mass has noticeably better resolution in the central region ($\abs{\etaG}\lesssim 0.5$) than in the forward regions ($\abs{\etaG}\gtrsim 1$), as expected. The regressed mass distribution also shows good continuity with respect to $\etaG$. The dependence of the regressed mass on the same kinematic quantities for simulated $\hag$ events is presented in Appendix~\ref{app:h4g-stability}, with similar conclusions.

\subsection{Mass dependence on detector conditions}
\label{subsec:Stability:conditions}

A critical question surrounding the end-to-end ML technique is how well
it accommodates mismodeled detector conditions, or even conditions
unseen in the training set; in particular, whether exposure to the full granularity of the detector data results in an algorithm that is overly dependent on detector conditions or the amount of pileup.

\begin{figure*}[!htbp]
\centering
\includegraphics[height=.4\linewidth]{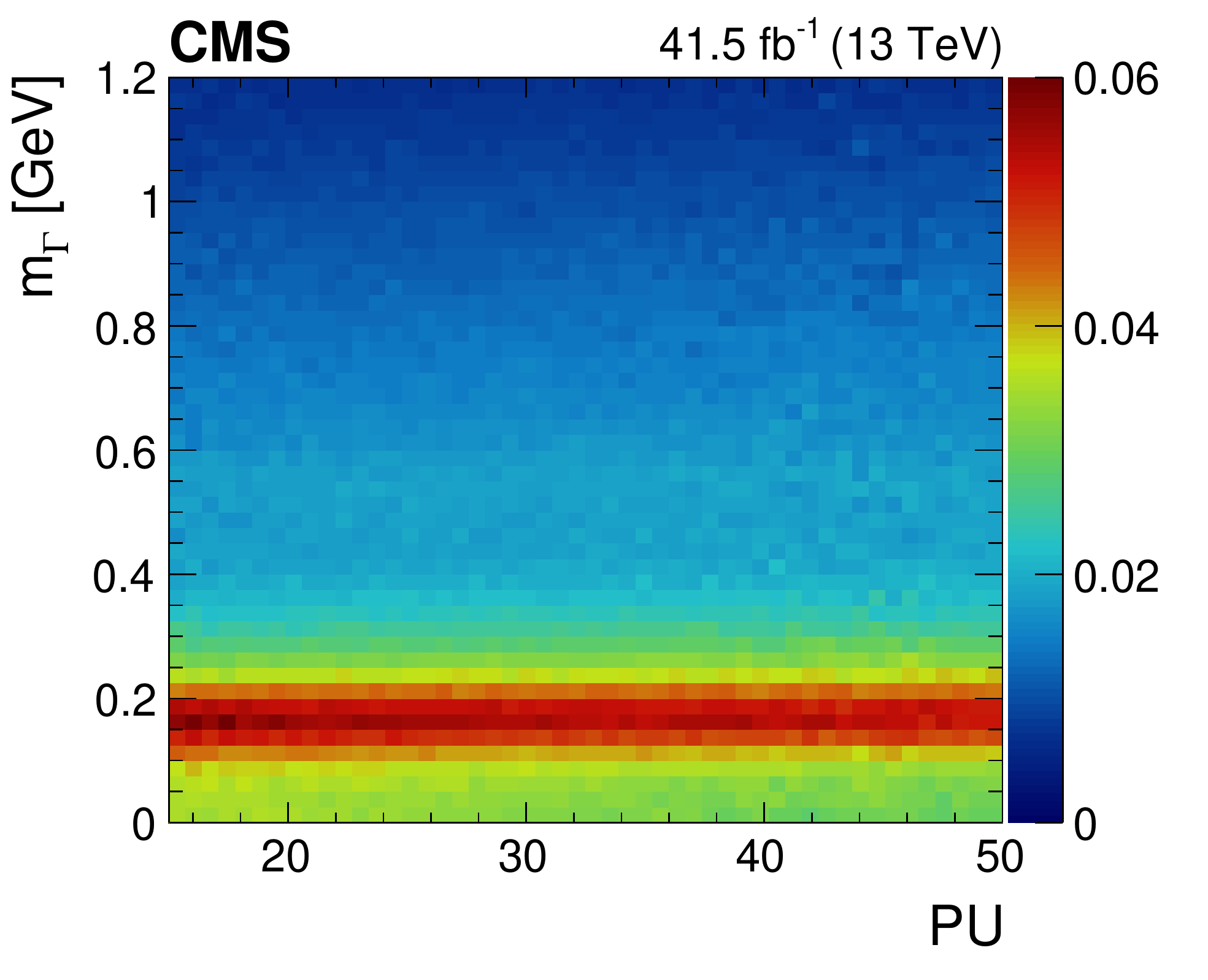}
\\
\includegraphics[height=.45\linewidth]{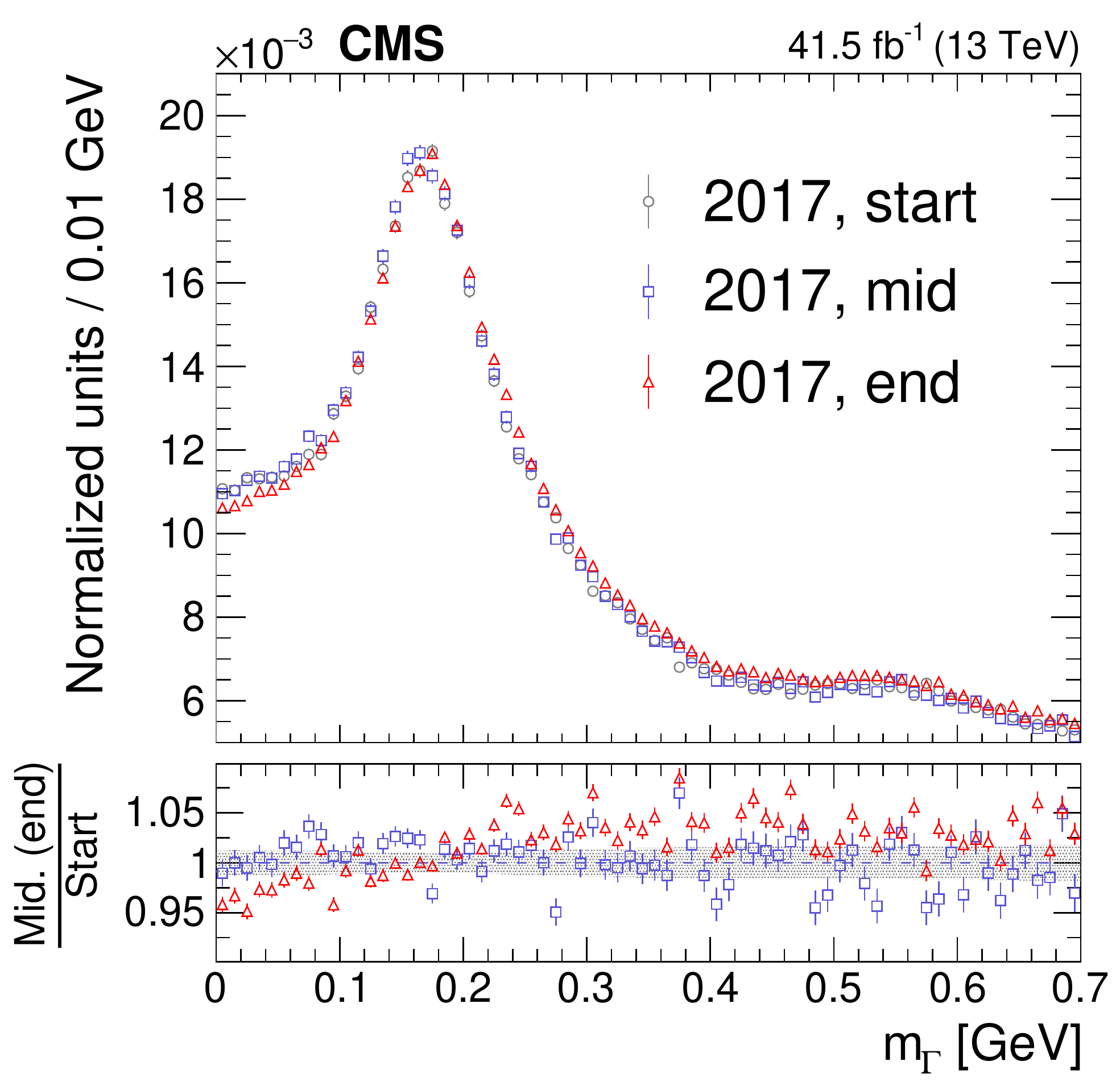}
\qquad
\includegraphics[height=.45\linewidth]{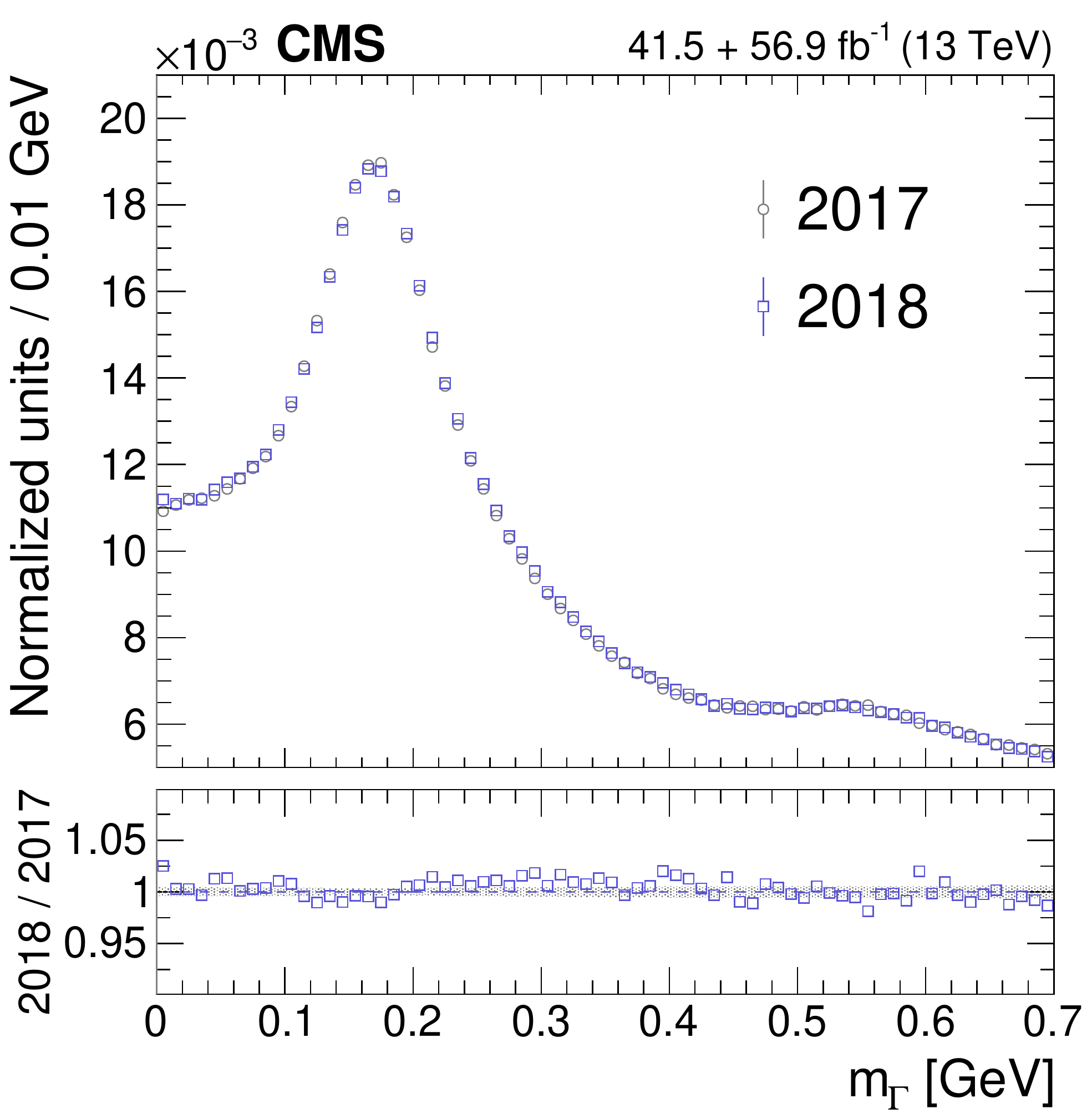}
\caption{Upper: two-dimensional plot of the regressed mass for $\pgg$ data events vs. the amount of pileup (PU). The mass distribution is normalized in vertical slices of the amount of pileup. The relative contribution over each vertical slice is given by the color scale to the right of the plot. Lower left: the regressed mass distributions for the start (gray circles), middle (blue squares), and end (red triangles) of data taking during the year 2017. Lower right: the regressed mass distributions for the 2017 (gray circles) and 2018 (blue squares) data-taking periods. The lower two plots are normalized to unity and the vertical bars on the points show the statistical uncertainties. The lower panel for the lower left plot gives the ratio of distributions for the middle to the start (blue squares) and the end to the start (red triangles) of the 2017 data-taking period. The lower panel for the lower right plot gives the ratio (blue squares) for the 2018 and 2017 data-taking periods. The vertical bars on the points in both lower panels show the statistical uncertainties in the numerator quantity, and the gray bands give the similar uncertainty in the denominator quantity.}
\label{fig:pi0-stability-conditions}
\end{figure*}

To address these questions using the same $\pgg$ data, we study the
regressed mass versus the mean number of interactions per bunch
crossing. Importantly, because of challenges with the LHC during the
data-taking period used in this study, the proton bunch scheme was
significantly altered in the latter half of the year. This resulted in
the pileup distribution being significantly skewed toward a higher number of interactions per bunch crossing. This effect was not fully modeled in the simulated data used to train the mass regressor. In spite of this, as shown in Fig.~\ref{fig:pi0-stability-conditions} (upper), we observe that both the peak and width of the regressed masses are stable versus the amount of pileup. For various slices of data in the pileup range from 15 to 50, we fit the regressed mass distribution in each slice to a Gaussian-plus-quadratic function in the vicinity of the mass peak. The variation in the position and width of the fitted mass peaks is less than 3\% and 9\%, respectively.

We next investigate the stability of the regressed mass with respect to changing detector conditions by comparing the regressed $\pgg$ mass spectrum at different times. In Fig.~\ref{fig:pi0-stability-conditions} (lower left), the mass spectrum is plotted for the start, middle, and end of the 2017 data-taking period. During these time segments, the electronic noise in the ECAL barrel increased by 25\%. Using the same fitting procedure described earlier, the positions and widths of the peak in the latter two time segments are within 1 (middle) and 4\% (end). These differences are similar to those measured in these datasets using established calibration techniques. As an additional check, in Fig.~\ref{fig:pi0-stability-conditions} (lower right), we show the invariant mass spectrum separately for the entire 2017 and 2018 data-taking periods. Between the ends of these two data-taking periods, the electronic noise in the ECAL barrel increased by about 10\%. The positions and widths of the $\Pgpz$ peak are consistent within uncertainties using a similar fitting procedure.

Robustness to changes in detector conditions is thus a principal strength of the end-to-end ML technique when using minimally processed detector data. However, such robustness is notably degraded when training on clustered data, as discussed in Appendix~\ref{app:aodvminiaod}.

\subsection{Mass dependence on data versus simulation}
\label{subsec:Stability:datavmc}

Finally, we compare the dependence of the mass regressor for events from data versus simulation. If the relative mass resolution in data is worse than it is in simulation, the significance of an observed mass peak in data would be adversely affected. If there is a bias in the position of the mass peak in data versus simulation, an inaccurate mass measurement would be made unless the bias is corrected.

First, it is important to decouple data-versus-simulation mismodeling caused by detector-related effects from that
attributable to the collimation of jets described in Sec.~\ref{subsec:Benchmarks:Data}. An unbiased assessment of the mass regressor requires a measurement of the detector-related mismodeling alone. Because of the computational and logistical challenge of obtaining minimally processed simulated QCD events with energetic $\pgg$, we analyze instead the agreement between data and simulation using electrons from $\zee$ events. Electrons from $\zee$ decays are produced in abundance, and are reconstructed with good isolation and high purity. The tag-and-probe method is used to select the electrons for both data and simulation within the range $\abs{\eta_{\Pe}} < 1.44$.

Although the electron is effectively massless in this energy regime, radiation is produced from its bending in the CMS magnetic field. This causes the energy in the seed crystal of the electron shower to be slightly smeared toward neighboring crystals in the shower~\cite{acat}, similar to the pattern seen in instrumentally merged $\agg$ decays. As a result, the regressed electron mass spectrum displays a peak at $\mapred \approx 0.1\GeV$ that can be used to measure differences in the mass spectrum.

\begin{figure*}[tbp]
\centering
\includegraphics[height=.45\linewidth]{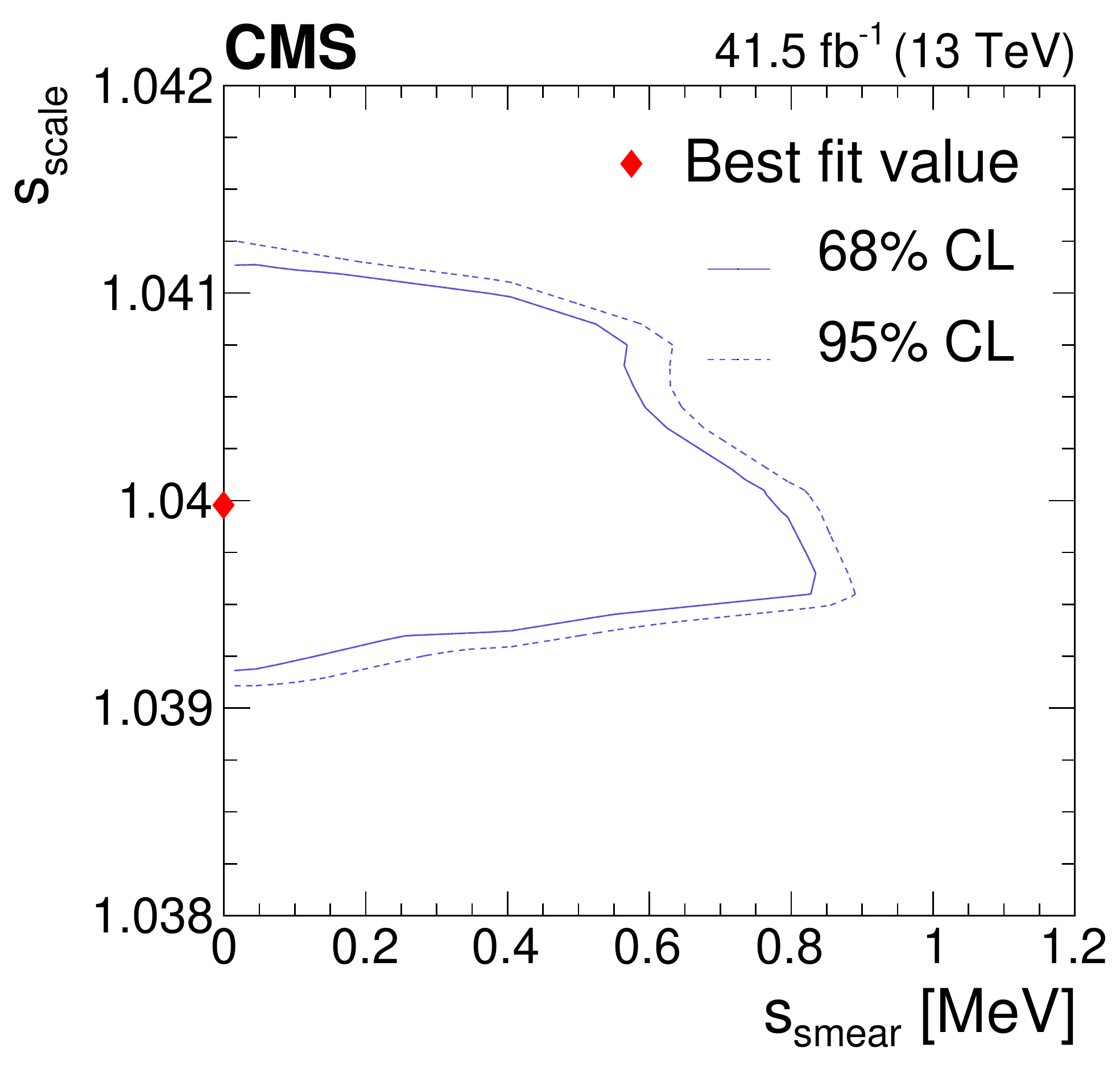}
\qquad
\includegraphics[height=.45\linewidth]{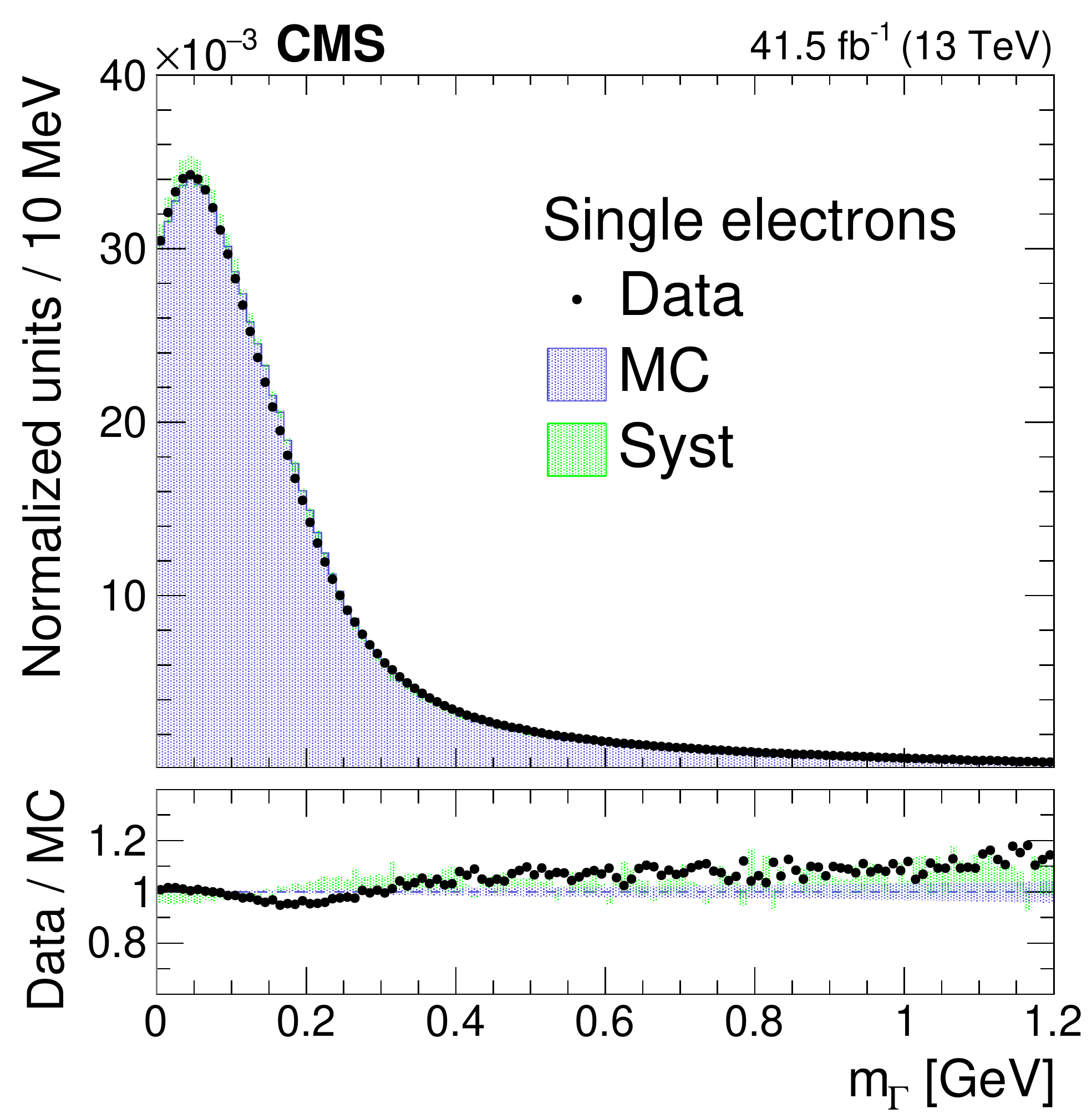}
\caption{The agreement in the regressed $\mG$ spectrum between electrons in data versus simulation. Left: contours of 68\% (solid line) and 95\% (dotted line) confidence level (\CL) in the $\chi^2$ test statistic as a function of the ($\scale$, $\smear$) hypothesis. The best fit point ($\scale=1.040$, $\smear=0\MeV$) is indicated by the red diamond. Right: the regressed mass distributions in data (points) and the best fit Monte Carlo (MC) simulation (blue region) for electrons from $\zee$ events. The difference between the simulated distribution under the null scale and smearing hypothesis versus the best fit hypothesis (Syst) is plotted as a green band. Each of the distributions are normalized to unity, including samples outside $\maroi$. Statistical uncertainties in the data distribution are negligible. The lower panel shows the ratio of the data to the simulation under the best fit hypothesis (points). The statistical uncertainties in the latter are plotted as a blue band. The ratio of the simulated distribution for the null to the best fit hypothesis is displayed as a green band.
}
\label{fig:ele-datavmc}
\end{figure*}

We parametrize the differences between data and simulation in terms of a Gaussian relative mass scale $\scale$ and smearing difference $\smear$. A scan is performed over different ($\scale$, $\smear$) hypotheses, in steps of ($\Delta\scale=4\times10^{-3}$, $\Delta\smear=0.4\MeV$).
At each hypothesis, the value of $m_{\Gamma,\,\mathrm{i}}$ for each electron candidate in the simulated sample is then smeared using the probability distribution $\mathcal{N}(\scale \times m_{\Gamma,\,\mathrm{i}},\,\smear)$, where $\mathcal{N}(\mu, \sigma)$ is a Gaussian function parametrized by mean $\mu$ and standard deviation $\sigma$. The best fit mass scale and smearing between data and simulation is then defined as the ($\scale$, $\smear$) hypothesis for which the chi-square ($\chi^2$) test statistic between the mass distribution in data and the transformed simulated sample is at a minimum. The definition $\chi^2 = \sum_i (h_\mathrm{data,i} - h_\mathrm{MC,i})^2 / h_\mathrm{MC,i}$ is used, where $h_\mathrm{data,i}$ and $h_\mathrm{MC,i}$ denote the normalized data and transformed simulation counts, respectively, at bin $i$ of $\mG$. We determine the best fit value for the relative mass scale to be $1.040$ and the best fit value for the mass difference is less than $\Delta\smear$. The contours of the $\chi^2$ scan over the scale and smearing hypotheses are displayed in Fig.~\ref{fig:ele-datavmc} (left). The resulting regressed mass distributions for electrons in data versus simulation under the best fit hypothesis are shown in Fig.~\ref{fig:ele-datavmc} (right).

After the difference in mass scale is taken into account, we find the
core of the mass distribution to be well modeled in the simulation, as seen in Fig.~\ref{fig:ele-datavmc} (right). Although there are some systematic differences in the high-side tail of the mass distribution between data and simulation, these deviations will be less relevant in an application where data are limited in the tails and statistical uncertainties are larger. Indeed, the significance of an observed mass peak
is primarily driven by the core of the peak and not the tail contribution. Thus, the lack of any significant mass smearing implies the full mass resolution of the regressor is preserved in data.

\section{Summary}
\label{sec:Conclusions}

A novel end-to-end particle reconstruction technique is introduced that is able to serve as a general strategy for reconstructing decays of boosted particles. The method involves the use of deep learning algorithms that do not rely on particle-flow objects, but are trained directly on minimally processed detector-level data to reconstruct particle properties of interest. Using simulated $\agg$ decays in the CMS electromagnetic calorimeter, where $\Pa$ is a hypothetical scalar particle, the technique is used to reconstruct the diphoton invariant mass over a wide range of photon-merging scales, corresponding to Lorentz boosts $\boost=$ 60--600. Furthermore, when domain continuation is incorporated in the training, the most challenging parts of the $\agg$ phase space ($\boost > 150$) are made accessible. The resulting end-to-end mass regressor, in addition to being a highly sensitive tool, also has a robust response. Based on studies using simulated samples and collision data, a stable mass response is observed in various kinematic, beam, and detector conditions.

Studies are under way to employ the end-to-end mass regressor in $\pgg$ reconstruction and in searches for new physics, such as $\PH/\PX \to \Pa\Pa \to 4\gamma$, where $\PH$ is the Higgs boson and $\PX$ is some new heavy resonance. Furthermore, although demonstrated for the specific case of mass reconstruction of a boosted particle decaying to photons, the end-to-end deep learning technique can be used for arbitrary decay modes by including additional subdetector information. The technique is not restricted to mass reconstruction; other particle properties, particularly those that are currently resolution constrained, such as the lifetime of particles in long-lived decays, stand to benefit significantly. It can potentially be used instead of particle-flow techniques to reconstruct the four-momenta of resolved decays. Sensitivity gains in these cases, however, are likely to be more modest, and must be balanced against the challenges of accessing the wider event content of the CMS datasets.

The technique of training via domain continuation can be exploited independently of the end-to-end method. Indeed, it is applied to the training of the photon neural network used as a benchmark. The application of this technique is not specific to high energy physics. It should be applicable in any machine learning regression task that seeks to regress a quantity near a boundary, physical or otherwise, that is close in scale to its resolution.

When end-to-end particle reconstruction is combined with domain continuation, diphoton showers that are completely unresolved can now be reconstructed. This is a regime inaccessible to existing reconstruction techniques, and it is the first time a technique has been developed to achieve this important goal.

\begin{acknowledgments}
  We congratulate our colleagues in the CERN accelerator departments for the excellent performance of the LHC and thank the technical and administrative staffs at CERN and at other CMS institutes for their contributions to the success of the CMS effort. In addition, we gratefully acknowledge the computing centers and personnel of the Worldwide LHC Computing Grid and other centers for delivering so effectively the computing infrastructure essential to our analyses. Finally, we acknowledge the enduring support for the construction and operation of the LHC, the CMS detector, and the supporting computing infrastructure provided by the following funding agencies: BMBWF and FWF (Austria); FNRS and FWO (Belgium); CNPq, CAPES, FAPERJ, FAPERGS, and FAPESP (Brazil); MES and BNSF (Bulgaria); CERN; CAS, MoST, and NSFC (China); MINCIENCIAS (Colombia); MSES and CSF (Croatia); RIF (Cyprus); SENESCYT (Ecuador); MoER, ERC PUT and ERDF (Estonia); Academy of Finland, MEC, and HIP (Finland); CEA and CNRS/IN2P3 (France); BMBF, DFG, and HGF (Germany); GSRI (Greece); NKFIH (Hungary); DAE and DST (India); IPM (Iran); SFI (Ireland); INFN (Italy); MSIP and NRF (Republic of Korea); MES (Latvia); LAS (Lithuania); MOE and UM (Malaysia); BUAP, CINVESTAV, CONACYT, LNS, SEP, and UASLP-FAI (Mexico); MOS (Montenegro); MBIE (New Zealand); PAEC (Pakistan); MES and NSC (Poland); FCT (Portugal); MESTD (Serbia); MCIN/AEI and PCTI (Spain); MOSTR (Sri Lanka); Swiss Funding Agencies (Switzerland); MST (Taipei); MHESI and NSTDA (Thailand); TUBITAK and TENMAK (Turkey); NASU (Ukraine); STFC (United Kingdom); DOE and NSF (USA).  

  \hyphenation{Rachada-pisek} Individuals have received support from the Marie-Curie program and the European Research Council and Horizon 2020 Grant, contract Nos.\ 675440, 724704, 752730, 758316, 765710, 824093, 884104, and COST Action CA16108 (European Union); the Leventis Foundation; the Alfred P.\ Sloan Foundation; the Alexander von Humboldt Foundation; the Belgian Federal Science Policy Office; the Fonds pour la Formation \`a la Recherche dans l'Industrie et dans l'Agriculture (FRIA-Belgium); the Agentschap voor Innovatie door Wetenschap en Technologie (IWT-Belgium); the F.R.S.-FNRS and FWO (Belgium) under the ``Excellence of Science -- EOS" -- be.h project n.\ 30820817; the Beijing Municipal Science \& Technology Commission, No. Z191100007219010; the Ministry of Education, Youth and Sports (MEYS) of the Czech Republic; the Hellenic Foundation for Research and Innovation (HFRI), Project Number 2288 (Greece); the Deutsche Forschungsgemeinschaft (DFG), under Germany's Excellence Strategy -- EXC 2121 ``Quantum Universe" -- 390833306, and under project number 400140256 - GRK2497; the Hungarian Academy of Sciences, the New National Excellence Program - \'UNKP, the NKFIH research grants K 124845, K 124850, K 128713, K 128786, K 129058, K 131991, K 133046, K 138136, K 143460, K 143477, 2020-2.2.1-ED-2021-00181, and TKP2021-NKTA-64 (Hungary); the Council of Science and Industrial Research, India; the Latvian Council of Science; the Ministry of Education and Science, project no. 2022/WK/14, and the National Science Center, contracts Opus 2021/41/B/ST2/01369 and 2021/43/B/ST2/01552 (Poland); the Funda\c{c}\~ao para a Ci\^encia e a Tecnologia, grant CEECIND/01334/2018 (Portugal); the National Priorities Research Program by Qatar National Research Fund; MCIN/AEI/10.13039/501100011033, ERDF ``a way of making Europe", and the Programa Estatal de Fomento de la Investigaci{\'o}n Cient{\'i}fica y T{\'e}cnica de Excelencia Mar\'{\i}a de Maeztu, grant MDM-2017-0765 and Programa Severo Ochoa del Principado de Asturias (Spain); the Chulalongkorn Academic into Its 2nd Century Project Advancement Project, and the National Science, Research and Innovation Fund via the Program Management Unit for Human Resources \& Institutional Development, Research and Innovation, grant B05F650021 (Thailand); the Kavli Foundation; the Nvidia Corporation; the SuperMicro Corporation; the Welch Foundation, contract C-1845; and the Weston Havens Foundation (USA).

\end{acknowledgments}

\clearpage
\bibliography{auto_generated}  

\appendix
\numberwithin{figure}{section}
\section{Supplementary studies}
\label{sec:Appendix}

\renewcommand{\thefigure}{A.\arabic{figure}}

\subsection{Minimally processed versus clustered data}
\label{app:aodvminiaod}

\begin{figure*}[!htbp]
\centering
\includegraphics[height=.38\linewidth]{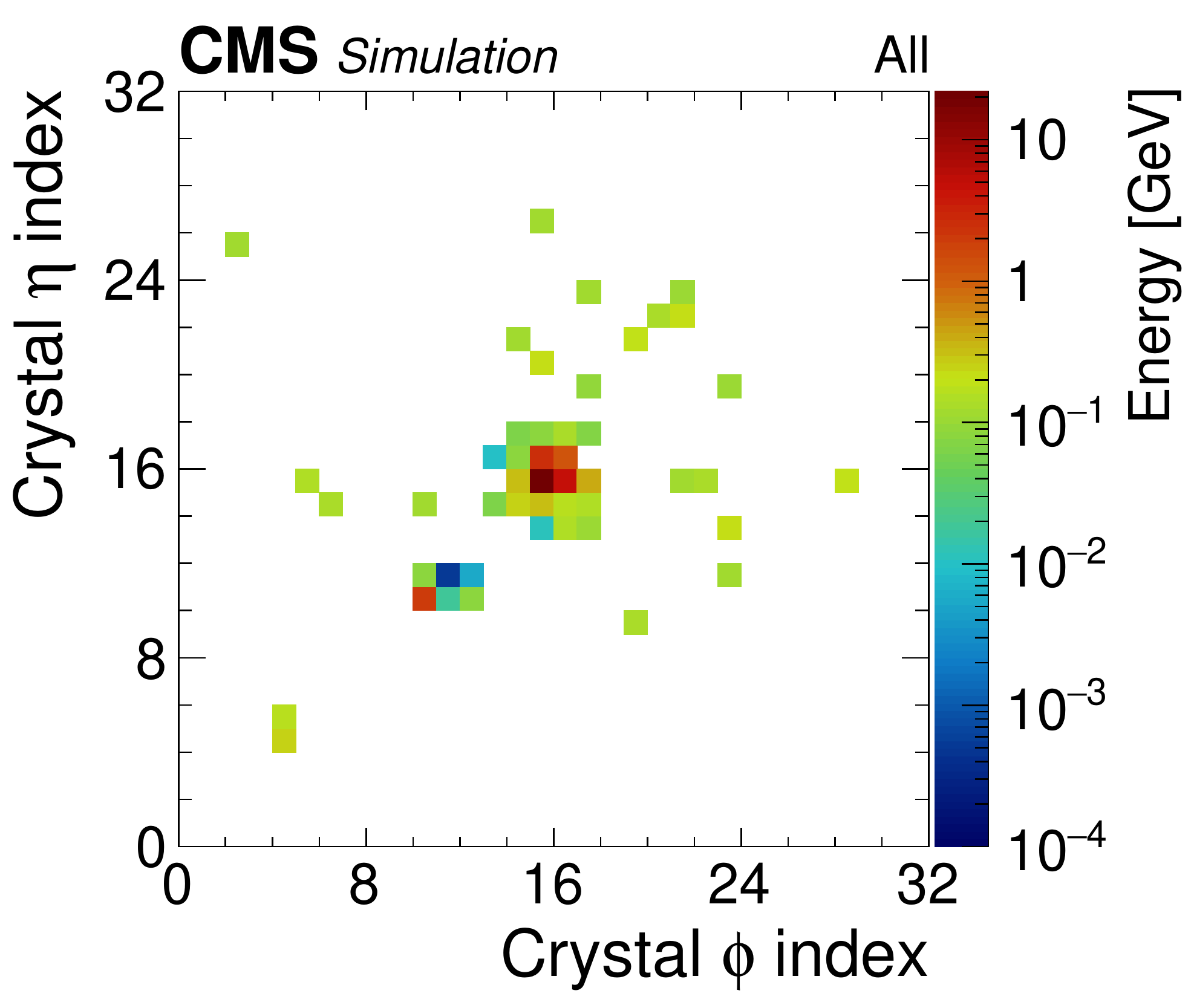}
\qquad
\includegraphics[height=.38\linewidth]{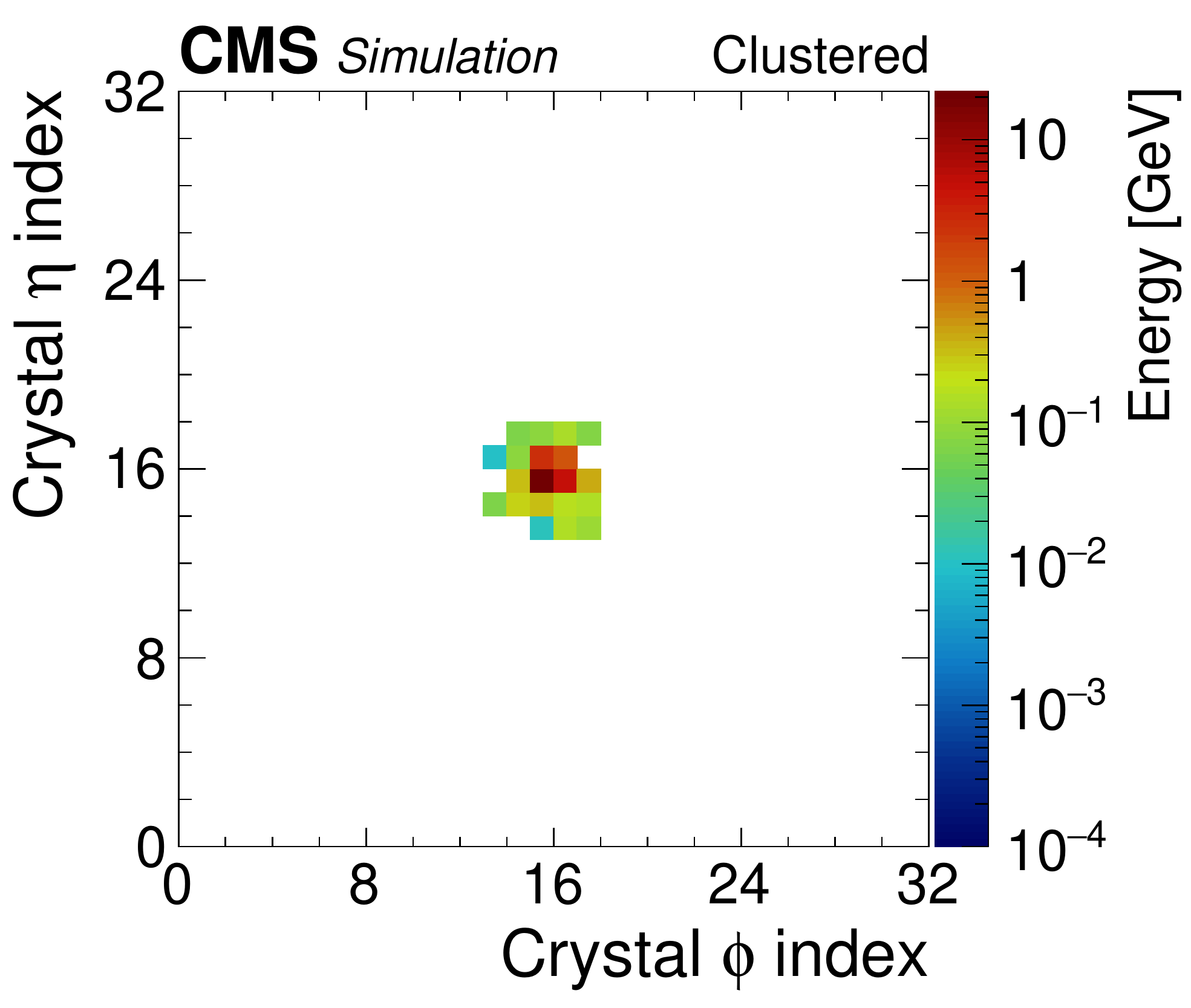}
\caption{A typical $\agg$ decay using minimally processed (left) and clustered (right) data. The energy distributions are plotted in relative ECAL crystal index coordinates of the pseudorapidity $\eta$ versus the azimuthal angle $\phi$ and color coded by energy.}
\label{fig:droppedhits}
\end{figure*}

We attribute the robustness of the end-to-end ML mass regressor (discussed in Sec.~\ref{sec:Stability}) to the use of minimally processed (\textit{all}) rather than clustered (\textit{clustered}) detector data. As shown in Fig.~\ref{fig:droppedhits}, the PF clustering algorithm filters out low-energy deposits and, under certain situations, may completely remove all the deposits associated with the lower-energy photon from the $\agg$ decay. For example, in Fig.~\ref{fig:droppedhits} (left), showing the minimally processed data, the lower-energy photon is visible on the lower left of the core photon, at a distance of approximately $7$ ECAL crystals. In the clustered data on the right plot, the deposits associated with the lower-energy photon have been dropped, along with other, isolated low-energy deposits.

\begin{figure*}[!htbp]
\centering
\includegraphics[width=.4\linewidth]{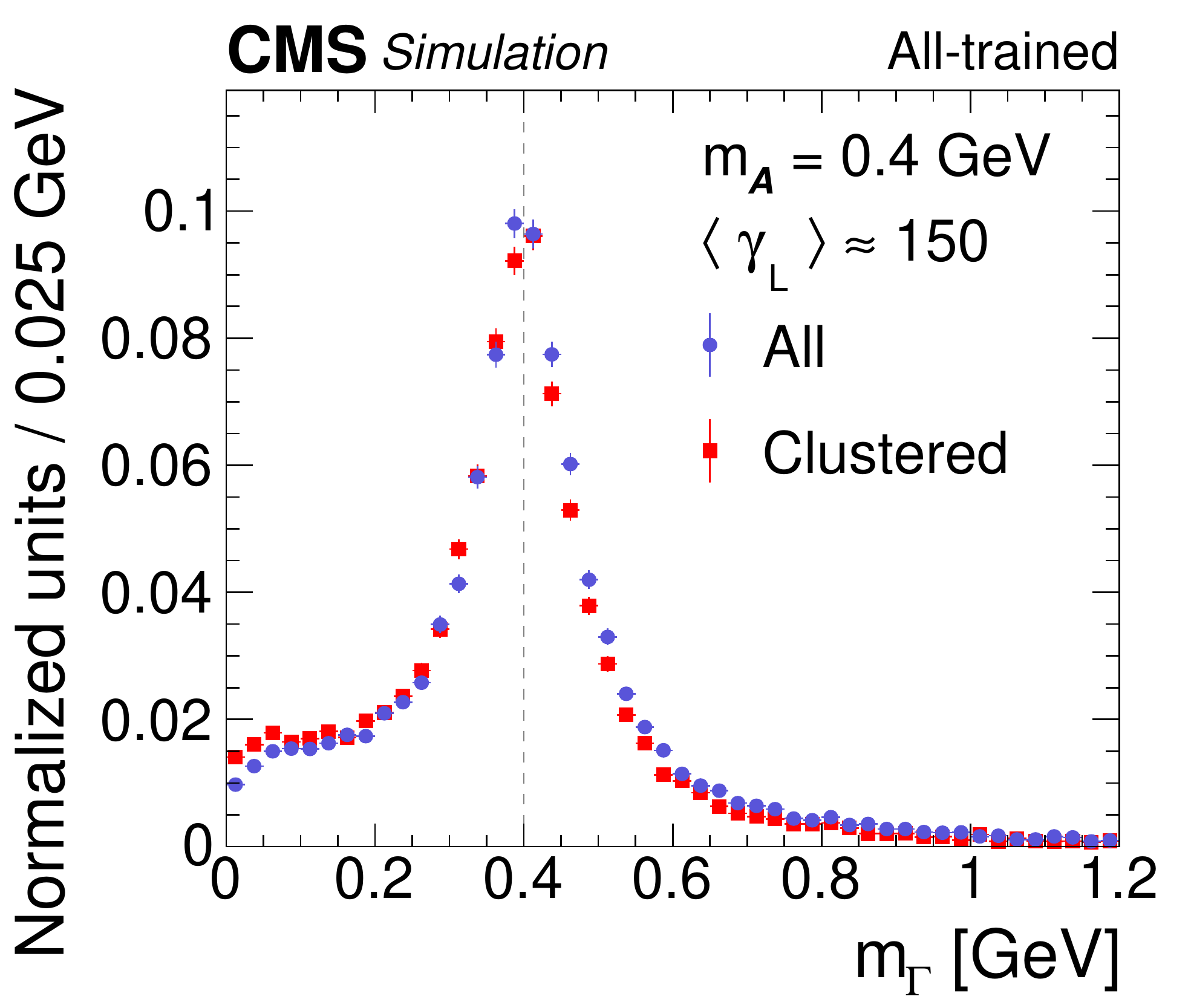}
\qquad
\includegraphics[width=.4\linewidth]{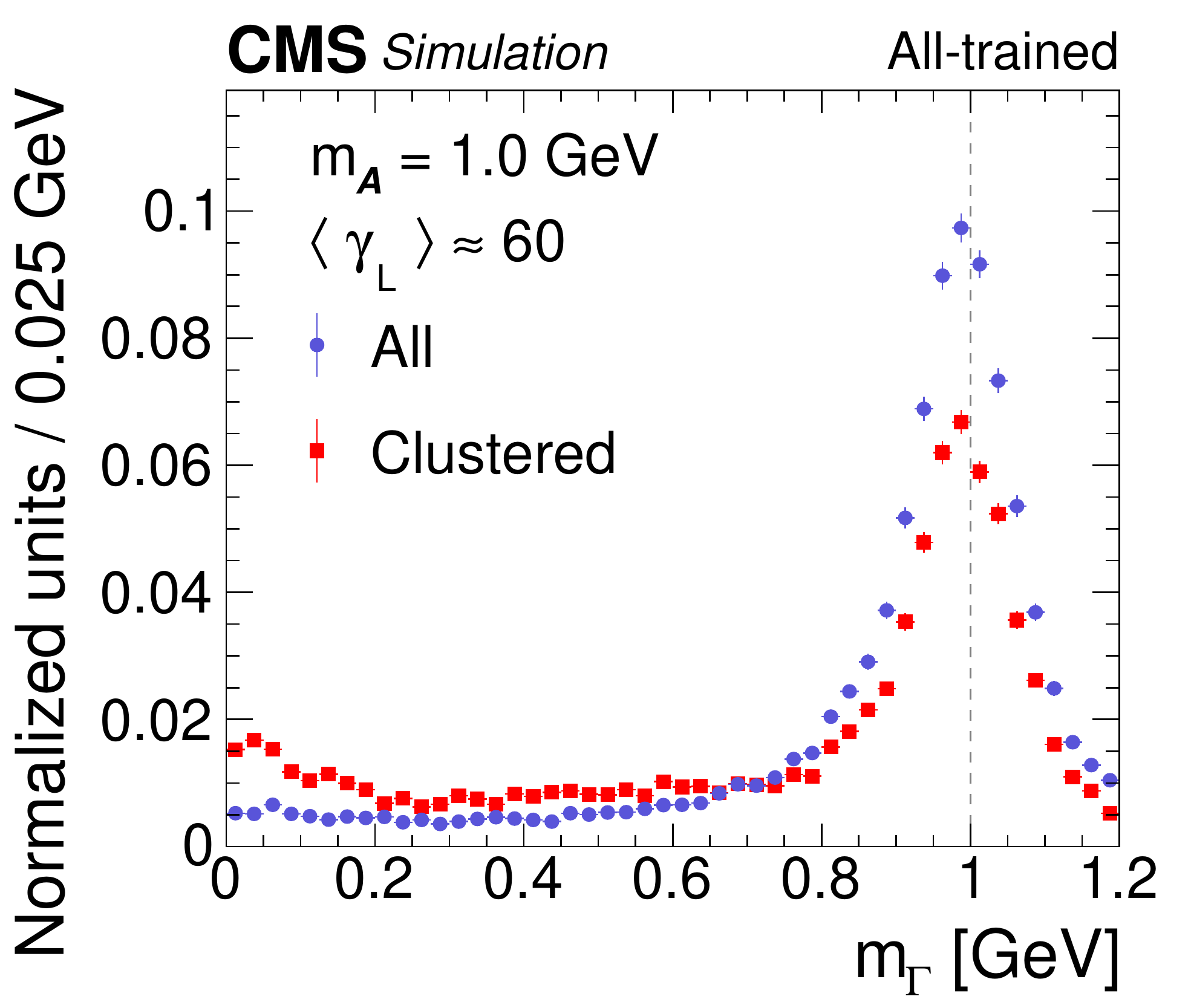}
\\
\includegraphics[width=.4\linewidth]{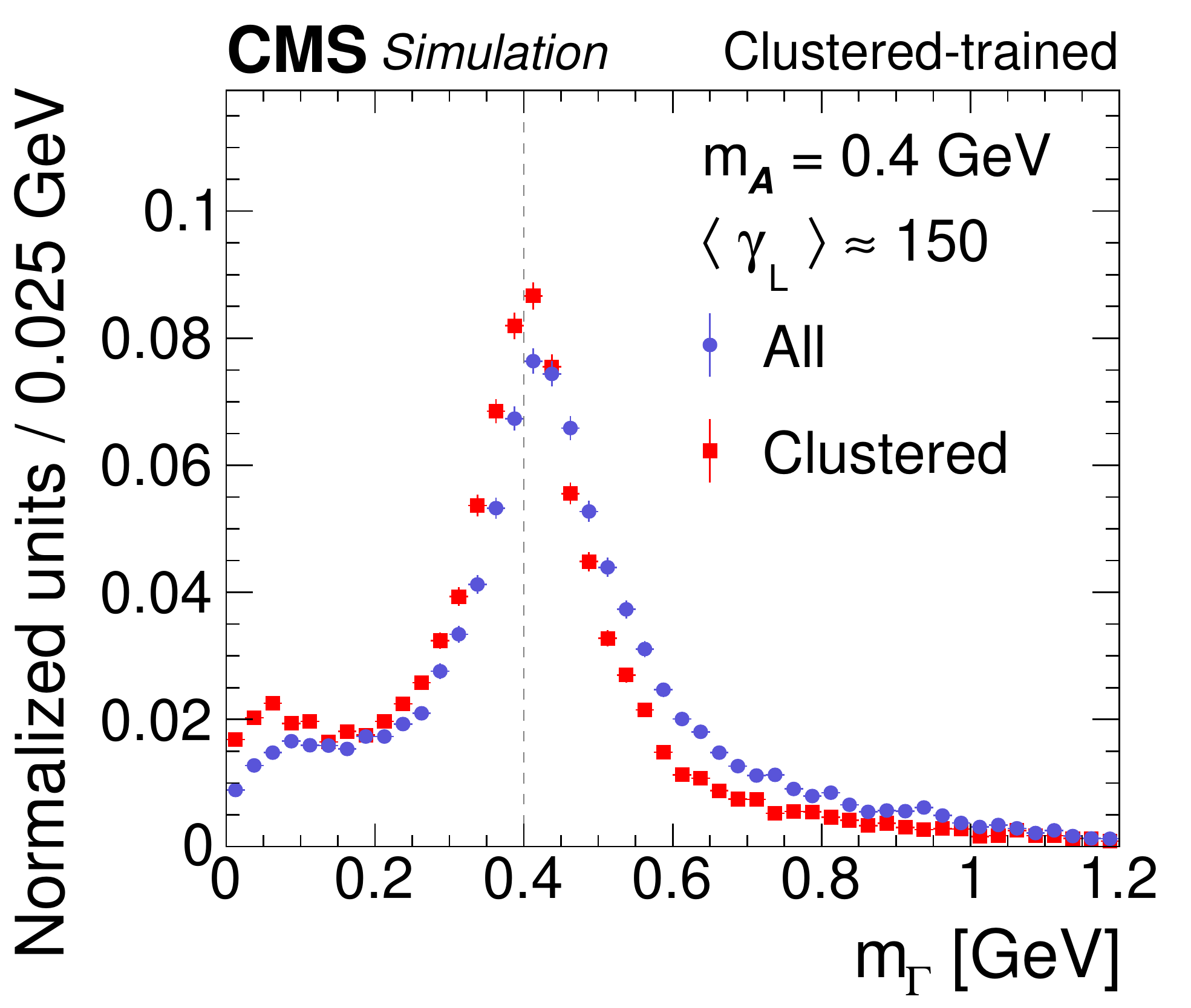}
\qquad
\includegraphics[width=.4\linewidth]{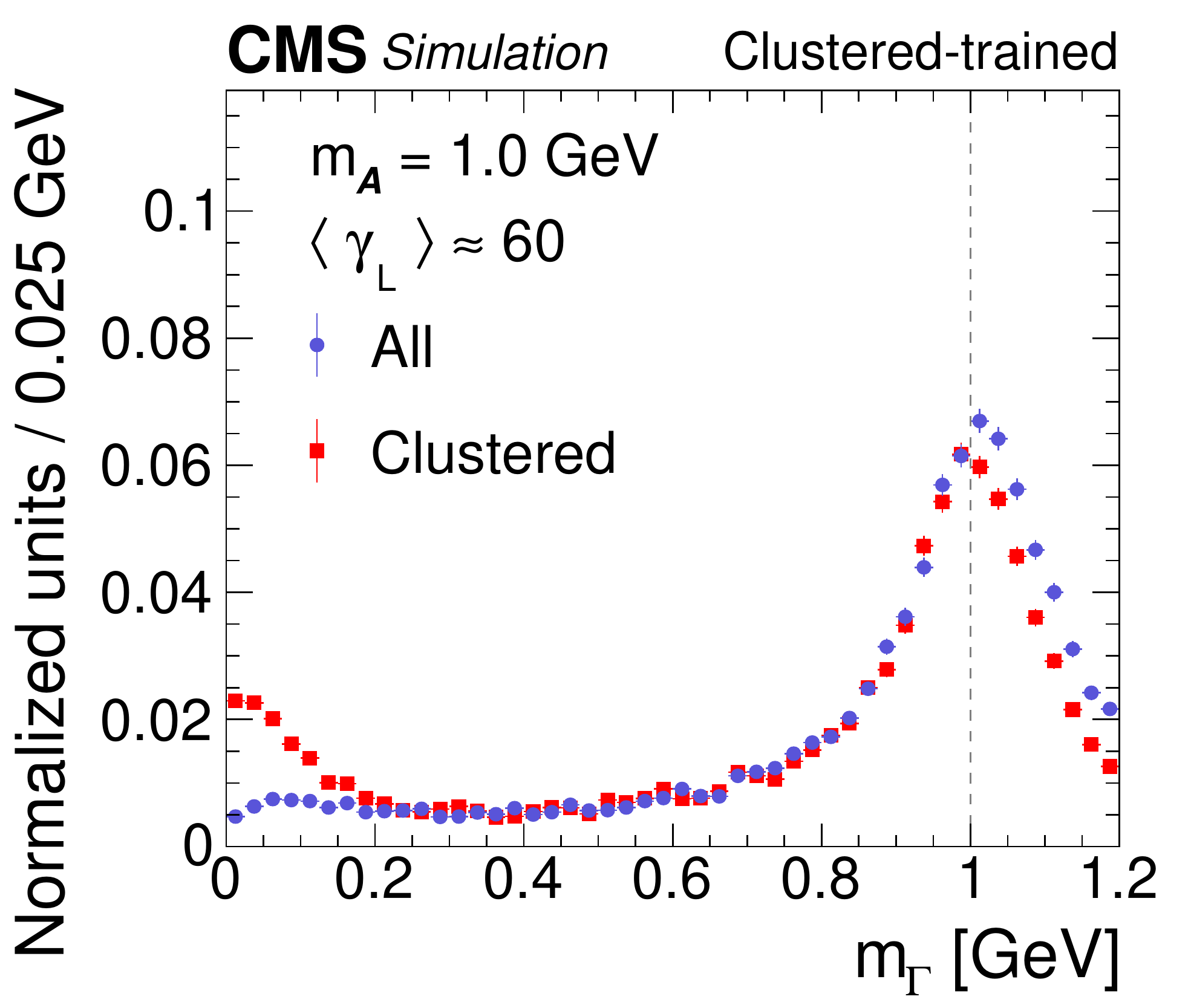}
\caption{Regressed mass spectra for the mass regressor trained on minimally processed data (upper) versus clustered data (lower) at shower-merged boosts (left) and barely resolved boosts (right). For each scenario, the mass regressor is run on the same set of $\agg$ decays, composed either of minimally processed (blue circles, all) or clustered (red squares, clustered) data. The vertical bars on the points give the statistical uncertainties and the vertical dotted line shows the input $\ma$ value.
}
\label{fig:aodvsminiaod-trained}
\end{figure*}

The impact of using minimally processed versus clustered data is seen in Fig.~\ref{fig:aodvsminiaod-trained}, which compares the effect of training on all (upper row) versus clustered data (lower row). The regressed mass spectra for shower-merged and barely resolved boosts are displayed in the left and right columns, respectively. For each scenario, we regress the mass of a sample constructed from all (blue circles) versus clustered (red squares) data to compare how well each mass regressor extrapolates to the other's domain. For the mass regressor trained on minimally processed data (upper row), despite the differences in input image (cf. Fig.~\ref{fig:droppedhits}), we see no evidence of a shift in the mass peak in either boost regime. This is not the case for the mass regressor trained on clustered data (lower row), which exhibits a shift when applied outside of its domain. This suggests it is desirable not to filter the detector data beforehand so that the mass regressor learns to suppress low-energy deposits. Without this opportunity, the mass regressor becomes more susceptible to variations in the data.

In addition, the loss of the lower-energy photon at resolved boosts (right column) leads to $\agg$ decays being incorrectly reconstructed as photons, causing both a drop in reconstruction efficiency at the correct mass value ($\mapred = 1\GeV$) and a buildup of samples at the wrong mass ($\mapred \approx 0\GeV$). Even if minimally processed data are presented to the mass regressor trained on clustered data (lower right, blue circles), the lost efficiency at the correct mass value is still not recovered. The use of minimally processed data is thus vital to maximizing the capabilities of the end-to-end ML-based technique.

\subsection{Hadronization effects}
\label{app:shower3b3}

\begin{figure*}[!htbp]
\centering
\includegraphics[height=.38\linewidth]{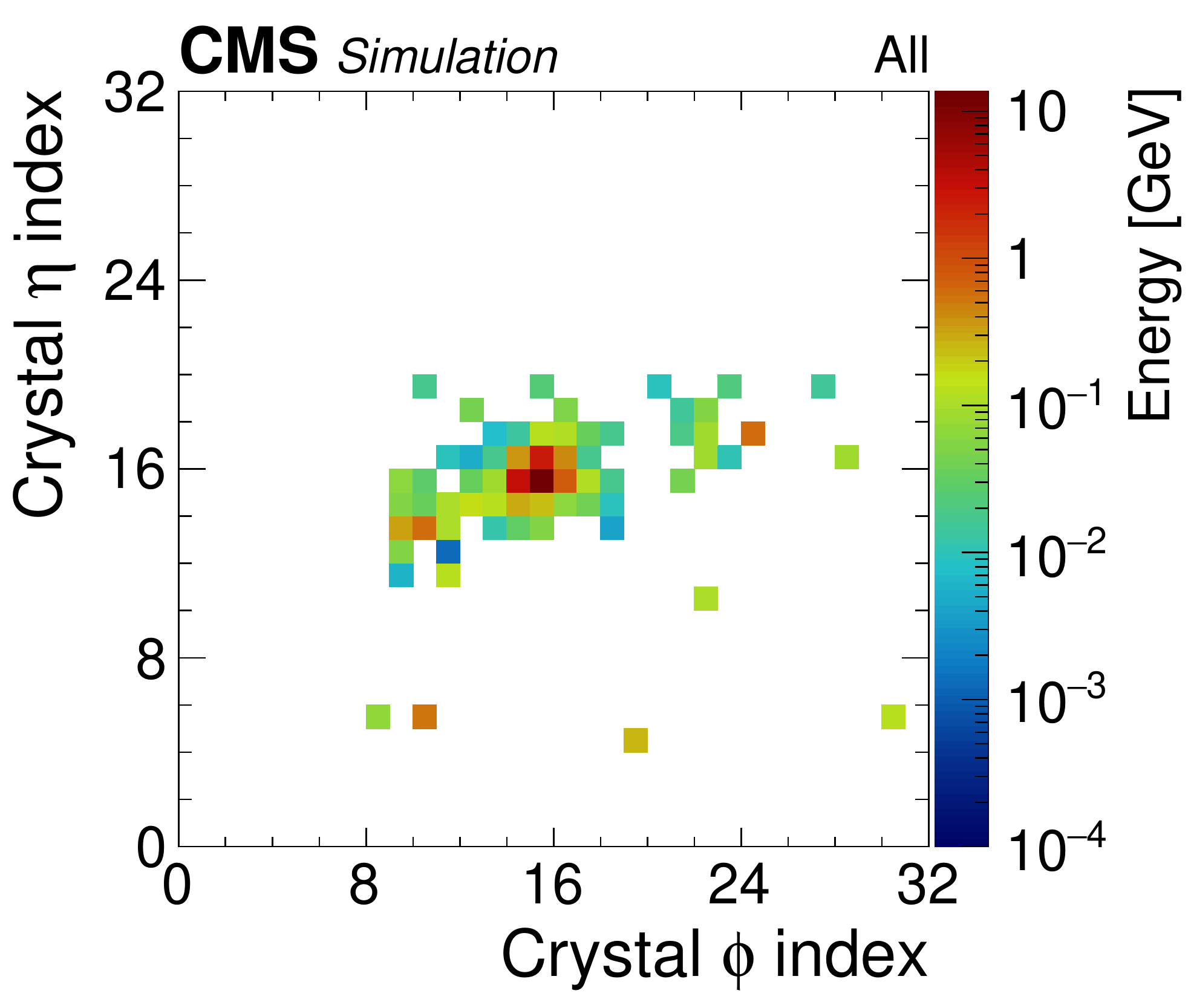}
\qquad
\includegraphics[height=.38\linewidth]{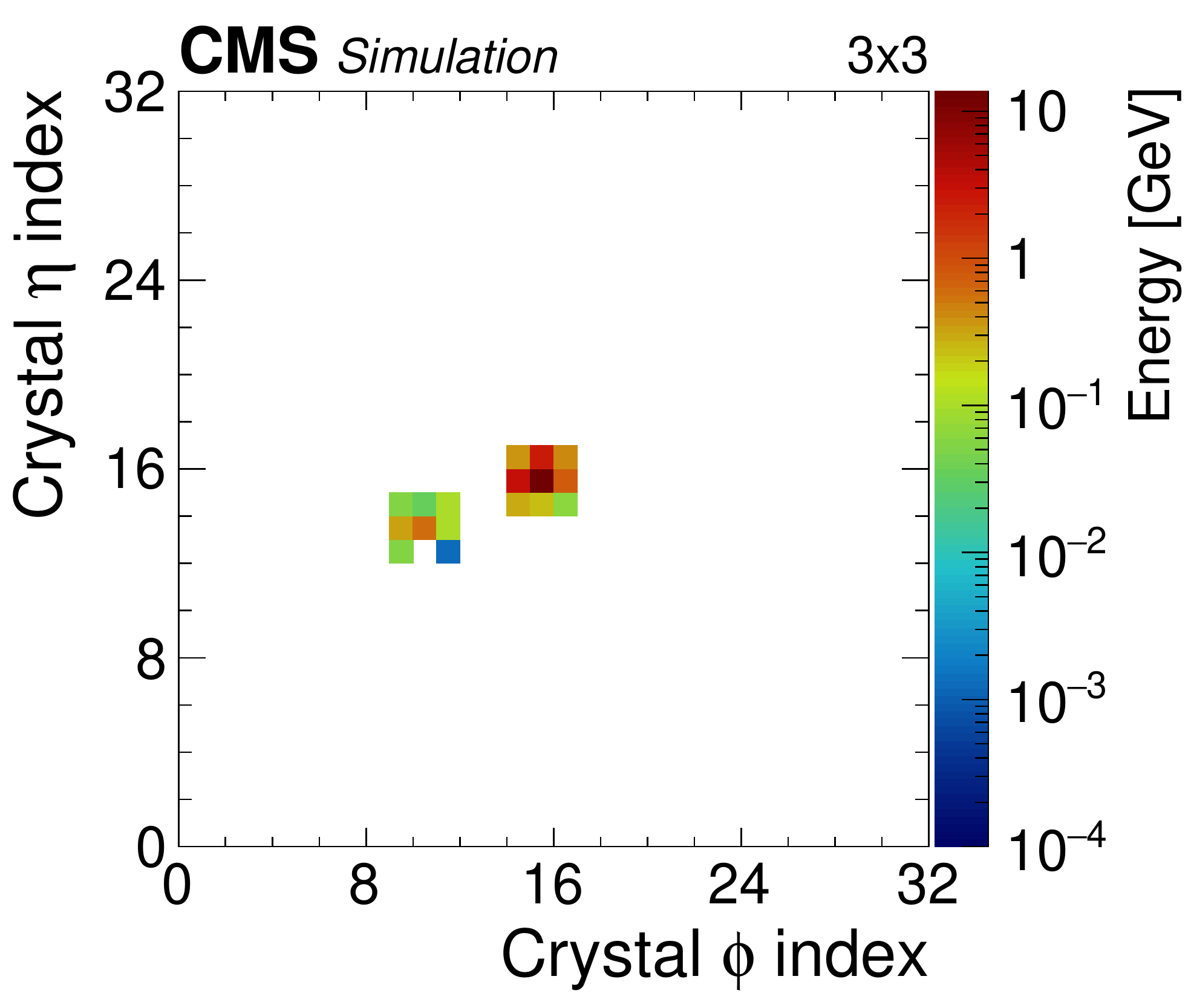}
\caption{A typical hadronic jet sample reconstructed by the \tbt algorithm with a mass near the $\Pgh$ meson peak, using minimally processed (\textit{all}, left) and \tbt clustered ($\tbtit$, right) data. The energy distributions are plotted in relative ECAL crystal index coordinates of the azimuthal angle $\phi$ versus the pseudorapidity $\eta$ and color coded by energy.}
\label{fig:allv3b3}
\end{figure*}

Although the end-to-end method is able to clearly reconstruct $\Pgpz$ candidates in hadronic jets, the $\Pgh$ resonance in jets appears much less pronounced compared with that reconstructed by the \tbt method (cf. Fig.~\ref{fig:pi0-benchmark}). At the particle energies that we study, the weaker $\Pgh$ resonance from the end-to-end method is due to the presence of additional particles collimated in the jet. As noted in Sec.~\ref{subsec:Benchmarks:Data}, because the \tbt method uses only a window of \tbt crystals, it is less exposed to these effects. To illustrate this point, in Fig.~\ref{fig:allv3b3}, we select a representative merged photon candidate reconstructed by the \tbt algorithm with a mass close to the $\Pgh$ resonance.

The ECAL energy pattern associated with the minimally processed detector data (\textit{all}), as used by the end-to-end technique, is plotted on the left. The corresponding energy pattern, after applying the \tbt clustering ($\tbtit$), is shown on the right. We have verified that, by applying the end-to-end method on candidates within a mass window of the $\Pgh$ peak built using only \tbt clusters, we are able to reconstruct an $\Pgh$ mass peak of similar size as for the \tbt algorithm. The effects of hadronization are also relevant for $\Pgpz$ reconstruction in jets. However, the higher production rate of $\Pgpz$ mesons permits the use of tighter photon identification criteria (cf. Sec.~\ref{subsec:Benchmarks:Data}), in order to maximize the isolated $\Pgpz$-like component.

\subsection{Mass dependence in simulated samples}
\label{app:h4g-stability}

\begin{figure*}[!htbp]
\centering
  \includegraphics[width=.328\linewidth]{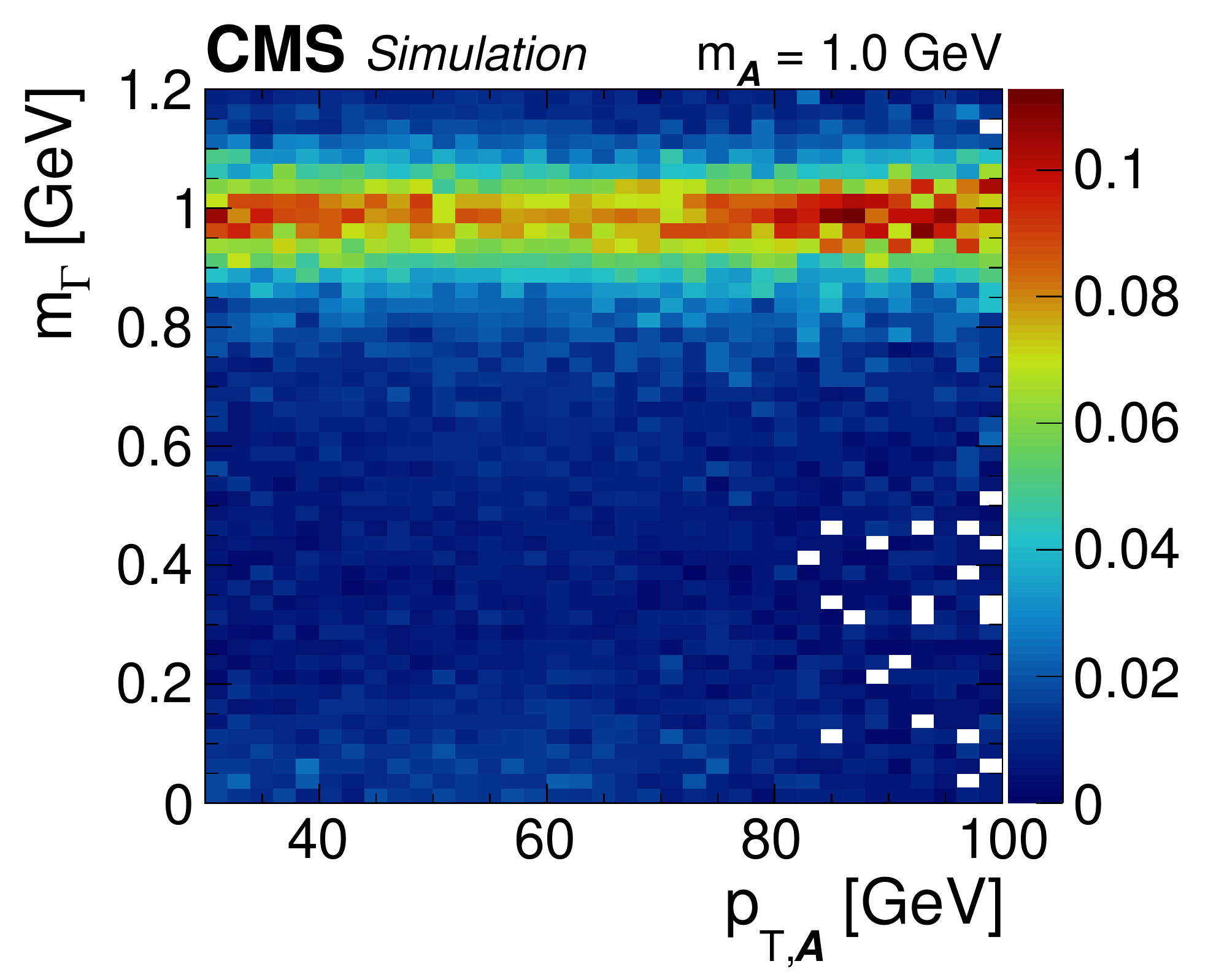}
  \includegraphics[width=.328\linewidth]{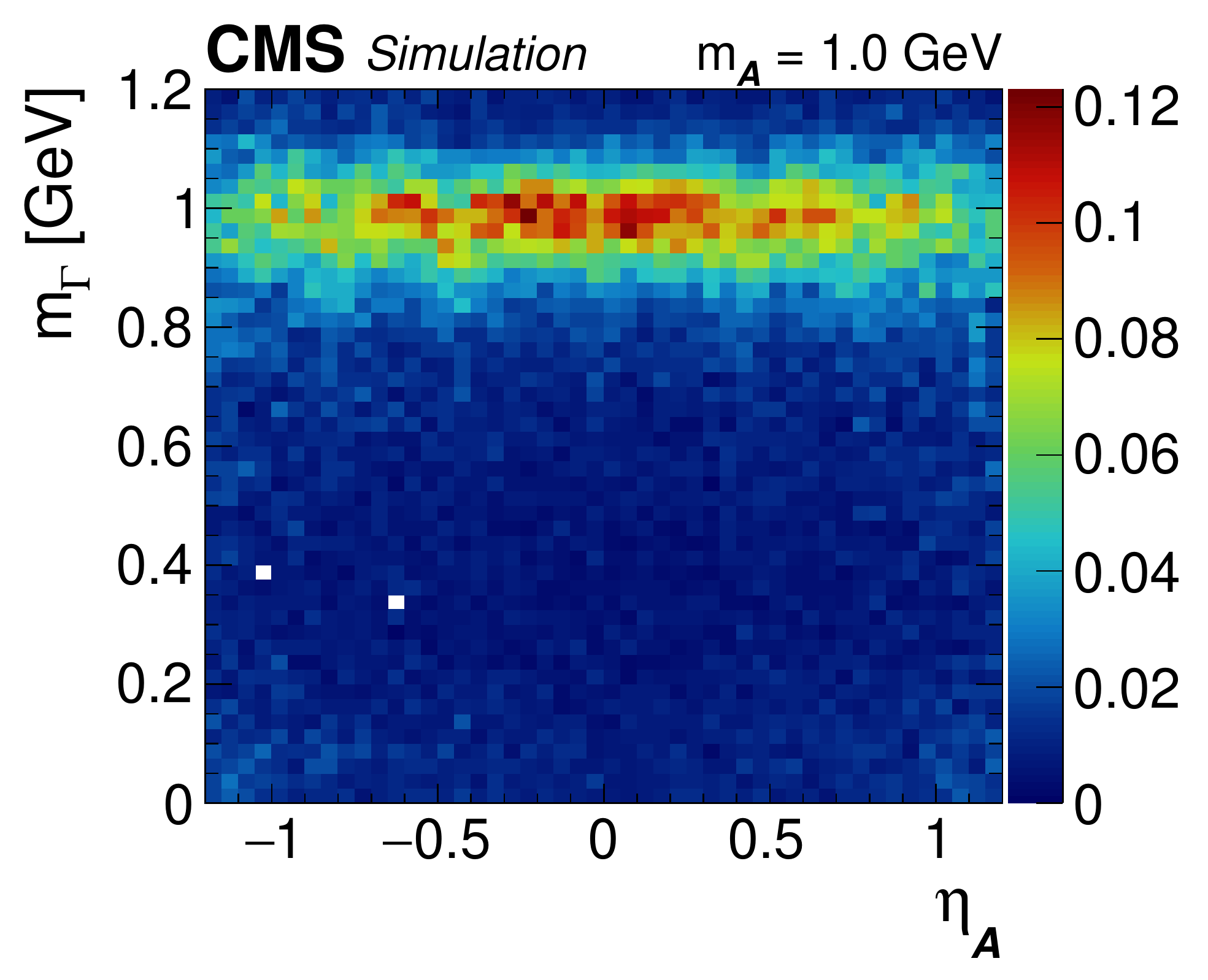}
  \includegraphics[width=.328\linewidth]{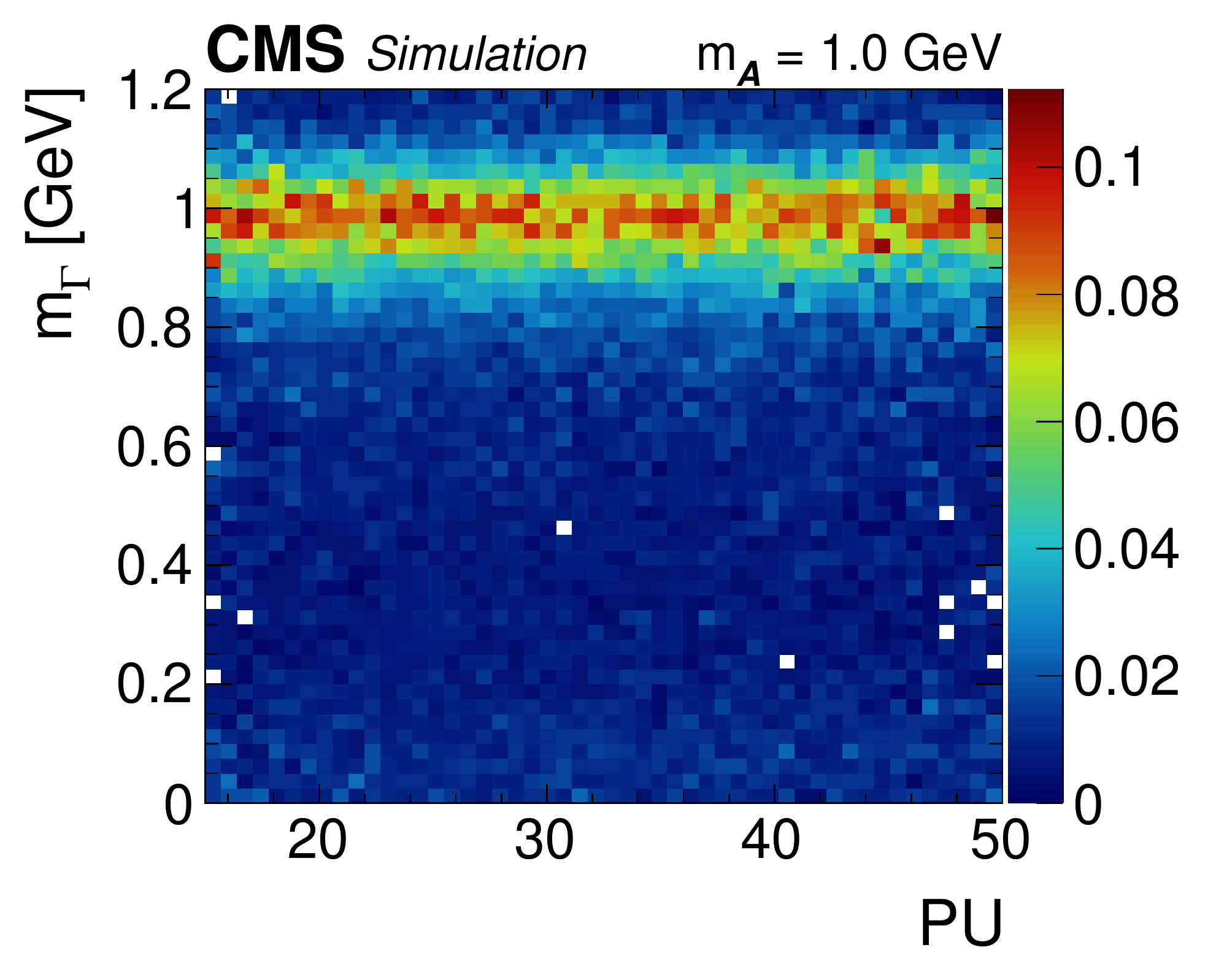}
\\
  \includegraphics[width=.328\linewidth]{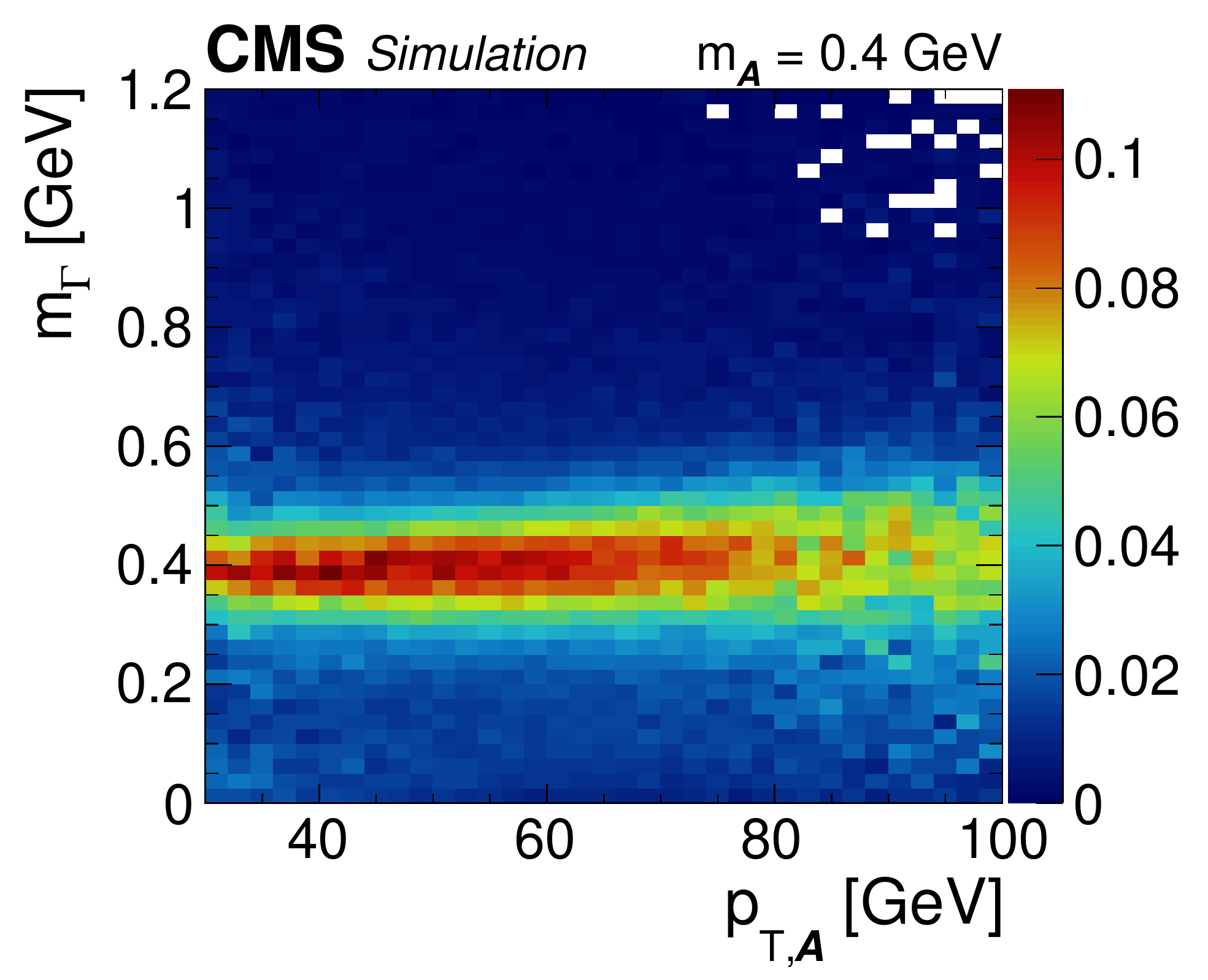}
  \includegraphics[width=.328\linewidth]{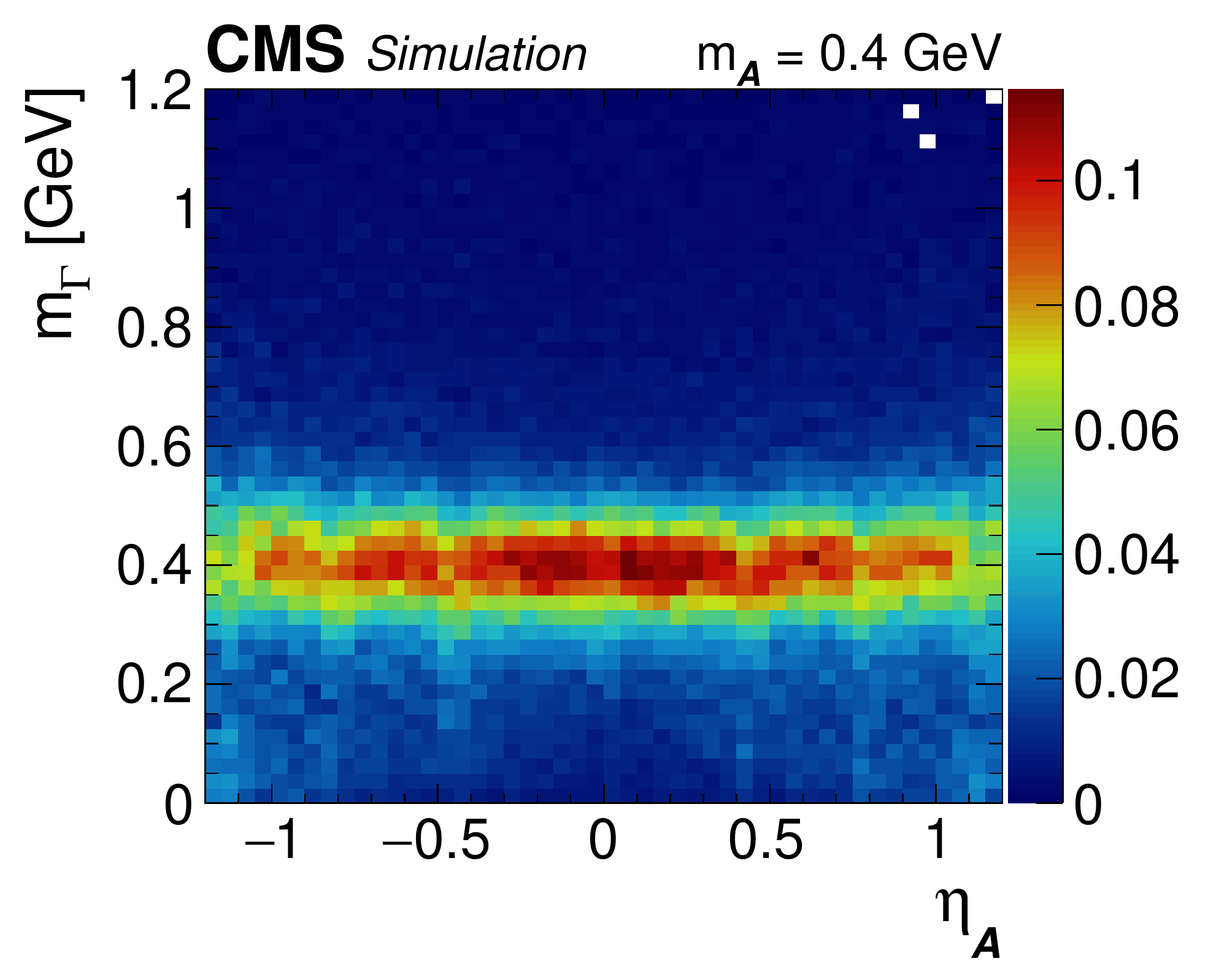}
  \includegraphics[width=.328\linewidth]{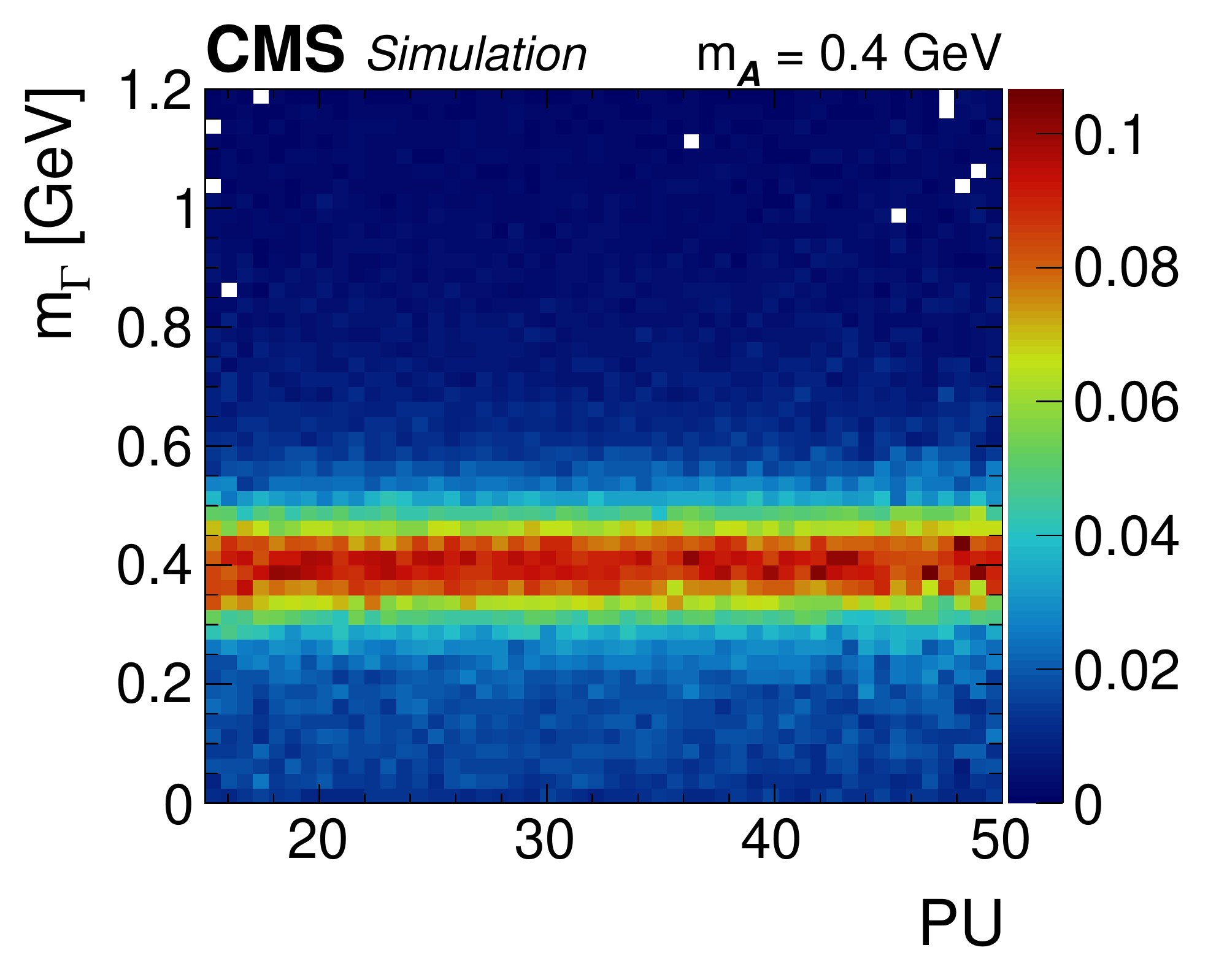}
\\
  \includegraphics[width=.328\linewidth]{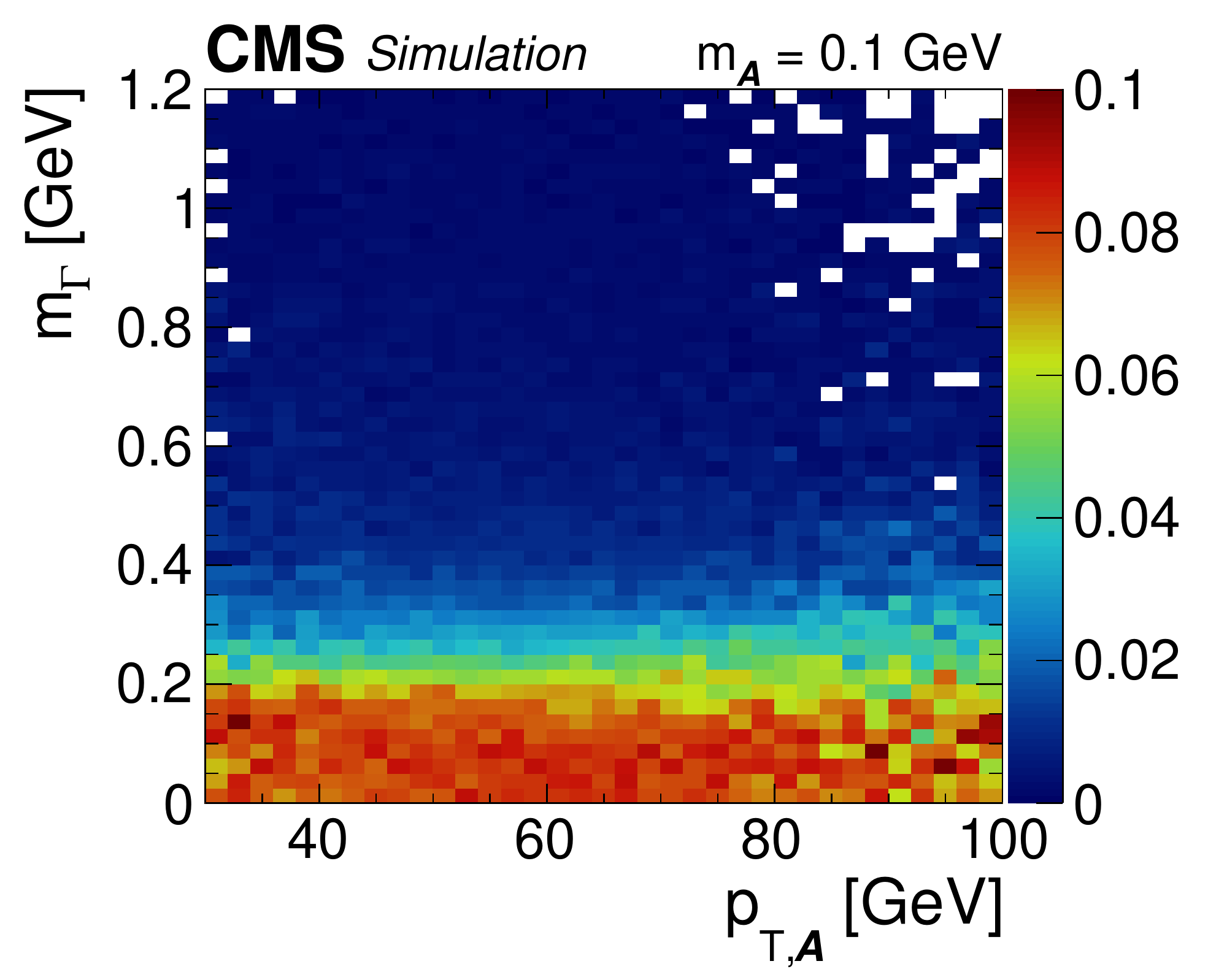}
  \includegraphics[width=.328\linewidth]{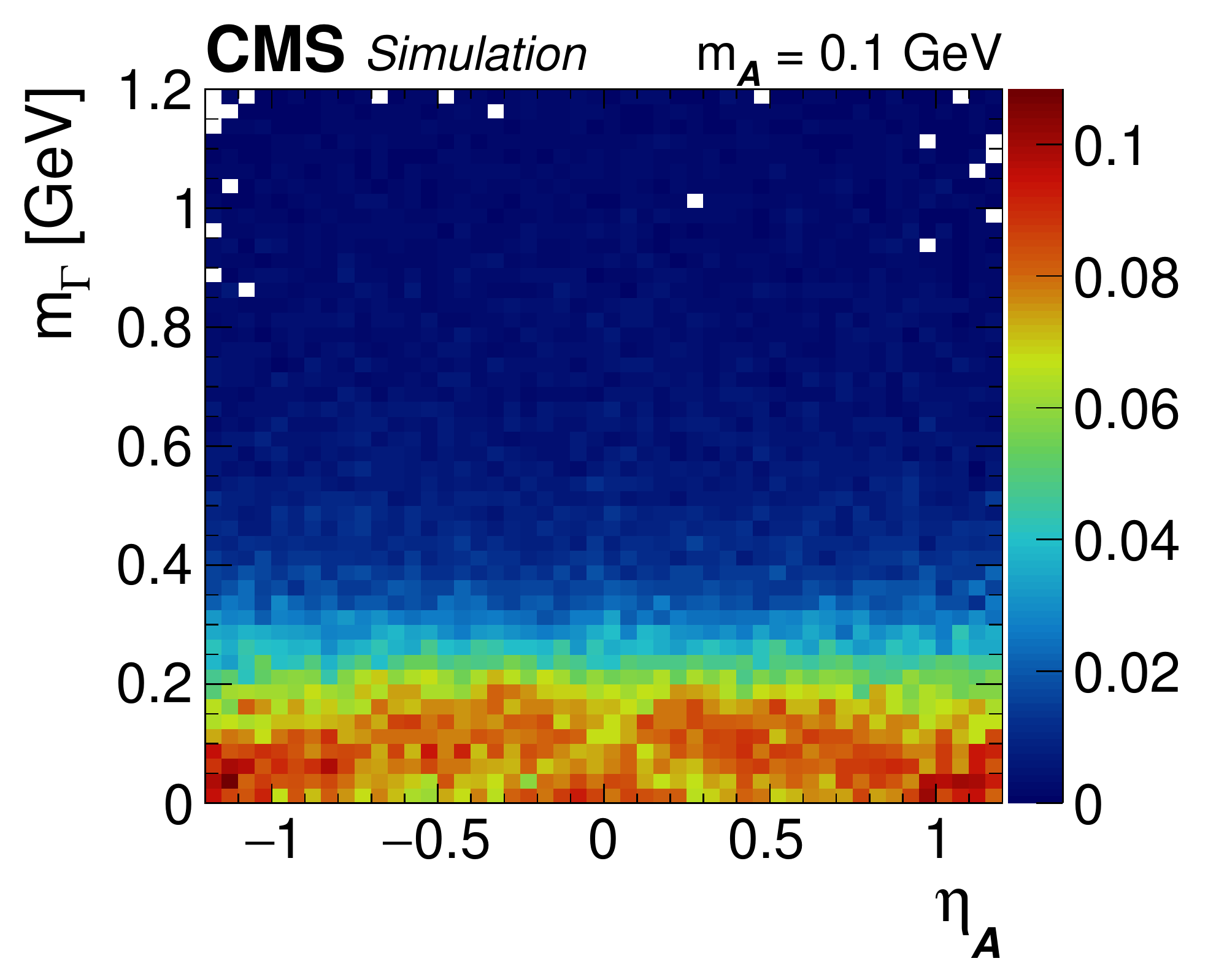}
  \includegraphics[width=.328\linewidth]{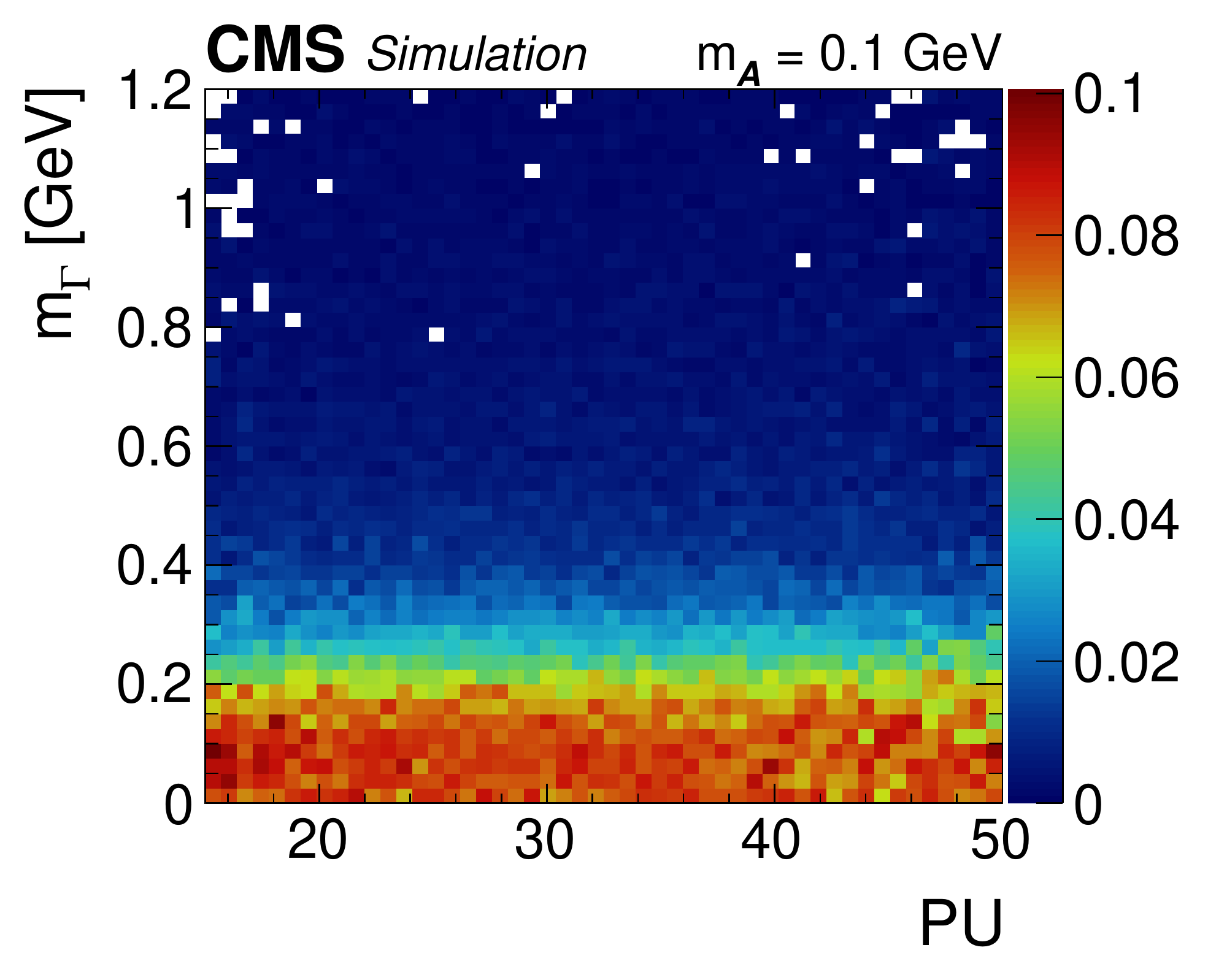}
\\
\caption{Regressed mass spectra~vs.~generated $\ptagen$ (left), generated $\etaagen$ (center), and amount of pileup (right), for $\agg$ decays that are barely resolved (upper), shower merged (middle), or instrumentally merged (lower). In all plots, the regressed mass distribution is normalized in vertical slices of the quantity of interest. The relative contribution over each vertical slice is given by the color scale to the right of each plot.}
\label{fig:h4g-stability}
\end{figure*}

To complement the mass dependence studies performed on data using $\pgg$ decays and electrons, we present in Fig.~\ref{fig:h4g-stability} similar studies for $\agg$ decays from simulated $\hag$ events passing our event and photon selection criteria. Two-dimensional plots are shown of the regressed mass versus the generated $\ptagen$ (left), $\etaagen$ (center), and amount of pileup (right) for
barely resolved (upper), shower merged (middle), and instrumentally merged (lower) decay products. There is good stability in the regressed mass response throughout the explored phase space and the varying beam conditions.

\cleardoublepage \section{The CMS Collaboration \label{app:collab}}\begin{sloppypar}\hyphenpenalty=5000\widowpenalty=500\clubpenalty=5000
\cmsinstitute{Yerevan~Physics~Institute, Yerevan, Armenia}
{\tolerance=6000
A.~Tumasyan\cmsAuthorMark{1}\cmsorcid{0009-0000-0684-6742}
\par}
\cmsinstitute{Institut~f\"{u}r~Hochenergiephysik, Vienna, Austria}
{\tolerance=6000
W.~Adam\cmsorcid{0000-0001-9099-4341}, J.W.~Andrejkovic, T.~Bergauer\cmsorcid{0000-0002-5786-0293}, S.~Chatterjee\cmsorcid{0000-0003-2660-0349}, K.~Damanakis\cmsorcid{0000-0001-5389-2872}, M.~Dragicevic\cmsorcid{0000-0003-1967-6783}, A.~Escalante~Del~Valle\cmsorcid{0000-0002-9702-6359}, R.~Fr\"{u}hwirth\cmsAuthorMark{2}\cmsorcid{0000-0002-0054-3369}, M.~Jeitler\cmsAuthorMark{2}\cmsorcid{0000-0002-5141-9560}, N.~Krammer\cmsorcid{0000-0002-0548-0985}, L.~Lechner\cmsorcid{0000-0002-3065-1141}, D.~Liko\cmsorcid{0000-0002-3380-473X}, I.~Mikulec\cmsorcid{0000-0003-0385-2746}, P.~Paulitsch, F.M.~Pitters, J.~Schieck\cmsAuthorMark{2}\cmsorcid{0000-0002-1058-8093}, R.~Sch\"{o}fbeck\cmsorcid{0000-0002-2332-8784}, D.~Schwarz\cmsorcid{0000-0002-3821-7331}, S.~Templ\cmsorcid{0000-0003-3137-5692}, W.~Waltenberger\cmsorcid{0000-0002-6215-7228}, C.-E.~Wulz\cmsAuthorMark{2}\cmsorcid{0000-0001-9226-5812}
% ...
\par}
\vskip\cmsinstskip
\dag:~Deceased\\
$^{1}$Also at YEREVAN STATE UNIVERSITY, Yerevan, Armenia\\
$^{2}$Also at TU Wien, Vienna, Austria\\
% ...
\end{sloppypar}
\end{document}